\pdfoutput=1

\documentclass[twocolumn,times,tighten,numberedappendix,appendixfloats]{aastex63}
\usepackage{epstopdf}	
\usepackage{graphicx}	
\usepackage{latexsym}	
\usepackage{bm}		
\usepackage[english]{babel}
\usepackage{booktabs}
\usepackage{amsmath}	
\usepackage[Symbol]{upgreek}	
\usepackage{natbib}	
\usepackage{longtable}
\usepackage{array}
\usepackage{blindtext}

\newlength{\txw}
\newlength{\txh}

\newcommand{\fesc}   {\ensuremath{f_{\rm esc}}}

\newcommand{\fescrel}{\ensuremath{f_{\rm esc}^{\rm rel}}}

\newcommand{\ie}     {i.e.,}

\newcommand{\HST}    {\emph{HST}}
\newcommand{\JWST}   {\emph{JWST}}
\newcommand{\GALEX}  {\emph{GALEX}}
\newcommand{\Chandra}{\emph{Chandra}}
\newcommand{\AV}     {\ensuremath{A_{V}}}
\newcommand{\NH}     {\ensuremath{N_{H}}}

\newcommand{\Lya}    {\ensuremath{\mbox{\textrm{Ly}}\alpha}}

\newcommand{\mAB}    {\ensuremath{m_{\text{\tiny AB}}}}
\newcommand{\MAB}    {\ensuremath{M_{\rm AB}}}
\newcommand{\zmean}  {\ensuremath{\langle z\rangle}}

\newcommand{\igm}    {\textnormal{\tiny \textsc{IGM}}}
\newcommand{\lyc}    {\textnormal{\fontsize{6}{6} \textsc{L}\text{y}\textsc{C}}}
\newcommand{\uvc}    {\textnormal{\tiny \text{UVC}}}
\newcommand{\SExtractor}{\textsc{SExtractor}}
\newcommand{\PreserveBackslash}[1]{\let\temp=\\#1\let\\=\temp}
\newcolumntype{C}[1]{>{\PreserveBackslash\centering}p{#1}}
\newcolumntype{R}[1]{>{\PreserveBackslash\raggedleft}p{#1}}
\newcolumntype{L}[1]{>{\PreserveBackslash\raggedright}p{#1}}

\defcitealias{Smith2018}{S18}

\accepted{April 7, 2020}
\submitjournal{ApJ}

\begin{document}

\title{The Lyman Continuum Escape Fraction of Galaxies and AGN in the GOODS Fields}

\correspondingauthor{Brent M. Smith}\author[0000-0002-0648-1699]{Brent M. Smith}\affiliation{School of Earth \& Space Exploration, Arizona State University, Tempe, AZ 85287-1404, USA}\email{bsmith18@asu.edu}
\author[0000-0001-8156-6281]{Rogier A. Windhorst}\affiliation{School of Earth \& Space Exploration, Arizona State University, Tempe, AZ 85287-1404, USA}
\author[0000-0003-3329-1337]{Seth H. Cohen}\affiliation{School of Earth \& Space Exploration, Arizona State University, Tempe, AZ 85287-1404, USA}
\author[0000-0002-6610-2048]{Anton M. Koekemoer}\affiliation{Space Telescope Science Institute, Baltimore, MD 21218, USA}
\author[0000-0003-1268-5230]{Rolf A. Jansen}\affiliation{School of Earth \& Space Exploration, Arizona State University, Tempe, AZ 85287-1404, USA}
\author[0000-0002-0486-0222]{Cameron White}\affiliation{Department of Astronomy and Steward Observatory, The University of Arizona, Tucson, AZ 85721, USA}
\author[0000-0002-2724-8298]{Sanchayeeta Borthakur}\affiliation{School of Earth \& Space Exploration, Arizona State University, Tempe, AZ 85287-1404, USA}
\author[0000-0001-6145-5090]{Nimish Hathi}\affiliation{Space Telescope Science Institute, Baltimore, MD 21218, USA}
\author[0000-0003-4176-6486]{Linhua Jiang}\affiliation{The Kavli Institute for Astronomy and Astrophysics, Peking University, Beijing, 100871, People’s Republic of China} 
\author[0000-0003-3527-1428]{Michael Rutkowski}\affiliation{Department of Physics and Astronomy, Minnesota State University Mankato, Mankato, MN 56001, USA}
\author[0000-0003-0894-1588]{Russell E. Ryan Jr.}\affiliation{Space Telescope Science Institute, Baltimore, MD 21218, USA} 
\author[0000-0002-7779-8677]{Akio K. Inoue}\affiliation{Department of Physics, School of Advanced Science and Engineering, Waseda University, 3-4-1, Okubo, Shinjuku, Tokyo 169-8555, Japan}\affiliation{Waseda Research Institute for Science and Engineering, 3-4-1, Okubo, Shinjuku, Tokyo 169-8555, Japan}
\author[0000-0002-8190-7573]{Robert W. O'Connell}\affiliation{Department of Astronomy, University of Virginia, Charlottesville, VA 22904-4325, USA}
\author[0000-0001-6529-8416]{John W. MacKenty}\affiliation{Space Telescope Science Institute, Baltimore, MD 21218, USA} 
\author[0000-0003-1949-7638]{Christopher J. Conselice}\affiliation{School of Physics \& Astronomy, University of Nottingham, University Park, Nottingham, NG7 2RD, UK}
\author[0000-0002-1566-8148]{Joseph I. Silk}\affiliation{The Johns Hopkins University, Baltimore, MD 21218, USA}

\shortauthors{Smith, B., et al.} 
\shorttitle{LyC escape in GOODS}

\begin{abstract}

We present our analysis of the LyC emission and escape fraction of 111 spectroscopically verified galaxies with and without AGN from 2.26\,$<$\,$z$\,$<$4.3. We extended our ERS sample from \citet{Smith2018} with 64 galaxies in the GOODS North and South fields using WFC3/UVIS F225W, F275W, and F336W mosaics we independently drizzled using the HDUV, CANDELS, and UVUDF data. Among the 17 AGN from the 111 galaxies, one provided a LyC detection in F275W at \mAB\,=\,23.19\,mag (S/N\,$\simeq$\,133) and \GALEX\ NUV at \mAB\,=\,23.77\,mag (S/N\,$\simeq$\,13). We simultaneously fit \textit{SDSS} and \Chandra\ spectra of this AGN to an accretion disk and Comptonization model and find \fesc\ values of $\fesc^{\mathrm{F275W}}\!\simeq\!28^{+20}_{-4}$\% and $\fesc^{\mathrm{NUV}}\!\simeq\!30^{+22}_{-5}$\%. For the remaining 110 galaxies, we stack image cutouts that capture their LyC emission using the F225W, F275W, and F336W data of the GOODS and ERS samples, and both combined, as well as subsamples of galaxies with and without AGN, and \emph{all} galaxies. We find the stack of 17 AGN dominate the LyC production from \zmean$\simeq$2.3--4.3 by a factor of $\sim$10 compared to all 94 galaxies without AGN. While the IGM of the early universe may have been reionized mostly by massive stars, there is evidence that a significant portion of the ionizing energy came from AGN. 

\end{abstract}

\journalinfo{\textrm{(submitted to)} The Astrophysical Journal}

\section{Introduction} \label{sec:intro}
The most likely sources of ionizing photons responsible for the reionization of the intergalactic medium (IGM) --- specifically Lyman continuum (LyC; $\lambda_{\rm rest}$\,$<$912\AA) --- are believed to be massive stars in early galaxies and super-massive black holes (SMBH) in the centers of those galaxies. Actively accreting SMBHs, or active galactic nuclei (AGN), convert a portion of their gravitational potential energy into thermal energy of the accreting disk of infalling matter, which causes the accretion disk to emit black-body radiation that peaks in the UV \citep{Shields1978,Malkan1982}. The interaction between this radiation and the matter of the accretion disk, the dust torus, the surrounding gas and dust clouds of the AGN, and viewing angle determine the emitted spectrum.

Several physical mechanisms are understood to create the various features observed in AGN spectra \citep{Koratkar1999}. The non-ionizing and ionizing UV is thought to be created primarily by thermal emission from the accretion disk, although observations of AGN spectra exhibit a more complex, double power-law continuum with a break near $\lambda_{\rm rest}$\,$\sim$\,1000\AA\ \citep{Zheng1997,Telfer2002,Scott2004,Shull2012,Lusso2015}. This feature is believed to be the low energy wing of Comptonized photons produced by cooling electrons in the warm, optically thick, magnetized plasma of accretion disk coronae. The physical origin of this warm, optically thick component is not well determined \citep{Walton2013,Rozanska2015,Petrucci2018}, though it is proposed to be responsible for the observed soft X-ray excess seen in some AGN spectra \citep{Kaufman2017,Petrucci2018}. 

Ionizing photons from stars are produced by their photospheres, and massive O and B type stars are likely the most important contributors to the stellar ionizing radiation emitted by galaxies \citep{Barkana2001,Stark2016}. Because O and B type stars typically form from multiple open star clusters known as OB associations, in galactic nuclei, and/or starburst regions, the surrounding gas is transformed into giant \ion{H}{2} regions several pc in diameter \citep[e.g.,][]{Tremblin2014}. The stellar LyC flux that exceeds the recombination rate of hydrogen can escape into the IGM as long as it is not absorbed by intervening \ion{H}{1} gas in the interstellar medium (ISM) or circum-galactic medium. 

The fraction of ionizing radiation produced by stars and AGN that escapes into the IGM is known as the LyC escape fraction (\fesc). The literature often states that stellar LyC escaping from high-redshift, star-forming, possibly low-mass galaxies are likely the dominant sources of LyC that reionized the IGM at $z$\,$\sim$\,6--7 \citep[e.g.,][]{Bouwens2012,Wise2014,Duncan2015}, and require \fesc\,$\sim$10--30\% to complete this phase transition from observational constraints \citep[e.g.,][]{Finkelstein2012,Robertson2015,Bouwens2016}. The dearth of observed high \fesc\ values for more massive star-forming galaxies (SFGs) found throughout the literature (\citealt{Smith2018}, hereafter \citetalias{Smith2018}, and references therein; see also \citealt{Japelj2017,Grazian2017,Steidel2018,Iwata2019}), as well as the decline in the AGN luminosity function at 3$\lesssim$\,$z$\,$\lesssim$\,6 \citep{Aird2015,Kulkarni2019}, have led to conclusions that low mass, star-forming dwarf galaxies may be more likely candidates for the agents of reionization (\citealt{Finkelstein2012,Stark2016,Weisz2017}; but see \citealt{Naidu2019}). Simulations show that \fesc\ should increase with decreasing halo mass \citep{Yajima2011,Kimm2014,Wise2014}, and recent work have observed that low-mass, low-metallicity, compact star-forming galaxies with extreme [\ion{O}{3}] emission and [\ion{O}{3}]/[\ion{O}{2}] line ratios exhibit detectable LyC emission at low-redshift \citep[0\,$\leq$\,$z$\,$\lesssim$\,1;][]{Izotov2016,Izotov2017,Izotov2018} and at $z$\,$\simeq$\,3.1 \citep{Fletcher2019}.

However, \citet{Tanvir2018} constrain the average \fesc\ of low-mass SFGs at 2\,$<$\,$z$\,$<$5 to $\langle\fesc\rangle$\,$<$\,1.5\% using 138 gamma-ray burst afterglows, and present evidence that their \fesc\ does not change at $z$\,$>$\,5. They first determine the neutral hydrogen column-density (\NH) of these galaxies and infer an \fesc\ from their total sample and find no evolution of \NH\ with redshift. Typical GRB hosts show higher \NH\ column densities at $z$\,$>$\,2 from observation, similar to those of damped \Lya\ systems \citep{Jakobsson2006}. Two of the GRBs in \citet{Tanvir2018} do show sufficiently low \NH\ to allow more LyC radiation to escape, suggesting that stellar feedback can clear the ISM to allow higher \fesc\ in \emph{rarer} cases. Simulations show that \fesc\ is likely anisotropic in a galaxy \citep{Wise2009,Kim2013,Paardekooper2015}, therefore some lines-of-sight may have much higher \fesc\ near regions associated with SNe winds \citep{Fujita2003,Trebitsch2017,Herenz2017}. Long-duration GRBs (LGRBs), like those studied in \citet{Tanvir2018}, are known to reside exclusively in SFGs, and are strongly correlated with UV-bright regions in their hosts, although the nature short GRBs hosts are less clear \citep[e.g.,][]{Savaglio2009,Berger2014}. Since LGRB host galaxies are often observed to be dwarfs with high specific star-formation rates \citep{Svensson2010,Hunt2014,McGuire2016}, and the bulk of low-redshift dwarfs are observed to exhibit very low \fesc\ \citep[e.g., at $z$\,$\simeq$\,0.5, \fesc\,$<$\,3\%;][]{Rutkowski2016}, hypotheses that propose dwarf galaxies to be the main reionizers at $z$\,$>$\,6 may be in conflict with observation and additional sources of ionizing flux would be needed. 

\begin{figure*}[ht!]\centerline{
\includegraphics[width=.267\txw]{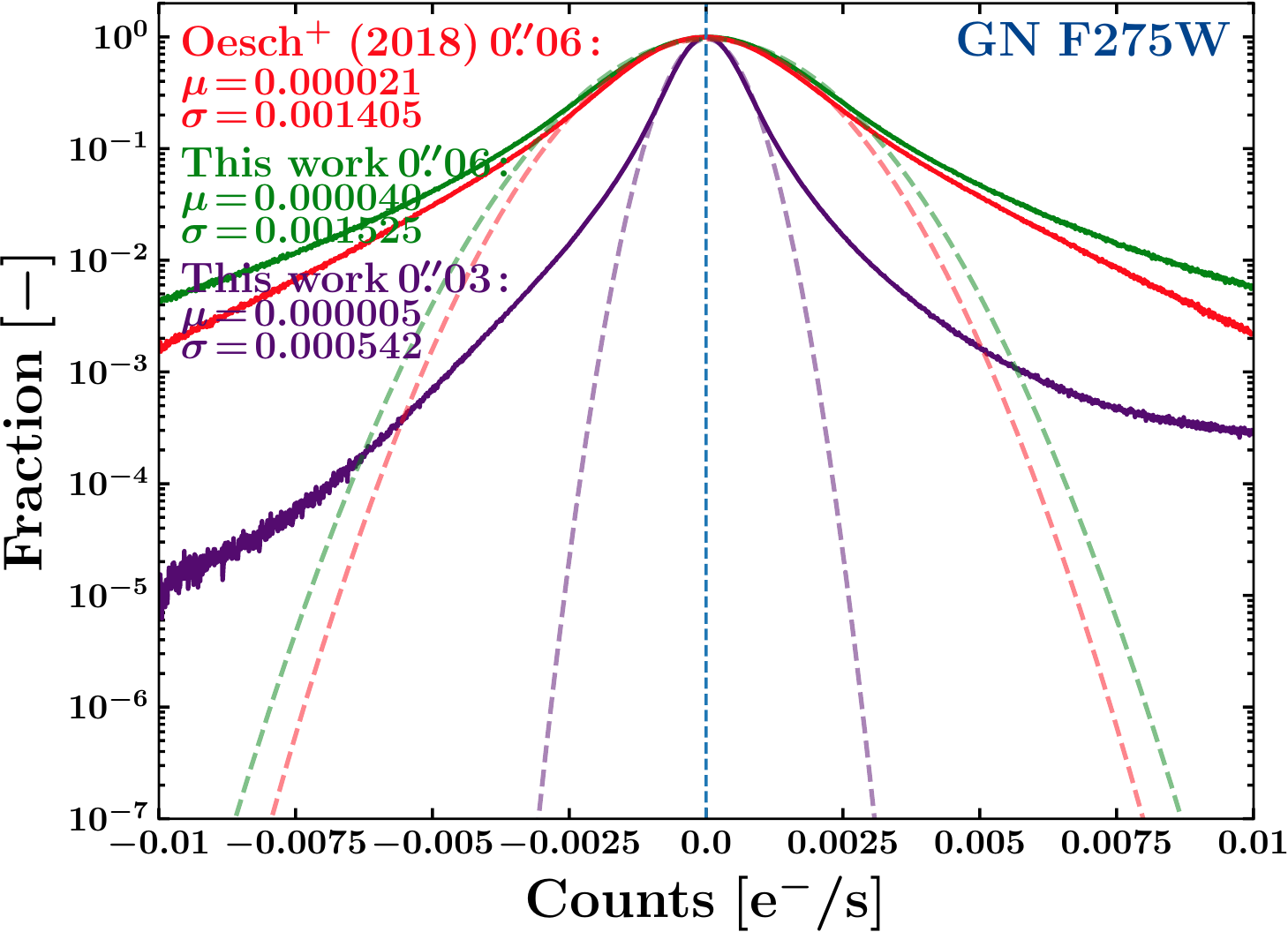}
\includegraphics[width=.245\txw]{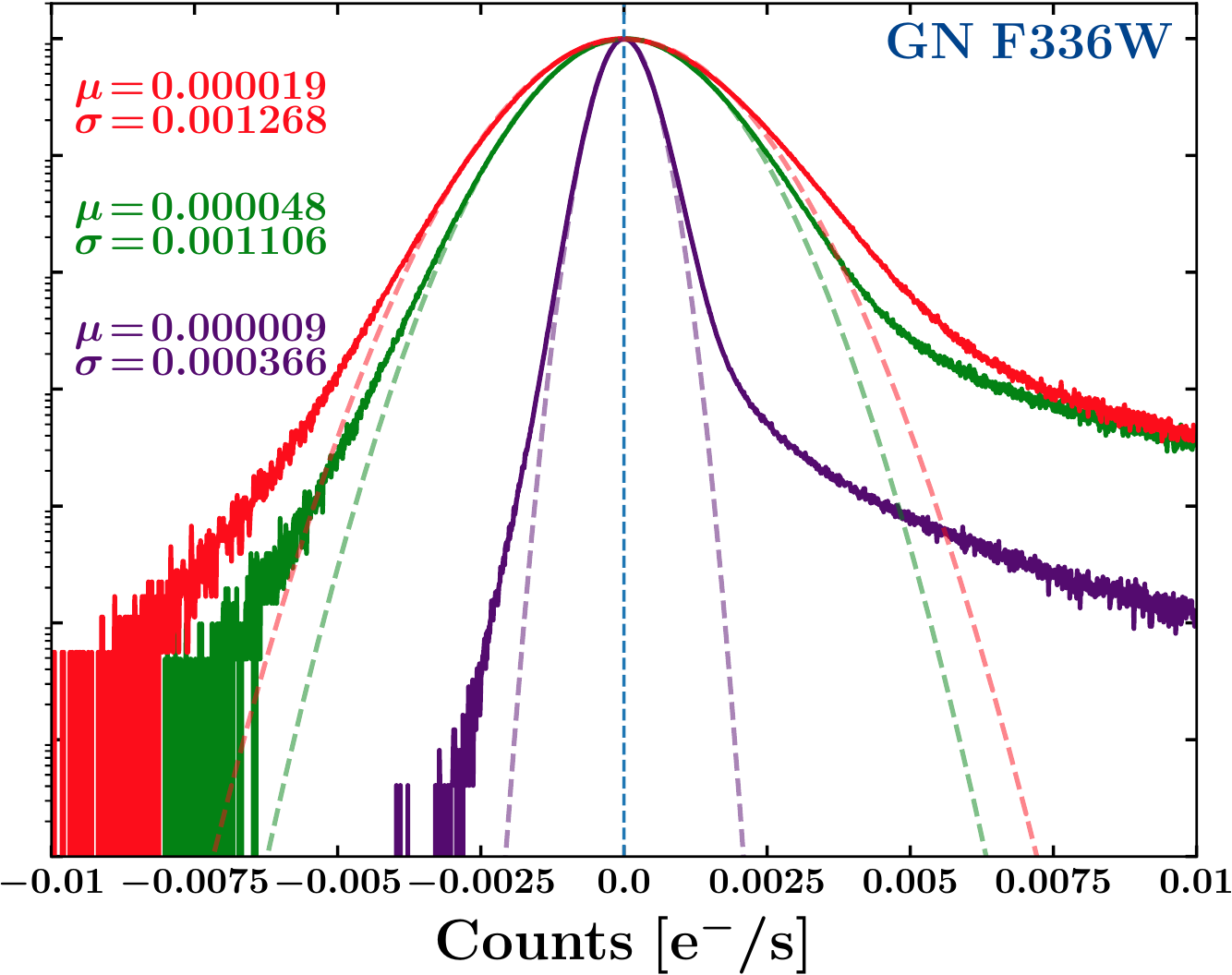}
\includegraphics[width=.245\txw]{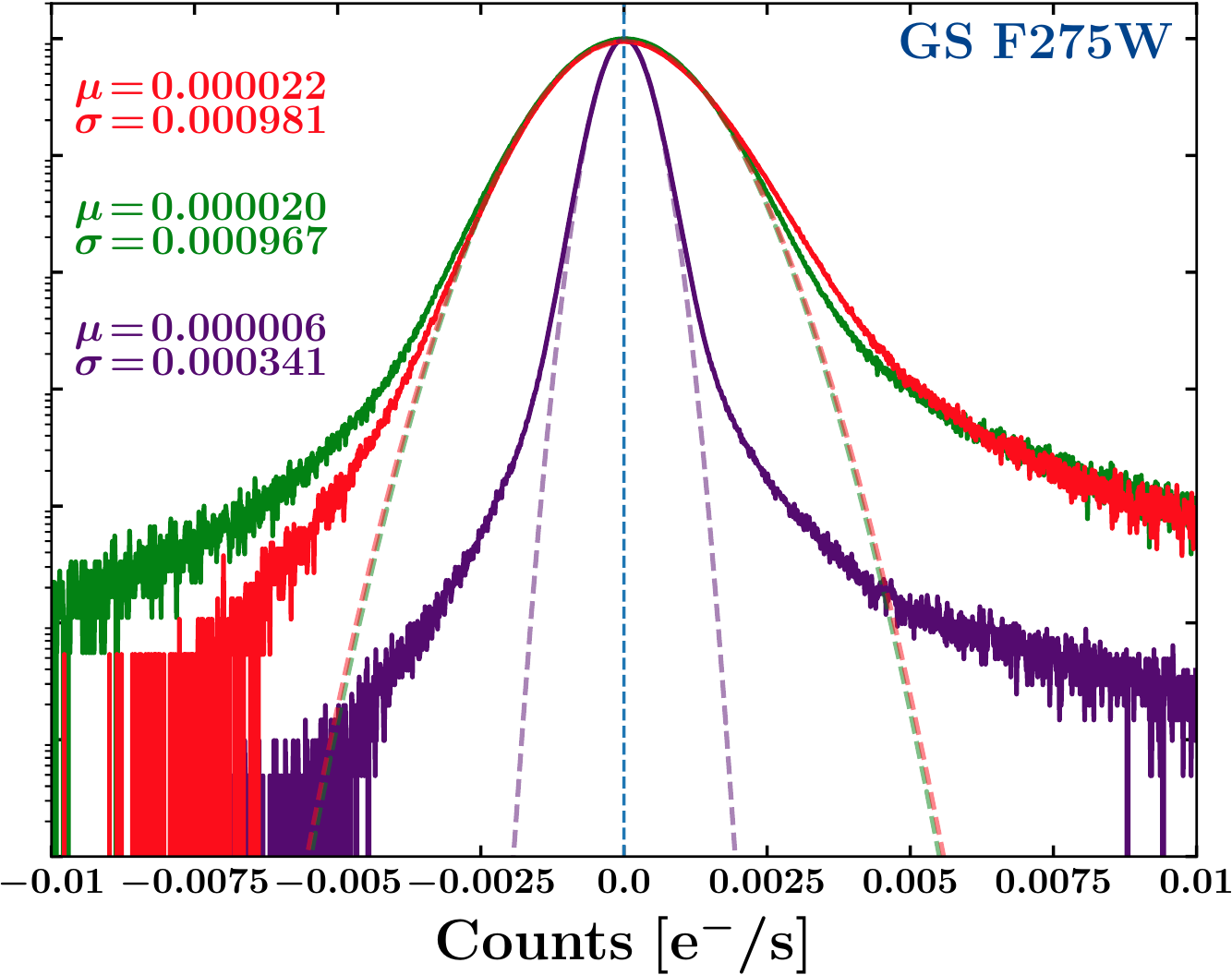}
\includegraphics[width=.245\txw]{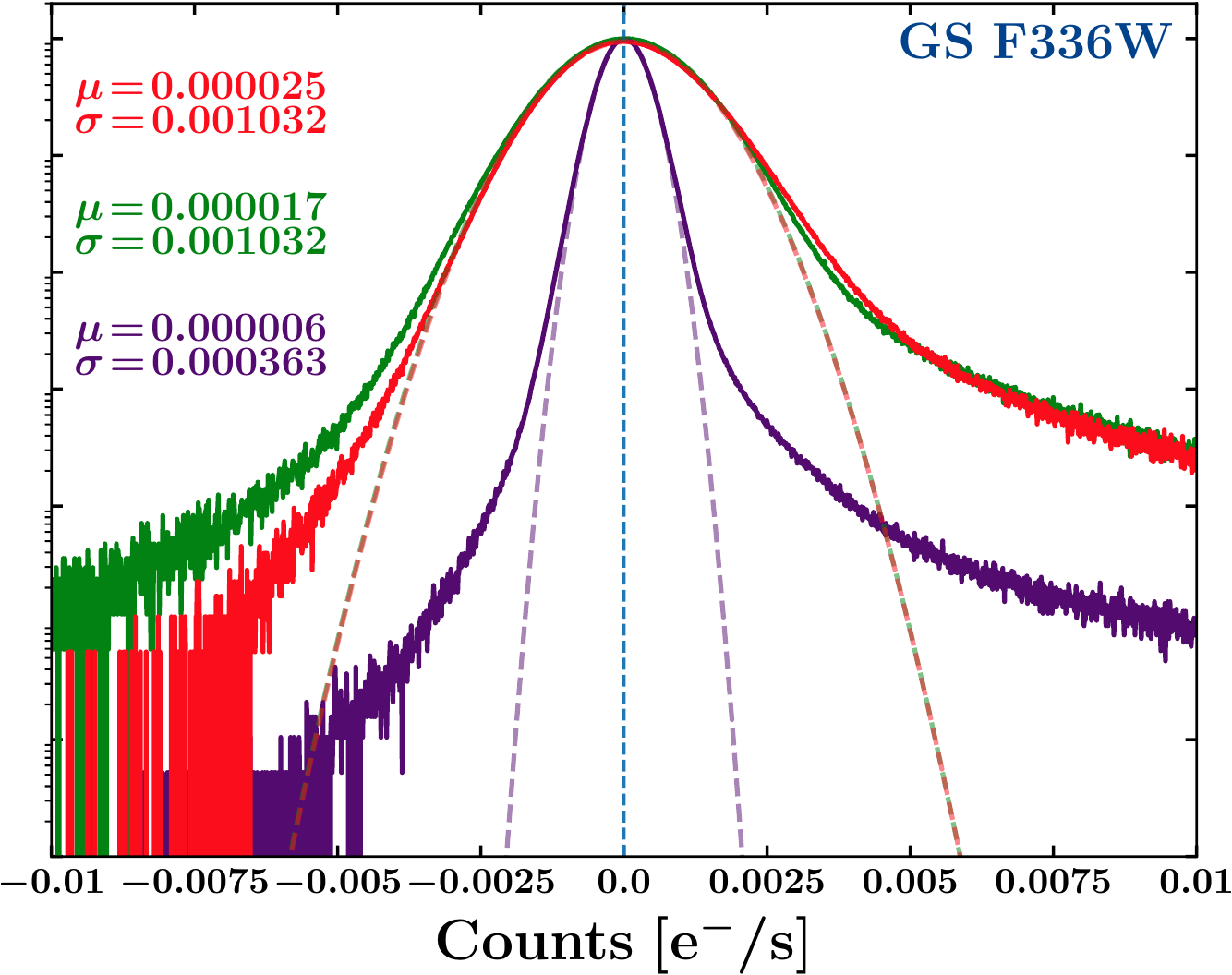}
}
\caption{\noindent\small
Pixel histograms of the GOODS/HDUV mosaics from \citet{Oesch2018} and this work in the WFC3/UVIS F275W and F336W filters for the GOODS North (GN) and GOODS South (GS) fields (solid lines), along with Gaussian functions fit to each histogram (lighter dotted lines). Our mosaics were sky-subtracted and drizzled to 0\farcs03 and 0\farcs06 pixels and the \citet{Oesch2018} data was drizzled to 0\farcs06 pixels. The mean ($\mu$) and dispersion ($\sigma$) of the histograms is indicated in each plot for all three versions of the GOODS/HDUV mosaics. The 0\farcs06 mosaic pixel-value distributions are generally consistent in the four cases, and our mosaics show modest improvement in $\mu$ and $\sigma$ in three. The 0\farcs03 mosaic pixel-value distributions have the smallest $\mu$ and $\sigma$ in all four images.\label{skyhists}}
\end{figure*} 

\citet{Grazian2018} discuss how several well-studied, nearby dwarf galaxies with measured \fesc\ and \emph{relative} LyC escape fractions ($f_{\rm esc,rel}$) have been detected in X-rays, potentially indicating an AGN component. The relative LyC escape fraction is defined as ratio of the escaping LyC to the observed LyC, relative to the escaping non-ionizing UV-continuum (UVC), or
\begin{equation}
f_{\rm esc,rel}=\frac{(f_{1500}/f_{\lyc})_{\rm int}}{(f_{1500}/f_{\lyc})_{\rm obs}}\mathrm{e}^{\tau_{\igm}(z)}
\end{equation}
where $f_{1500}$ is the rest-frame 1500\AA\ flux, $f_{\lyc}$ is the LyC flux, and $\tau_{\igm}(z)$ is the optical depth of the IGM at redshift $z$ \citep{Steidel2001,Inoue2005,Siana2007,Siana2010}. This parameter is related to \fesc\ by \fesc\,=\,$f_{\rm esc,rel}10^{-0.4A_{1500}}$, where $A_{1500}$ is the galactic extinction at 1500\,\AA. 
\citet{Kaaret2017} detect point-source X-ray flux with \Chandra\ from the $z$\,=\,0.048 SFG Tololo 1247–232 ($f_{\rm esc,rel}$\,=\,21.6\%, 1.5\%$\pm$0.5, measured by \citealt{Leitherer2016} and \citealt{Puschnig2017}, respectively), and shows variability on the order of years, suggesting the presence of a low-luminosity AGN ($\mathrm{L_X}$\,$\simeq$\,$10^{41}$\,erg\,s$^{-1}$). \citet{Prestwich2015} detect a bright point-source ($\mathrm{L_X}$\,$\simeq$\,$10^{41}$\,erg\,s$^{-1}$) within Haro 11 \citep[$f_{\rm esc,rel}$\,$\sim$\,3\%;][]{Leitet2011} with a very hard spectrum (X-ray photon index $\Gamma$\,=\,1.2$\pm$0.2). \citet{Borthakur2014} find LyC flux emitted by a $z$\,=\,0.235 SFG (J0921+4509, \fesc\,$\simeq$\,20\%), which has been detected in hard X-rays with \textit{XMM-Newton} \citep[$\mathrm{L_X}$\,$\simeq$\,10$^{42}$\,erg\,s$^{-1}$;][]{Jia2011}, suggesting a possible AGN component as well. 

More recent studies on the sources of reionization have emphasized the role of AGN from observations \citep[e.g.,][]{Giallongo2015,Madau2015,Khaire2016,Mitra2017} and empirically-based numerical models \citep[e.g.,][]{Yoshiura2017,Bosch2018,Torres2020}. Their findings suggest that AGN display significant emission of ionizing flux, and stellar sources within SFG alone may not emit LyC at a sufficient rate required to complete reionization. If SNe winds in SFGs with no AGN component could clear enough channels in the ISM to allow more LyC to be emitted into the IGM, then SFGs should have higher \fesc\ values than observed throughout the literature. High mass X-ray binaries and X-rays produced by SNe have been proposed as sources of additional ionizing energy that could enhance \fesc\ in SFGs \citep[e.g.,][]{Mirabel2011,McQuinn2012,Bluem2019}. 

In this work, we analyze the escaping LyC flux and estimate the \fesc\ of 111 massive SFG and AGN from 2.26\,$\leq$\,$z$\,$\leq$\,4.3 using \HST\ WFC3/UVIS imaging. Building on our previous work \citepalias{Smith2018}, we determine the LyC escape fractions of SFG and AGN with significantly improved statistical sampling, and directly compare their contributions to the total ionizing background at these redshifts.

This paper is organized as follows.  In \S\ref{sec:data}, we describe the data that we used for our analysis and how it was reduced.  In \S\ref{sec:sample}, we describe our sample of 111 galaxies selected to have accurate spectroscopic redshifts with no contaminating, non-ionizing flux present in their LyC images.  In \S\ref{sec:qphotfesc}, we outline the method we implemented to create the stacked LyC images of our samples of galaxies, how we perform photometry on the stacks, the LyC flux that we measure or constrain, and the statistical significance thereof.  In \S\ref{sec:results}, we determine the stacked LyC escape fraction, how we calculated the \fesc\ values, and their implications.  In \S\ref{sec:discussion} and \ref{sec:conclusion}, we discuss our results and present our conclusions. We use Planck (\citeyear{PlanckCollaboration2018}) cosmology throughout: $H_{0}$ = 67.4 km\,s$^{-1}$\,Mpc$^{-1}$, $\Omega_{\rm m}$=0.315 and $\Omega_\Lambda$=0.685.  All flux densities (referred to as ``fluxes'' throughout) quoted are in the AB magnitude system \citep{Oke1983}.

\section{Data} \label{sec:data}
\begin{deluxetable}{lcccr}
\centering\tablecaption{WFC3/UVIS GOODS/HDUV Image Parameters\label{phottab}}
\tablewidth{\txw}
\tabletypesize{\scriptsize}
\tablehead{\colhead{Filter} & \colhead{Pixel size} & \colhead{Sky $\mu$} & \colhead{Sky $\sigma$} & \colhead{5$\sigma$ Limit} \\[-4pt]
\null & [10$^{-3}$\,arcsec] & [counts/s] & [counts/s] & [mag] \\
\multicolumn{1}{c}{(1)} & (2) & (3) & (4) & (5)}
\startdata
\multicolumn{5}{l}{\sc \underline{\citet{Oesch2018}}:}\\
\multicolumn{5}{l}{\sc GOODS North:}\\
F275W & 60 & 2.081$\times 10^{-5}$ & 1.405$\times 10^{-3}$ & 27.69 \\
F336W & 60 & 1.854$\times 10^{-5}$ & 1.268$\times 10^{-3}$ & 28.09 \\
\multicolumn{5}{l}{\sc GOODS South:}\\
F275W & 60 & 2.194$\times 10^{-5}$ & 9.813$\times 10^{-4}$ & 28.14 \\
F336W & 60 & 2.512$\times 10^{-5}$ & 1.032$\times 10^{-3}$ & 28.62 \\
\hline
\multicolumn{5}{l}{\sc \underline{This Work}:}\\
\multicolumn{5}{l}{\sc GOODS North:}\\
F275W & 30 & 5.447$\times 10^{-6}$ & 5.416$\times 10^{-4}$ & 28.21 \\
F336W & 30 & 9.124$\times 10^{-6}$ & 3.662$\times 10^{-4}$ & 29.11 \\
F275W & 60 & 4.017$\times 10^{-5}$ & 1.525$\times 10^{-3}$ & 27.81 \\
F336W & 60 & 4.776$\times 10^{-5}$ & 1.106$\times 10^{-3}$ & 28.68 \\
\multicolumn{5}{l}{\sc GOODS South:}\\
F275W & 30 & 5.764$\times 10^{-6}$ & 3.411$\times 10^{-4}$ & 28.70 \\
F336W & 30 & 5.753$\times 10^{-6}$ & 3.627$\times 10^{-4}$ & 29.12 \\
F275W & 60 & 1.996$\times 10^{-5}$ & 9.672$\times 10^{-4}$ & 28.23 \\
F336W & 60 & 1.732$\times 10^{-5}$ & 1.032$\times 10^{-3}$ & 28.73 
\enddata
\vspace{5pt}
\begin{minipage}{.43\txw}{\footnotesize \textbf{Table columns:} (1) WFC3/UVIS filter used for mosaic; (2) Angular size of drizzled pixel on one side; (3) Average value of sky pixels; (4) Dispersion of sky pixel values; (5) Faintest 5$\sigma$ detection in a 0\farcs4 diameter aperture, using zeropoint magnitudes from the Cycle 28 WFC3 Instrument Handbook.} 
\end{minipage}
\end{deluxetable}

The archival \HST\ image data we used for our LyC study includes WFC3/UVIS data from the Early Release Science field \citep[ERS;][]{Windhorst2011}, the Cosmic Assembly Near-infrared Deep Extragalactic Legacy Survey \citep[CANDELS;][]{Grogin2011, Koekemoer2011}, the Hubble Ultraviolet Ultra Deep Field \citep[UVUDF;][]{Teplitz2013}, and the Hubble Deep UV Legacy Survey \citep[HDUV;][]{Oesch2018}, which we independently drizzled using Astrodrizzle in the DrizzlePac software\footnote{\url{http://www.stsci.edu/scientific-community/software/drizzlepac.html}}. The ERS WFC3/UVIS data is described in more detail in \citetalias{Smith2018}. We used optical ACS/WFC data in F606W, F775W, and F850LP from the Great Observatories Origins Deep Survey \citep[GOODS;][]{Dickinson2003, Giavalisco2004}, WFC3/IR F098M, F125W, and F160W imaging in the ERS field \citep{Windhorst2011}, CANDELS WFC3/IR F105W, F125W, F160W and WFC/ACS F814W, and 3DHST WFC3/IR F140W imaging \citep{Momcheva2016} for Spectral Energy Distribution (SED) fitting (see \S\ref{sec:sedfitting}) and for studying the rest-frame, non-ionizing UV continuum (UVC, $\mathrm{\lambda_{rest}}\!\simeq$1400-1800\AA) of galaxies.

Our new mosaics include all available CANDELS and HDUV data in F275W and F336W taken in the GOODS North field, as well as all available data in F225W, F275W, and F336W from the HDUV and UVUDF surveys that covered the GOODS South field. We refer to both mosaics as the ``GOODS/HDUV'' data, and the ERS imaging that \citetalias{Smith2018} was based on is referred to as the ``ERS'' data.  

The ERS data was collected less than four months after WFC3 was installed in \HST during Shuttle Servicing Mission SM4 in 2009. The cumulative effects of high energy particle interactions with the WFC3/UVIS detector degrade the charge transfer efficiency (CTE) over time. The ERS data was taken before the WFC3/UVIS CCDs suffered from significant degradation, so CTE correction was \emph{not} required during the creation of the ERS mosaic images with Astrodrizzle \citepalias{Smith2018}. The ERS data reaches $\sim$2 orbit depth (\mAB\,$<$\,26.4 at 5$\sigma$ for F275W) over a wide $\sim$50\,arcmin$^2$ area in the F225W, F275W, and F336W filters \citep{Windhorst2011}, while the HDUV imaging reaches 4--8 orbit depth in F275W and F336W (\mAB\,$<$\,27.6 at 5$\sigma$ for F275W) in a combined $\sim$100\,arcmin$^2$ area across the GOODS North and South fields \citep{Oesch2018}. The UVUDF data covered a single pointing in GOODS South for 16, 16, and 14 orbits in F225W, F275W, and F336W, respectively \citep[\mAB\,$<$\,27.8 at 5$\sigma$ for F275W;][]{Rafelski2015}. The CANDELS survey also observed GOODS North in F275W and reached a $\sim$6 orbit depth \citep[\mAB\,$<$\,27.1 at 5$\sigma$;][]{Koekemoer2011} that brought the F275W GOODS North data to a total depth of $\sim$10 orbits. The HDUV imaging required the use of post-flash at the time of observation to mitigate CTE degradation effects, such as the loss of faint flux in the raw image data during readout. 
\noindent\begin{figure*}[th!]\centerline{
\includegraphics[width=.99\txw]{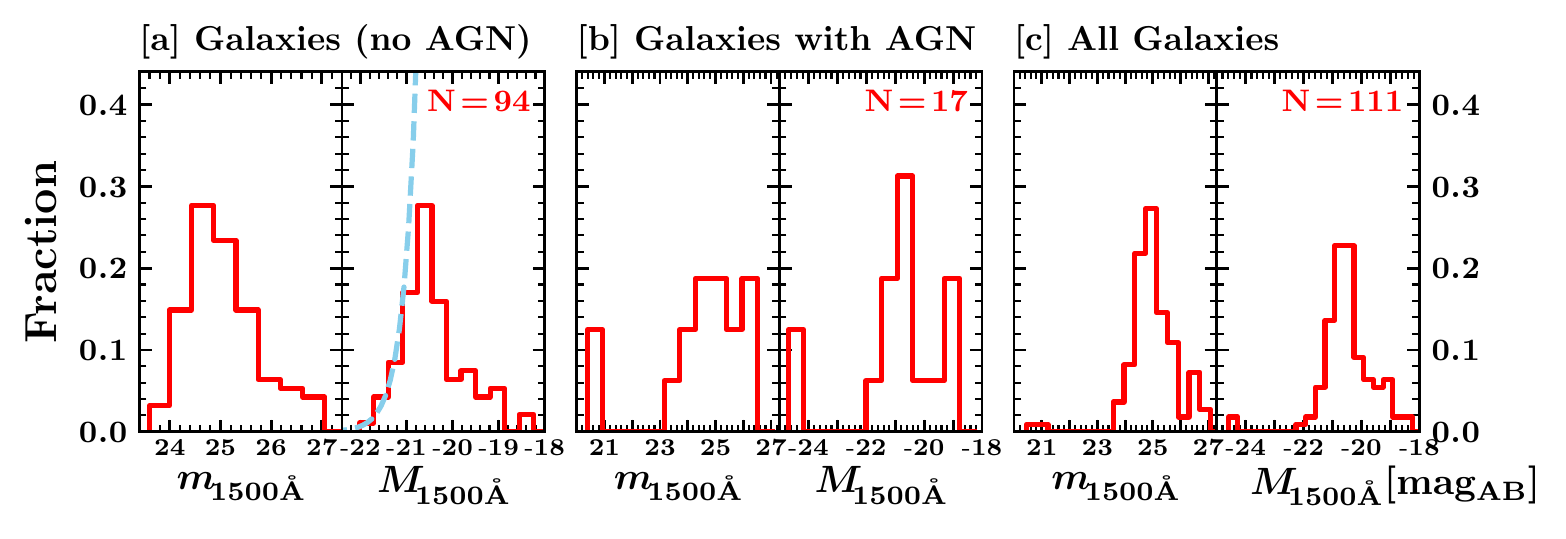}}
\caption{\noindent\small
[\emph{a}] Absolute and apparent magnitude distributions at rest-frame
1500\AA\ of the spectroscopic samples for galaxies \emph{without} AGN. [\emph{b}] Same, for galaxies \emph{with} evidence of AGN activity in their spectra. [\emph{c}] Same, for all galaxies in the sample. These magnitudes were derived from the observed SED fits at $\mathrm{\lambda_{rest}}\!=\!1500\!\pm\!100\mathring{A}$, and therefore do not require k-corrections. The blue dashed curve represents the faint-end slope of the luminosity function (LF) of \zmean$\simeq$2.97 galaxies with \MAB$^{\!\!\!\!\!\!\!*}$\,\, $\simeq$--20.63, and $\alpha\!\simeq$1.36\,dex/mag \citep{Parsa2016}. The Total sample is approximately representative of the galaxy LF at their \zmean\ to \MAB\,$\leq$\,-21\,mag and \mAB\,$\leq$\,24.5\,mag. \label{maghist}}
\end{figure*}

Fig.~\ref{skyhists} shows sky histograms comparing our versions of the GOODS/HDUV mosaics drizzled at 0\farcs03 and 0\farcs06 pixels to the public \citet{Oesch2018} mosaics drizzled at 0\farcs06 pixels. Our mosaics show modest improvements in the sky-subtraction level $\mu$ and the dispersion of the sky-background noise $\sigma$, with reduced values for $\mu$ and $\sigma$ in some cases. The 0\farcs03 pixel-value distributions show the most improvements in $\mu$ and $\sigma$, due to the smaller pixel size and associated lower correlated noise between pixels (see the Appendix of \citet{Casertano2000} for more details). The pixel distributions are generally consistent with \citet{Oesch2018}, and the slight improvements in the 0\farcs06 images are likely due to differences in processing the raw \HST\ frames and Astrodrizzle parameters used when constructing the mosaics. Our image processing steps before drizzling included subtracting a stacked dark current image from each frame to remove any thermal structure, more robust cosmic ray removal resulting in fewer bad pixels, and the removal of gradients caused by scattered background light. Each frame of the GOODS/HDUV mosaics was also CTE-corrected \citep{Anderson2010} and aligned to the same pixel grid as the GOODS ACS/WFC F435W v2 image\footnote{\url{https://archive.stsci.edu/prepds/goods/}}. We drizzled the GOODS/HDUV data to a 0\farcs06 pixel scale for comparing to the public \citet{Oesch2018} mosaics, and to 0\farcs03 for our LyC studies (i.e., our LyC photometric analysis and \fesc\ constraints). 

We chose to use the 0\farcs03 mosaics primarily because the smaller pixels increase the ability to resolve smaller features. This allowed for improved deblending of neighboring galaxies that can potentially contaminate LyC measurements with non-ionizing flux. Such compact galaxies can be detected at higher S/N ratios in optical \HST\ images and masked in the WFC3/UVIS mosaics. This also improved subsequent photometric estimates of LyC, which we based on the total count rate in the drizzled mosaic image within a measurement aperture, the local sky-background dispersion, and the RMS-values of the pixels within the aperture used for photometry. The improvements in the photometric statistics and contamination removal together improve the accuracy of the constraints on subsequent Monte Carlo (MC) analyses of the LyC escape fraction. 

\section{Sample Selection and Characteristics} \label{sec:sample}
\noindent\begin{figure*}[th!]\includegraphics[width=\txw]{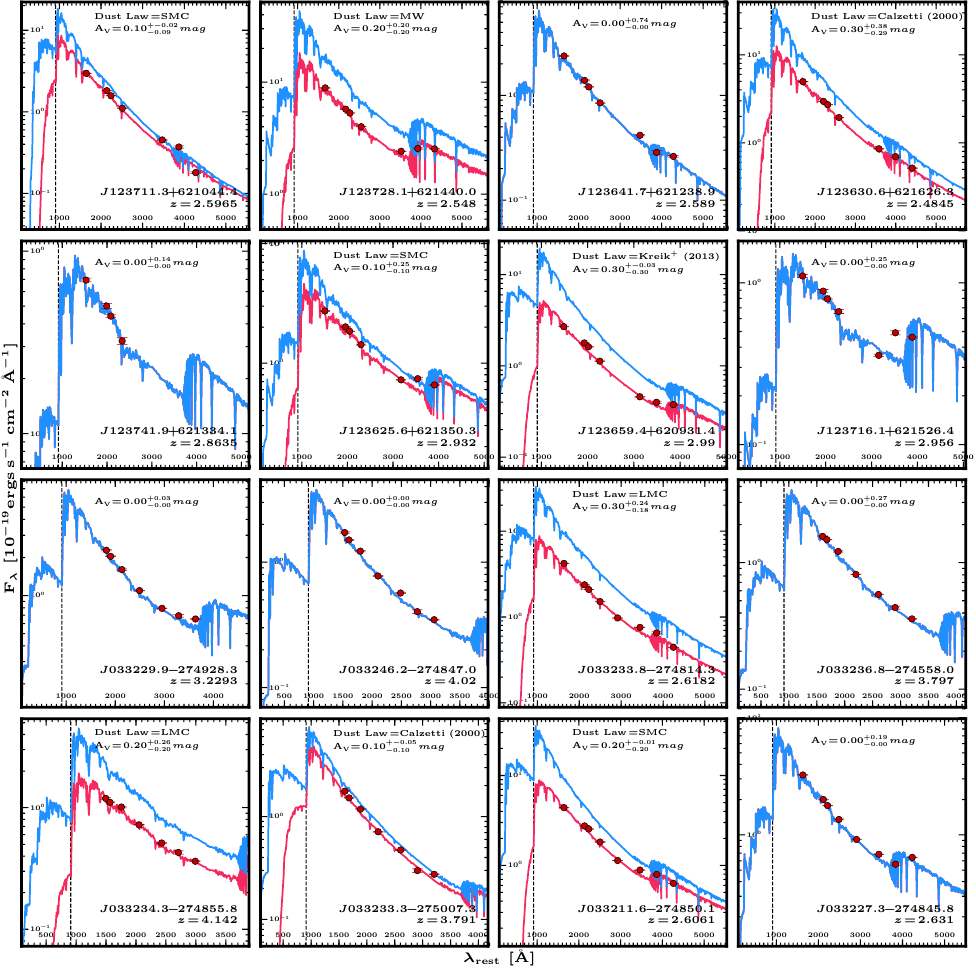}
\caption{Example best-fit BC03 SEDs from FAST \citep[red curve;][]{Kriek2009} fit to observed \HST\ ACS/WFC and WFC3/IR photometry (red filled circles), and the intrinsic BC03 SED with no extinction applied (blue). The best-fitting dust extinction law and $\mathrm{A_V}$ is indicated for each SED. The spectroscopic redshift is shown, along with the corresponding Lyman limit plotted as a black vertical dashed line. The remaining SEDs are shown in Appendix~\ref{app:sed}, and their age, mass, star-formation rate, and metallicity are listed in Table~\ref{objtab}. \label{galseds}}
\end{figure*}
\noindent\begin{figure*}[th!]\includegraphics[width=\txw]{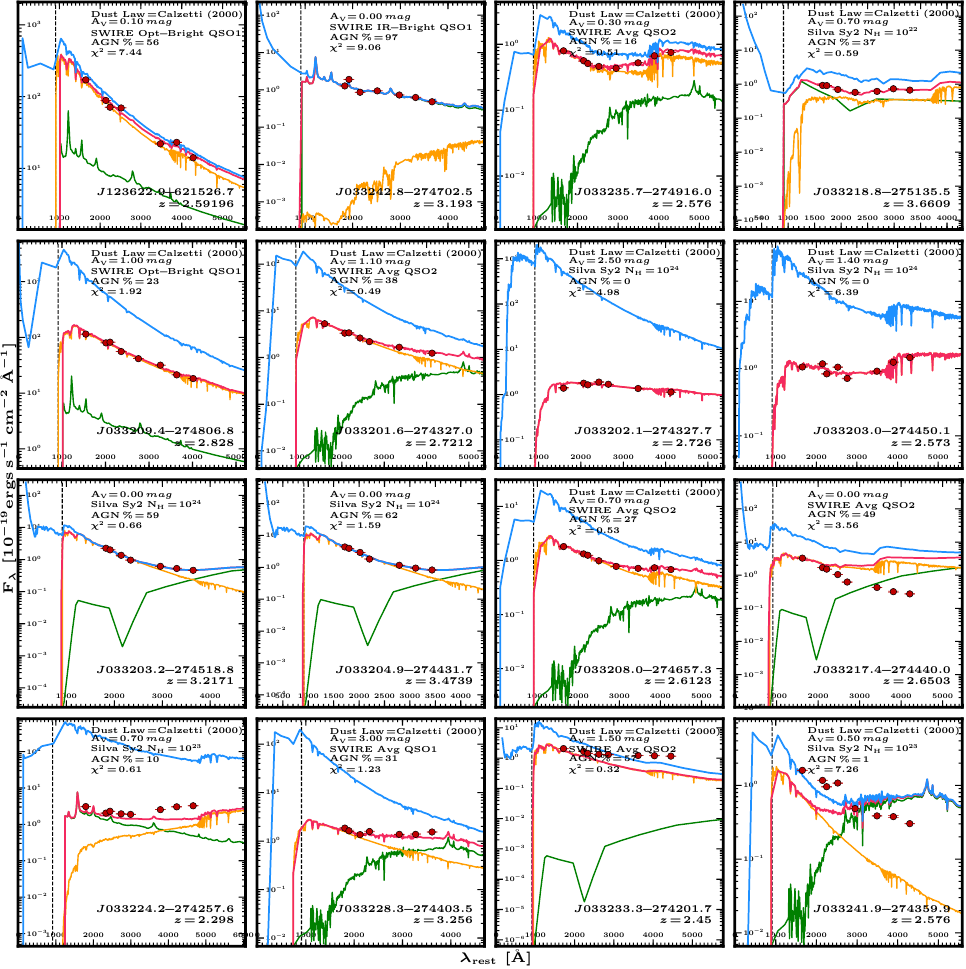}
\caption{Best-fit two-component SEDs from FAST \citep[red curve;][]{Aird2018}. The best-fit SEDs (red curve) were fit to the plotted \HST\ ACS/WFC and WFC3/IR photometry (red filled circles), and are the sum of a BC03 SED (orange) and the best-fit AGN template \citep[green;][]{Silva2004,Polletta2007}. $N_H$ indicates the neutral hydrogen absorption column density applied to the AGN template. The blue line is the un-reddened SED, i.e., equivalent to the red line without the \citet{Calzetti2000} extinction law applied. The best-fit BC03 SED parameters for dust extinction law and $\mathrm{A_V}$, AGN template, and AGN SED flux percentage at $\mathrm{\lambda_{rest}}$=5000\AA\ are indicated. Their age, mass, star-formation rate, and metallicity are listed in Table~\ref{objtab} \label{agnseds}}
\end{figure*}
Our sample used for LyC studies at 2.26\,$\leq$\,$z$\,$\leq$\,$4.3$ was selected from a compilation of spectroscopic surveys including the 3D-HST \citep{Brammer2012, Momcheva2016}, GMASS \citep{Kurk2013}, GOODS/FORS1 \citep{Cristiani2000}, GOODS/FORS2 \citep{Vanzella2006,Vanzella2008}, GOODS/VIMOS \citep{Popesso2009,Balestra2010}, K20 \citep{Mignoli2005}, MUSE-Wide \citep{Herenz2017,Urrutia2019}, SDSS DR14 \citep{Abolfathi2018}, TKRS \citep{Wirth2004}, TKRS2 \citep{Wirth2015}, VANDELS \citep{Pentericci2018}, VUDS \citep{Tasca2017}, VVDS \citep{LeFevre2013}, and the \citet{Szokoly2004}, \citet{Reddy2006}, \citet{Wuyts2009}, \citet{Silverman2010}, and \citet{Xue2016} surveys. This redshift range was selected so that the non-ionizing continuum ($\mathrm{\lambda_{rest}}$\,$>$\,912\,\AA) of a typical SFG SED would \emph{not} exceed more that 0.5\% of the total flux transmitted through the WFC3/UVIS filters. Including galaxies with redshifts lower than our defined redshift-bin ranges from table~1 of \citetalias{Smith2018} would introduce more ``red-leak'' of non-ionizing flux into the filter, and in some cases become comparable or dominate the LyC flux by several times. Since LyC flux can be 3--4\,mag fainter than the UVC (see Table~\ref{phottab} and table~2 of \citetalias{Smith2018}), we use the same redshift bins in \citetalias{Smith2018} to keep the percentage of red-leak to $\simeq$\,0.3\%.

\noindent\begin{figure*}[th!]\gridline{
\fig{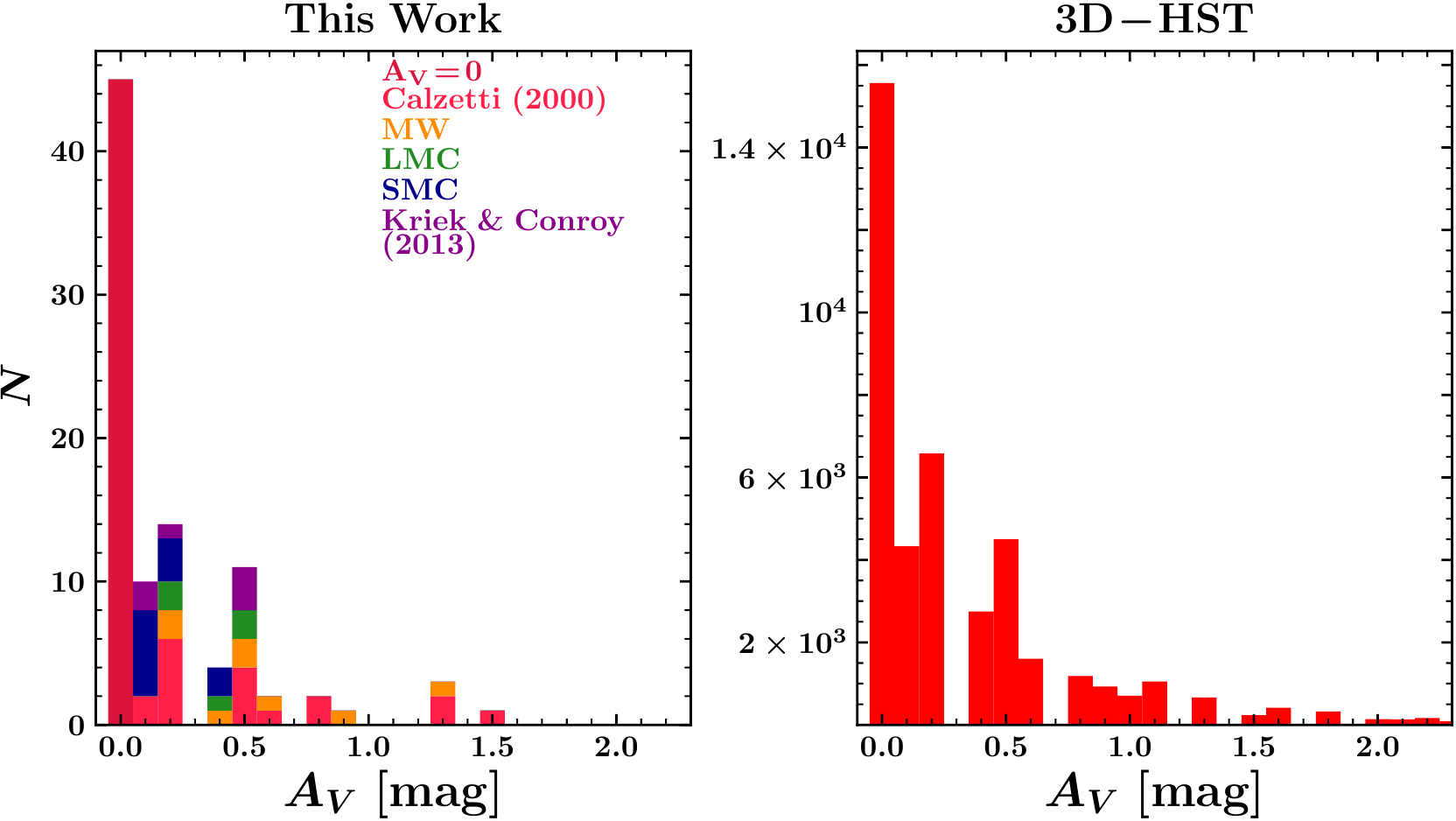}{.497\txw}{}\hspace{5pt}
\fig{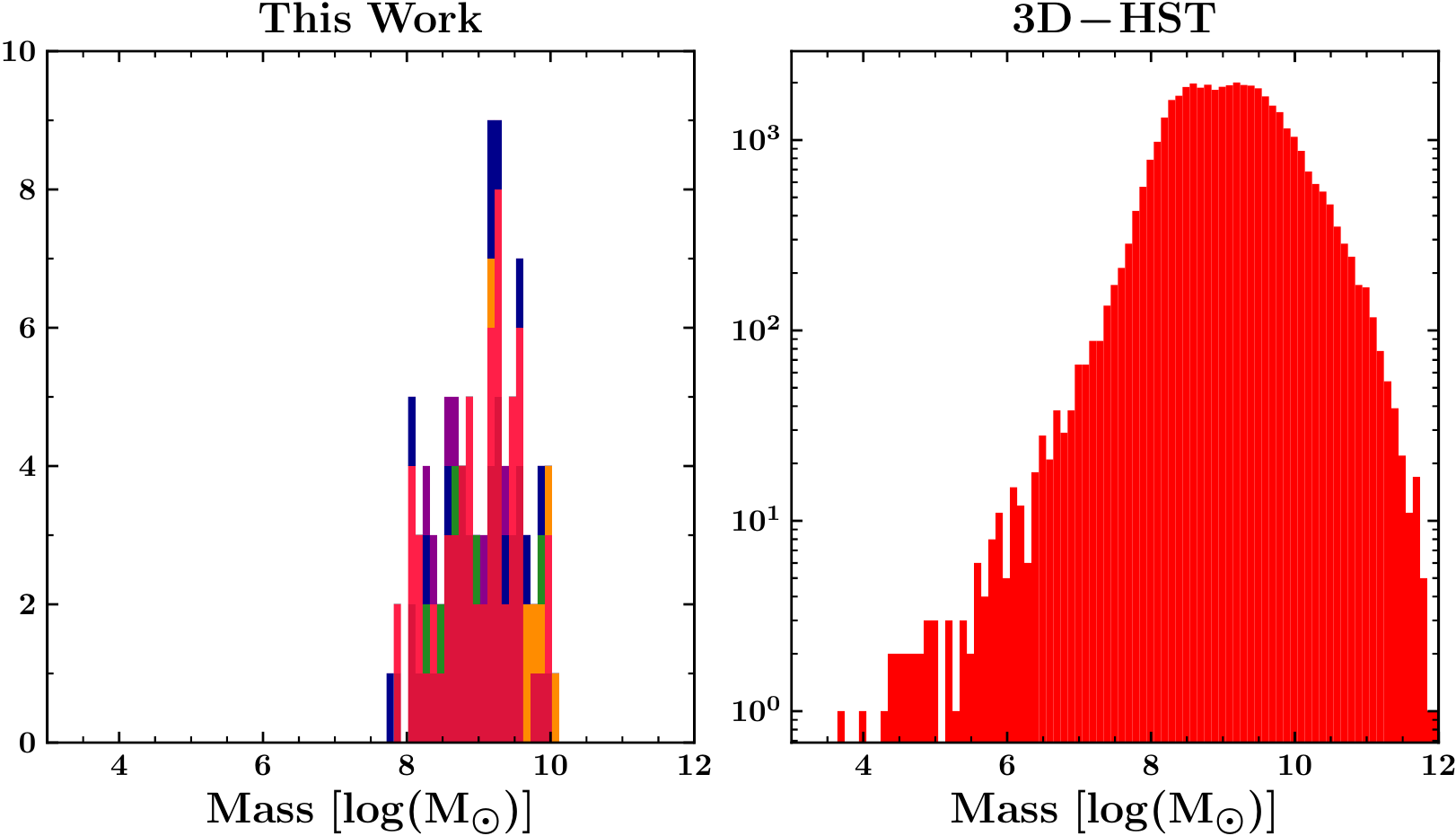}{.488\txw}{}}\vspace*{-20pt}
\gridline{
\fig{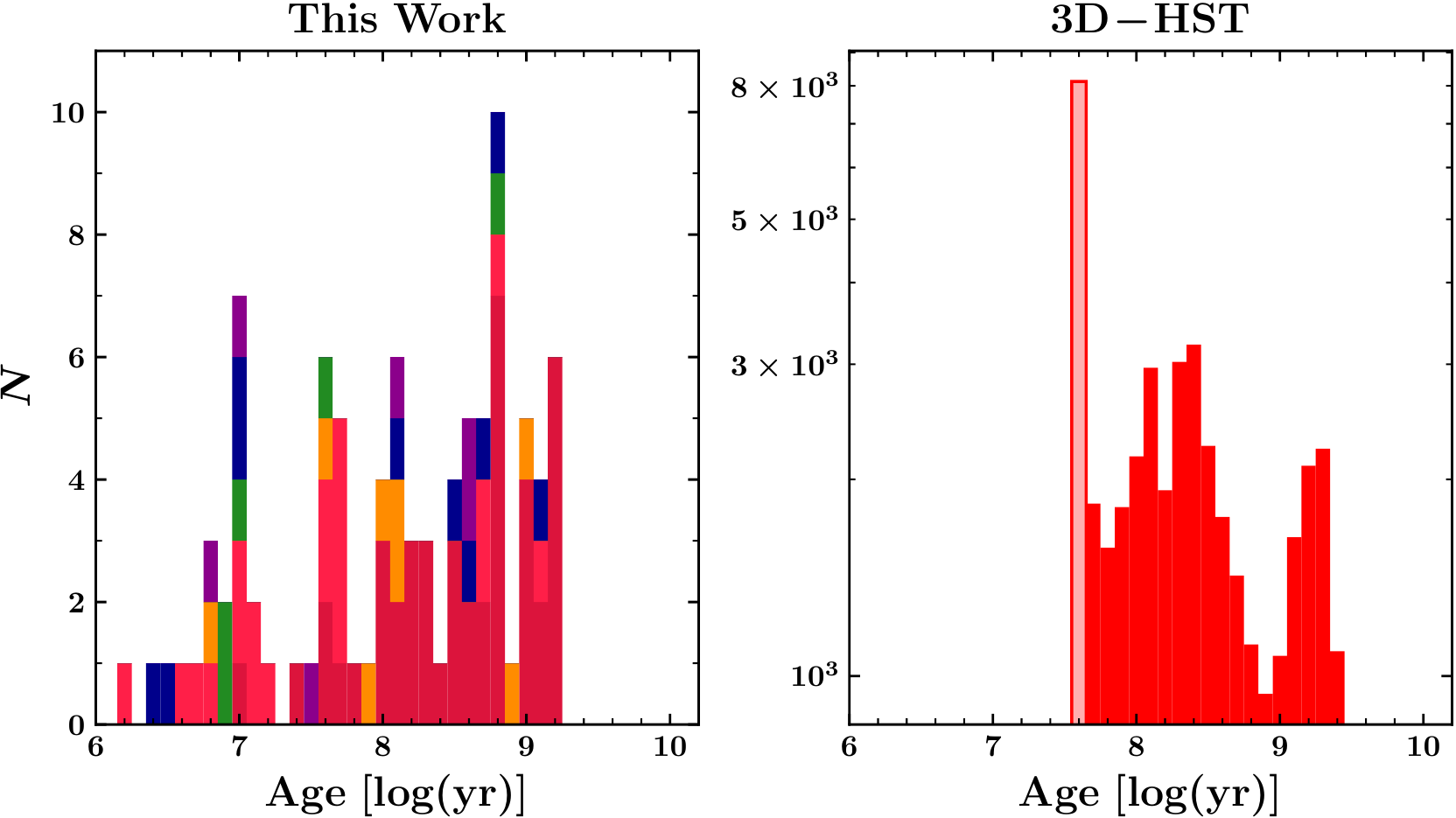}{.497\txw}{}\hspace{5pt}
\fig{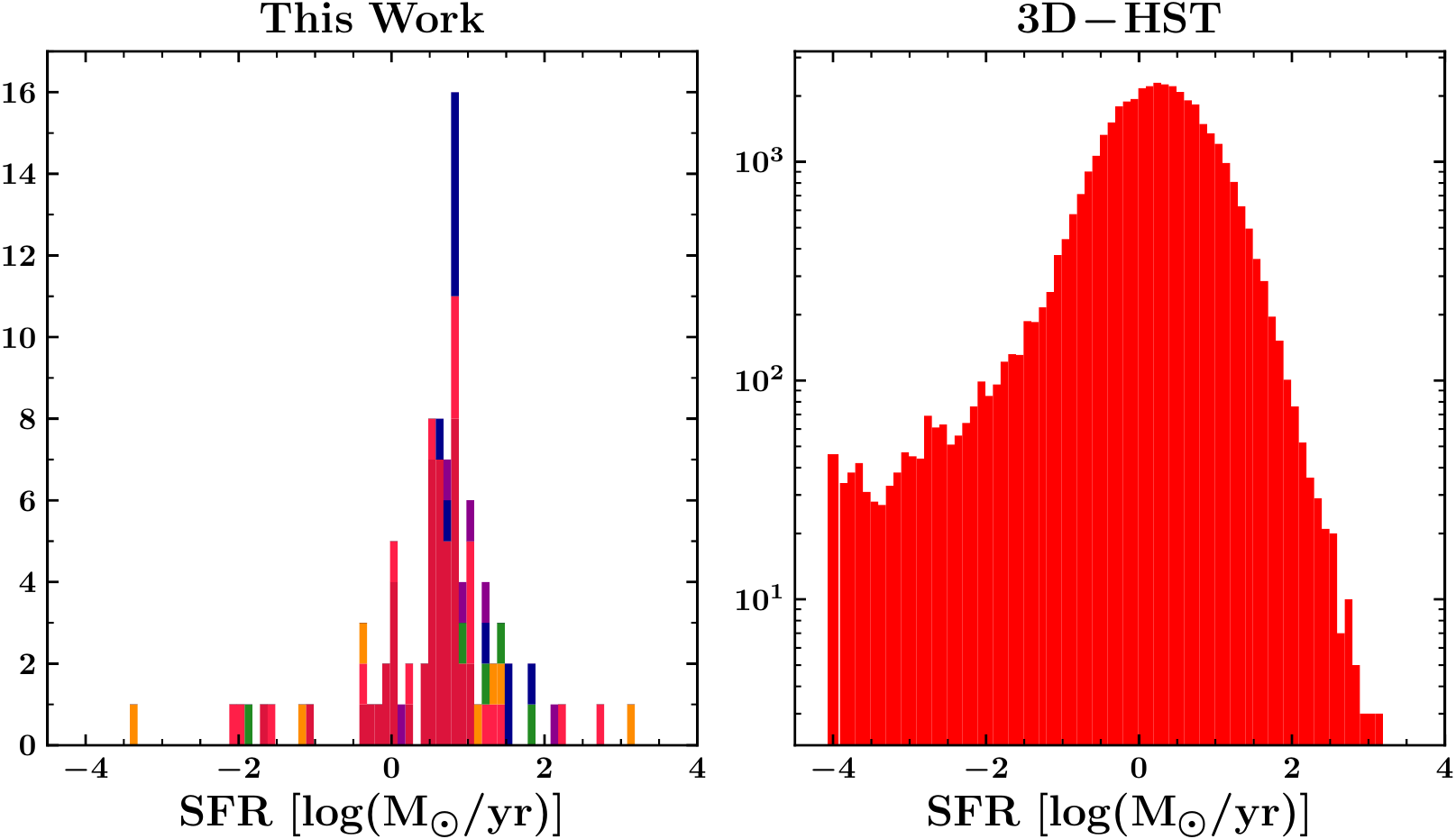}{.488\txw}{}}\vspace*{-20pt}
\caption{Histograms of best-fitting BC03 SED parameters of the GOODS/HDUV+ERS galaxies \emph{without} AGN that have reliable spectroscopic redshifts, compared to the SED parameters in the 3D-HST catalogs \citep{Brammer2012,Momcheva2016} within the same redshift-range (right panel in each pair). The applied SED dust attenuation law is indicated in each of the left panels. The light-red bin in the 3D-HST age histogram indicates that the authors did not fit to SEDs below log(Age/yr)$<$7.6. The y-axis of the 3D-HST mass, age, and SFR histograms are in log-scale and all others are linear. The general shape of the histograms from our sample is similar to the much larger 3D-HST sample of $\sim$42,000 galaxies. \label{paramhists}}
\end{figure*}
We repeated the same selection process of ranking 265 individual objects with 330 spectra as described in \citetalias{Smith2018} with our own quality assessment and criteria. The main objective was to select galaxies with spectra showing multiple, clearly visible emission/absorption lines that align with their expected positions at the stated redshift of the galaxy. These lines include the Lyman Break at 912\AA, \Lya\,1216\AA, \ion{Si}{2}\,1260\AA, \ion{O}{1}\,1304\AA, \ion{C}{2}\,1335\AA, \ion{Si}{4}\,1398\AA, \ion{C}{4}\,1549\AA, and \ion{C}{3}]\,1909\AA, and when present, \ion{C}{2}]\,2326\AA, \ion{Fe}{2}\,2344\AA, and sometimes \ion{N}{5}\,1240\AA, \ion{Fe}{2}\,2600\AA, \ion{Mg}{2}\,2798\AA, \ion{O}{2}\,3727\AA, [\ion{Ne}{3}]\,3869\AA, \ion{He}{2}\,4686\AA, \ion{H}{0}$\beta$\,4861\AA, or [\ion{O}{3}]\,4959+5007\AA. This was to ensure that all galaxies in our sample would not introduce any contaminating, non-ionizing flux into our LyC analyses from erroneous redshift determinations. This selection criterion does bias our sample towards predominantly luminous galaxies about as bright as $M^{*}$ at these redshifts (see Fig.~\ref{maghist}), and also towards galaxies with blue SEDs (see Fig.~\ref{galseds}--\ref{agnseds}). This should be taken into account when interpreting our subsequent LyC analyses on this sample.

Using the multi-band \HST\ image data, we also ensured that each galaxy had no nearby, potentially contaminating neighbors, and that the flux of the galaxy under consideration showed a drop-out in the expected band. Of those 265 unique objects, eight of us selected and agreed on 65 unique objects to have high quality spectra with accurate redshifts. Combining these with the 46 galaxies in the ERS field selected in \citetalias{Smith2018}, our total sample was increased to 111 galaxies with high quality spectra. This sample includes 17 galaxies \emph{with} AGN and 94 galaxies \emph{without} AGN. We identified AGN from typical (broad) emission lines in their spectra, for example \Lya, \ion{N}{5}, \ion{Si}{4}, \ion{C}{4}, \ion{He}{2}, \ion{C}{3}], and \ion{Mg}{2}. We also checked the positions of our spectroscopic sample against \Chandra\ 4 Ms and \textit{Very Large Array} 1.4 GHz source catalogs to identify possible obscured/type II AGN using their radio/X-ray luminosities and photon indices \citep[e.g.,][]{Xue2011,Fiore2012,Miller2013,Rangel2013,Xue2016}.

Our subsequent analyses are based upon the 111 galaxies, 46 of which are located in the ERS field, and 65 in GOODS/HDUV fields, with 56 in the South and 9 in the North. The much larger amount of data we collected in the South is primarily due to the greater efforts in releasing calibrated, publicly available spectra from users of the VLT, which can be accessed from the ESO archive or the survey's website. We perform our LyC analyses on the sample of galaxies selected from each of these fields separately and combined to determine if differences in LyC results are due to differences in the mosaics themselves. We also perform analyses on the sample of galaxies \emph{with} AGN \emph{separately} from galaxies \emph{without} AGN. We further sub-divide these galaxies by field, which results in 9 subsamples. These include galaxies \emph{without} AGN in ERS, GOODS/HDUV, and both fields, which we refer to as ``Total,'' galaxies \emph{with} AGN in ERS, GOODS/HDUV, and Total, and \emph{all} galaxies in our sample (referred to as ``All''), again in ERS, GOODS/HDUV, and the Total sample. 

In Fig.~\ref{maghist} we show the distributions of \mAB\ and \MAB\ sampled at $\mathrm{\lambda_{rest}}\!=\!1500\!\pm\!100$\AA\ for (a) galaxies without AGN signatures in the spectra, (b) galaxies \emph{with} AGN, and (c) the Total sample. The leftmost panel shows that our sample generally follows the luminosity function of galaxies to \MAB\,$\simeq$\,$-$21\,mag (\mAB\,$\simeq$\,24.5\,mag) at their average redshift of \zmean\,$\simeq$\,2.97, with $M^{*}\!=\!-20.63$\,mag and $\alpha\!\,=\,\!1.36$\,dex/mag. The Total sample is seen to be approximately representative of the galaxy LF at their average redshift to \MAB\,$\leq$\,-21\,mag and \mAB\,$\leq$\,24.5\,mag. These histograms are also consistent with the sample of \citetalias{Smith2018}. 

Two galaxies stand out in these histograms, both exceptionally bright galaxies with an AGN. One was found in the GOODS North field measured to be \mAB\,$\simeq$\,20.4\,mag in all observed optical+IR \HST\ bands, and the other was found in GOODS South measured at \mAB\,$\simeq$\,21\, mag in the optical \HST\ bands. The brighter QSO in GOODS North had a significant detection of LyC, while the other showed no detectable flux. We refer to this LyC-detected AGN at $z$\,=\,2.5920 as QSO\,$J$123622.9+621526.7. This QSO was originally detected in LyC by \citet{Jones2018}, and was shown to dominate the LyC emission in the GOODS-North field. We \textit{also} detect the LyC emitted by this AGN in the F275W band, which allows negligible redleak into the filter (see Appendix~B.1 in \citetalias{Smith2018}), at \mAB=23.19$\pm$0.01\,mag. \citet{Bianchi2017} detected this source at \mAB\,=\,23.77$\pm$0.08\,mag with the Galaxy Evolution Explorer (\GALEX) Near-Ultraviolet (NUV) detector as well. Their \GALEX\ Far-Ultraviolet (FUV) flux was determined to be \mAB=26.02\,mag with a signal-to-noise ratio S/N\,$\sim$\,1.8$\sigma$, centered on the flux detected in the NUV, and thus not considered to be a significant FUV detection. We study this object and its LyC emission in more detail in \S\ref{sec:qfesc}.

Since this object has the brightest LyC detection by far, with S/N\,$\simeq$\,133, it will likely dominate any LyC analyses that include it. We therefore study this object individually, combined with all other AGN in GOODS/HDUV, and combined with All galaxies in the ERS, GOODS/HDUV, and the Total samples. We refer to the sample of AGN excluding QSO\,$J$123622.9+621526.7 as the AGN$^{-}$ sample, and likewise the sample of all objects excluding this QSO as the All$^{-}$ sample. Measurements performed on samples that include or exclude this object are useful for understanding cosmological averages of LyC emission for galaxies and AGN at their average redshifts, and the impact of very rare, unusually bright AGN. 
\vspace*{20pt}

\subsection{SED Fitting} \label{sec:sedfitting}
We fit SEDs to each of the 111 objects in our sample of galaxies and AGN with reliable spectroscopic redshifts. We incorporated all available \HST\ WFC/ACS and WFC3/IR photometry longwards of \Lya\ at the fixed spectroscopic redshift of the galaxy. The photometry used for SED fitting was measured from these HST bands using \SExtractor\ \citep{Bertin1996} in dual image mode, with the $\chi^2$-image of \emph{all} HST bands used as the detection image \citep[see][for details]{Szalay1999}. The $\chi^2$-image allowed for source detection and exclusion of fainter interlopers than can be detected in the WFC3/UVIS images alone. The flux measured from the galaxy or AGN in each band was scaled to the the published zeropoint magnitude of the mosaic image where photometry was extracted \citep[see][]{Dickinson2003, Giavalisco2004, Windhorst2011, Momcheva2016}. Weight maps, or inverse-variance maps, and exposure maps were used to determine the uncertainties in the photometry when available. We used two versions of the FAST software \citep{Kriek2009} written in C++ and IDL to fit both galaxies \emph{with} and \emph{without} AGN.
\begin{figure*}[th!]\centerline{
\includegraphics[width=.3395\txw]{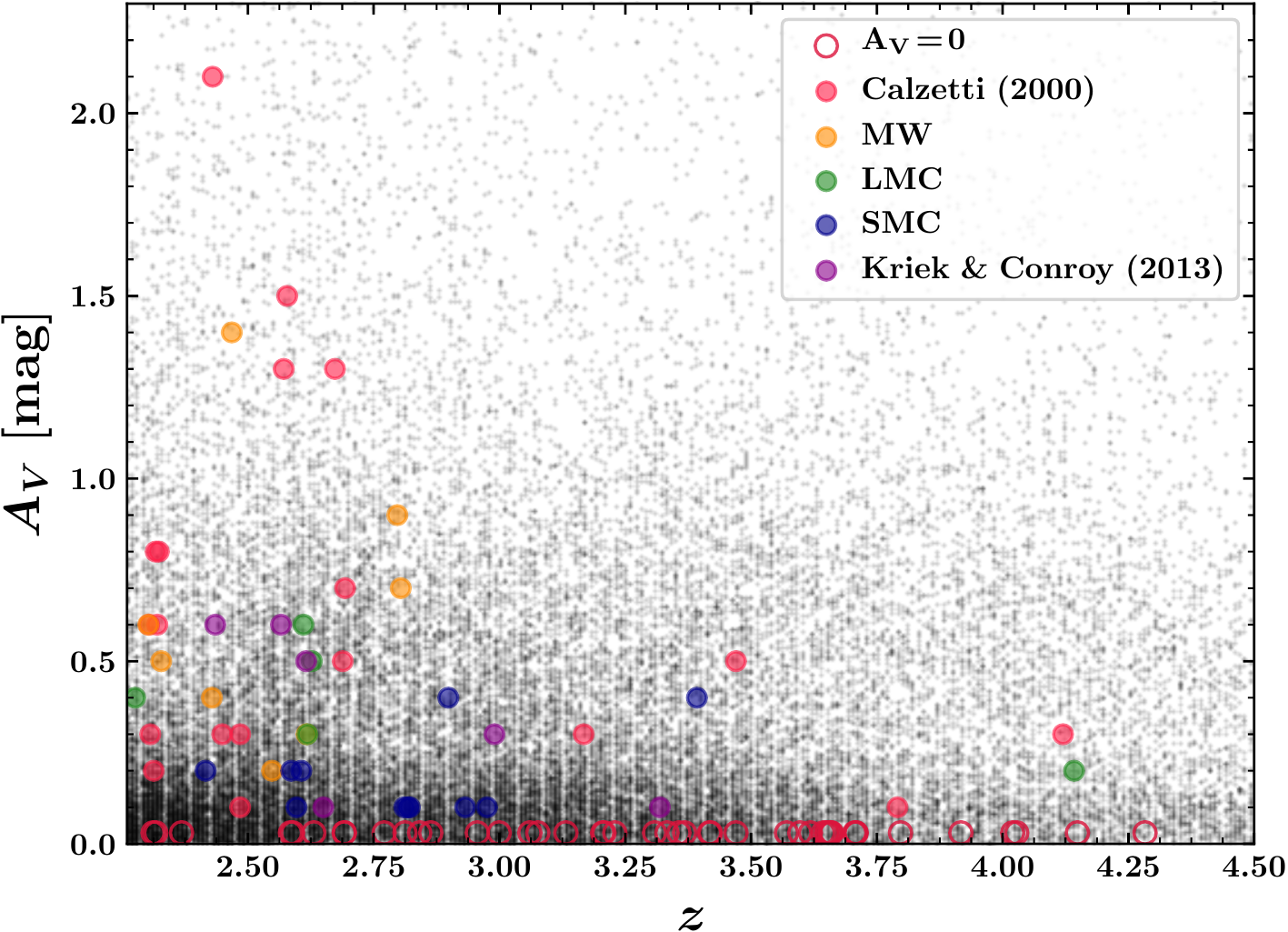}
\includegraphics[width=.3205\txw]{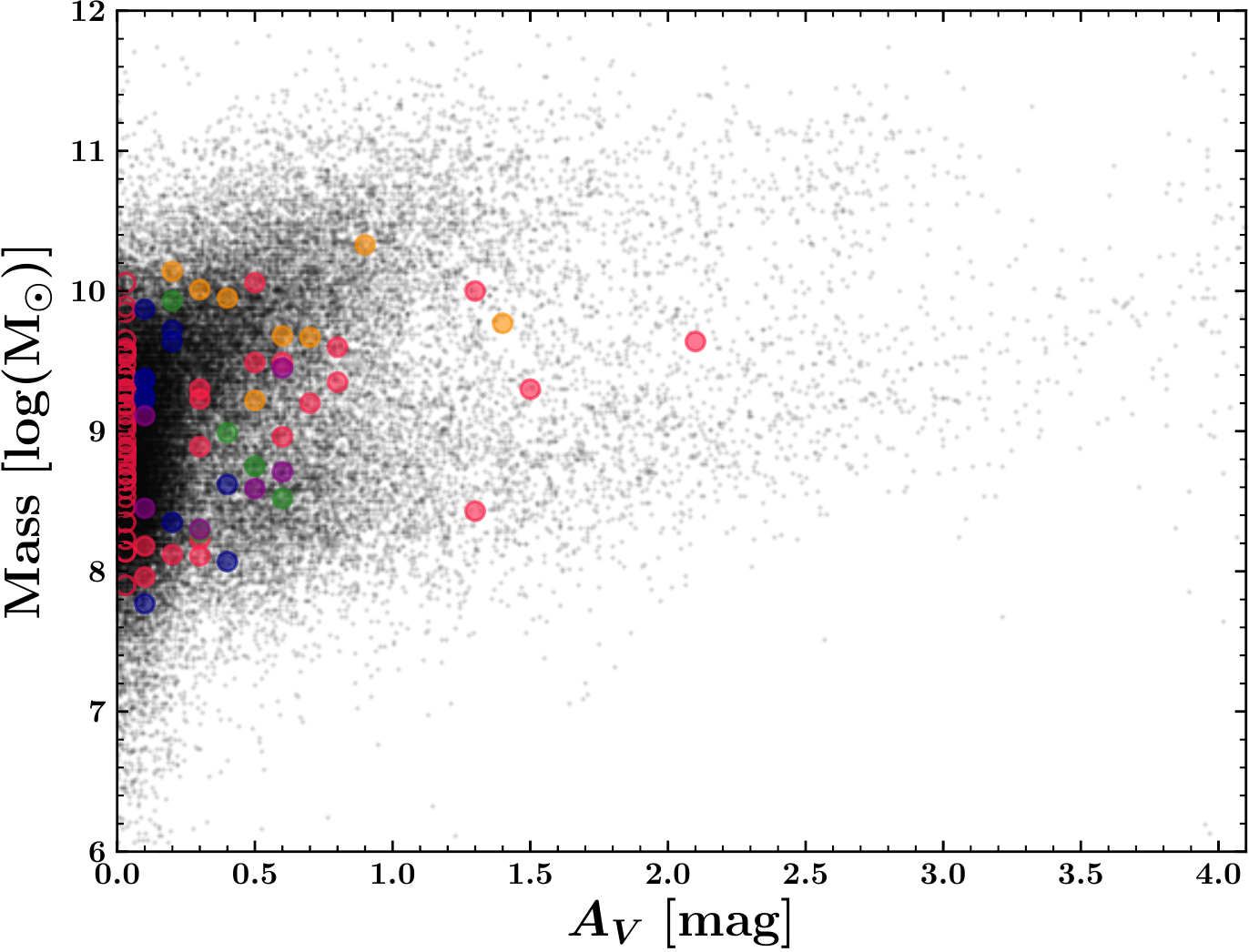}
\includegraphics[width=.3245\txw]{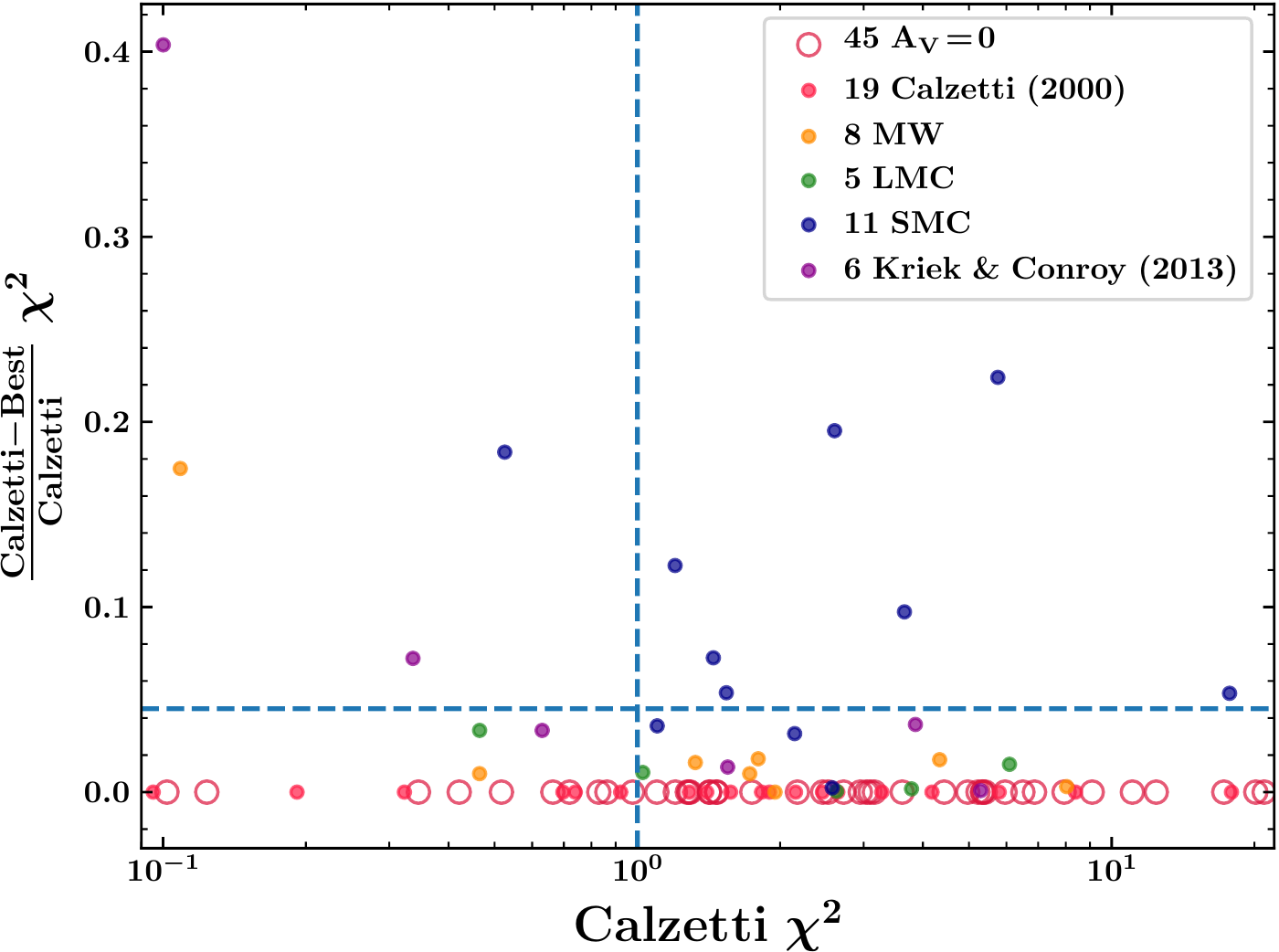}}
\caption{[\textbf{Left}]: $A_V$ vs. spectroscopic redshift of galaxies in our sample taken from their best-fitting BC03 SED, plotted according to their dust-attenuation law. The $A_V$ and redshift of galaxies in the 3D-HST catalog are plotted as black dots for comparison; [\textbf{Middle}]: Same, but comparing the SED log(mass) vs. $A_V$; [\textbf{Right}]: The normalized difference of the $\chi^2$ from the best-fit SED using the \citet{Calzetti2000} dust-attenuation law and the better fitting SED with a different dust-law vs. the original $\chi^2$ using a \citet{Calzetti2000} dust-attenuated SED, indicated by the color of the plotted dot. As indicated by the dashed lines, a subset of the SED fits using a SMC dust-attenuation law shows the greatest improvements in $\chi^2$ compared to fits using \citet{Calzetti2000}. \label{paramplots}}
\end{figure*} 

Due to our large sample of 94 galaxies \emph{without} AGN, we used the C++ based FAST++ program\footnote{\url{https://github.com/cschreib/fastpp}} \citep{Schreiber2018} for fitting their SEDs. FAST++ is advantageous for use with large galaxy samples as it has a faster runtime and supports multi-threading for fitting SEDs in parallel or for parallelizing MC simulations to estimate SED parameter uncertainties. For galaxies \emph{with} AGN, we use the IDL version of FAST\footnote{\url{https://github.com/jamesaird/FAST}} \citep{Kriek2009, Aird2018} since this software can simultaneously fit two-component SEDs, i.e. a SED with a stellar and an AGN component. Example best-fitting SEDs are shown in figs.~\ref{galseds}--\ref{agnseds}, where the best-fit is defined to be the SED fit with the lowest $\chi^2$ value between the available measured WFC/ACS and WFC3/IR photometry and the synthetic photometry calculated from the inner product of the corresponding filter curve and the SED.  

We fit our galaxies \emph{without} AGN to the synthetic stellar SEDs from the \citet{Bruzual2003} (BC03) program GALAXEV, and galaxies \emph{with} AGN to BC03 \emph{and} AGN templates from \citet{Silva2004} and \citet{Polletta2007}. Their best-fitting BC03 parameters are listed in Table~\ref{objtab}, and are also indicated in the example SEDs in Fig.~\ref{galseds}. We also list the best-fitting AGN template and the percentage of flux produced by the AGN component at 5000\AA\ in Fig.~\ref{agnseds} and Table~\ref{objtab}. The AGN template was allowed to vary from 0-100\% of the total SED, in increments of 1\%. 

\noindent\begin{figure*}[th!]\centerline{
\includegraphics[width=.33\txw]{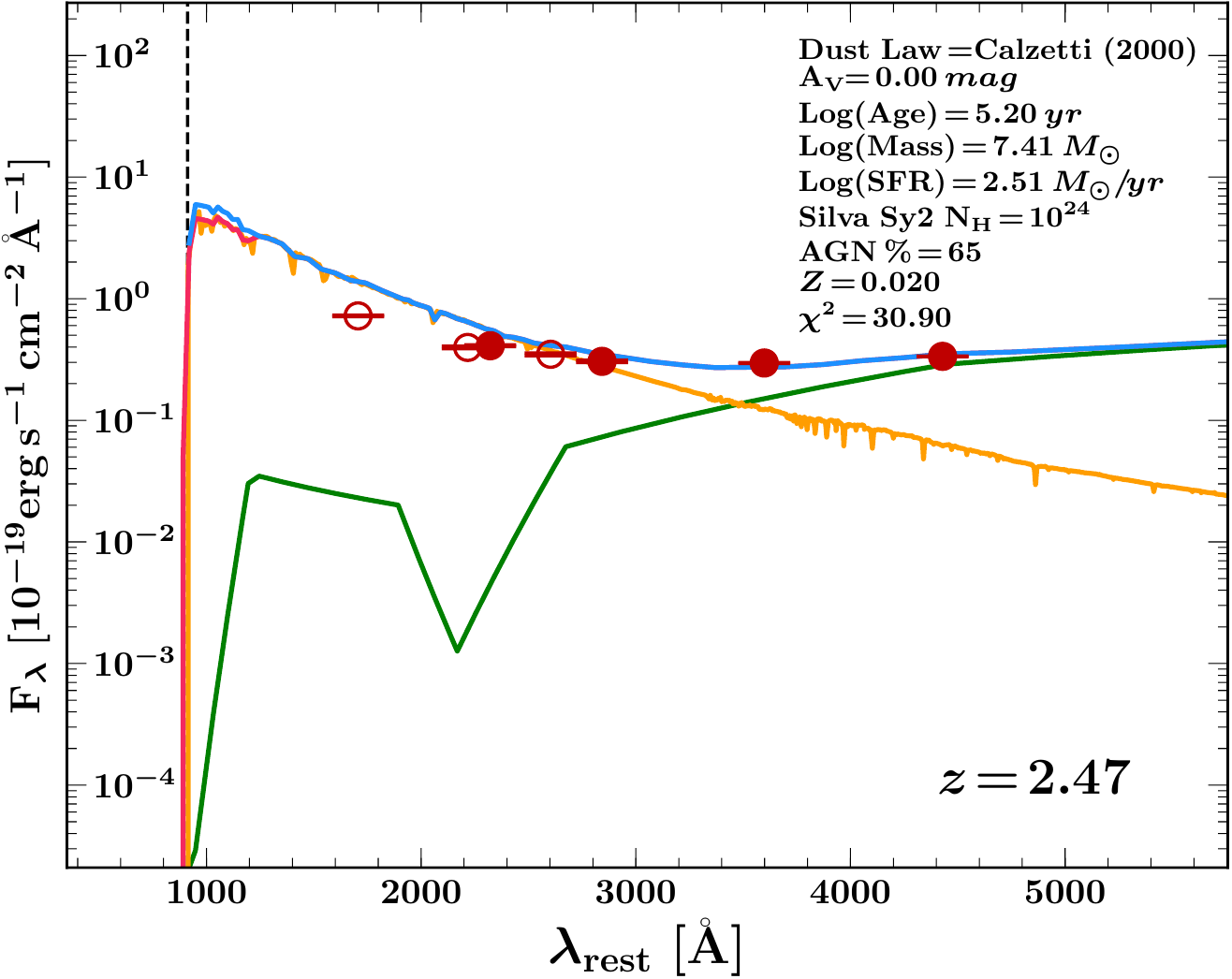}
\includegraphics[width=.33\txw]{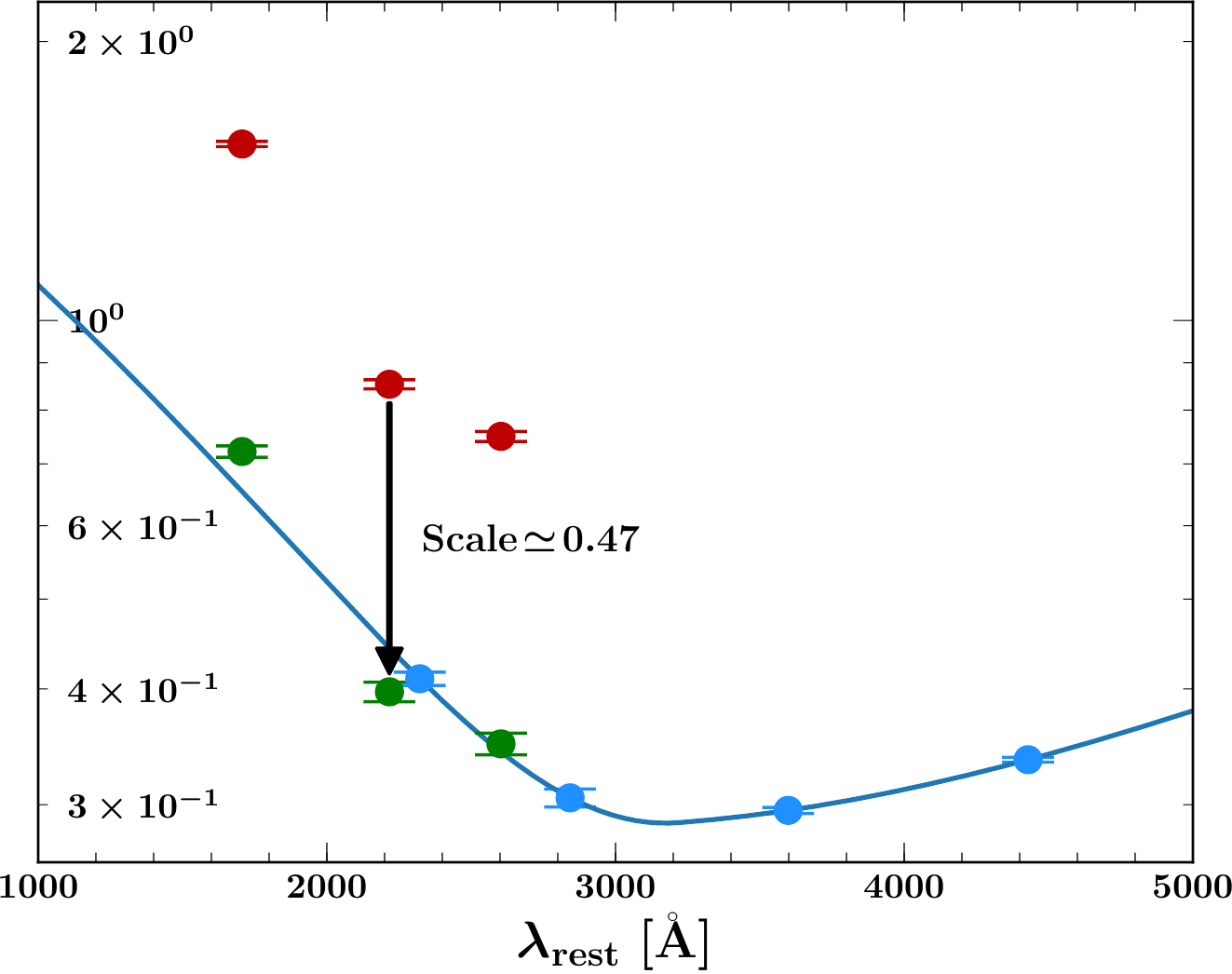}
\includegraphics[width=.311\txw]{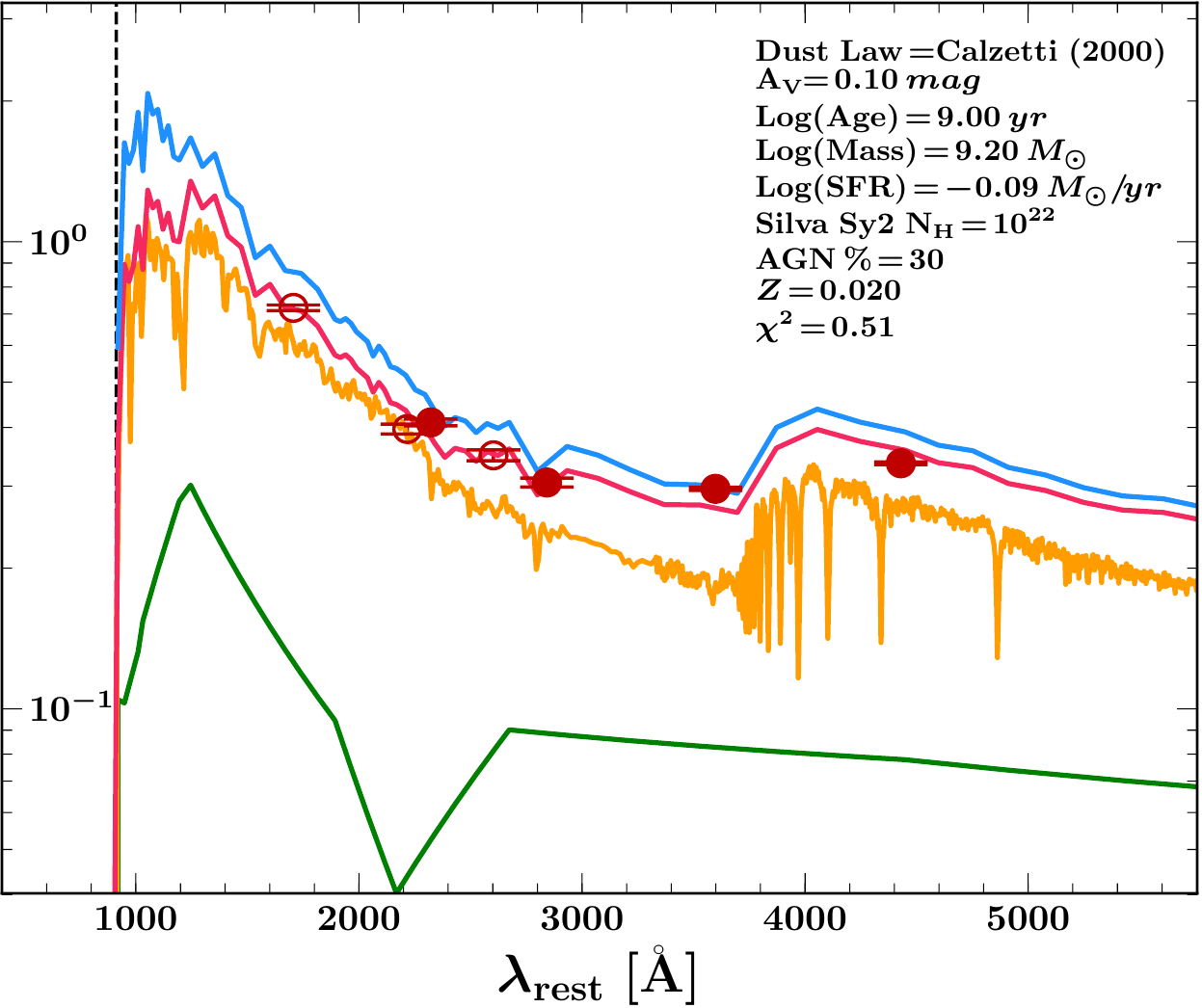}}
\caption{[\textbf{Left}]: Best-fit two-component SED from FAST \citep[red curve;][]{Aird2018} for variable AGN $J$033208.3-274153.6. The best-fit SEDs (red curves) were fit to the plotted \HST\ ACS/WFC and WFC3/IR photometry (red open circles indicate the WFC/ACS data observed during the 2002-2005 epoch, the red filled circles indicate the WFC3/IR data from the 2010-2012 epoch). The red curve is the sum of the best-fit BC03 SED (orange) and the best-fit AGN template \citep[green;][]{Silva2004,Polletta2007}, while the blue curve is equivalent to the red curve corrected for dust extinction. The best-fit BC03 SED parameters (dust extinction law, $\mathrm{A_V}$, age, mass, star-formation rate, and metallicity), AGN template, and AGN SED flux percents at 5000\AA\ are indicated. The difference in observation epoch appears to have provided enough time for $J$033208.3-274153.6 to display variability. [\textbf{Middle}]: The observed flux of $J$033208.3-274153.6 taken in the early epoch (2002-2005; blue points) and the cubic-spline interpolation through those points (solid blue line), compared to the observed flux taken in the later epoch (2010-2012; red points). The red points were scaled down by a factor of $\sim$0.47 to match the interpolated data as closely as possible (green points). [\textbf{Right}]: The best-fit two-component SED from FAST after scaling the later-epoch flux. The $\chi^2$ is seen to improve significantly, and the observed changes in SED parameters can be compared between the two fits. \label{variability}}
\end{figure*}
Histograms of the BC03 SED parameters $A_V$, mass, age, and star-formation rate (SFR) of our galaxies \emph{without} AGN are shown in Fig.~\ref{paramhists}. These parameters are more representative of the dominant UV-bright stellar population since the SEDs were fit to the rest-frame UV and optical photometry, and not necessarily representative of the bulk of the stellar population. We also show the same parameters from SED fits performed by the 3D-HST Collaboration \citep{Brammer2012,Skelton2014,Momcheva2016} for galaxies within the same redshift range $z\!=\!2.26-4.3$. We use a minimum SFR of $10^{-4}$\,$\mathrm{M_{\odot}/yr}$, and the 3D-HST survey restricted the age of their SEDs to $\geq$\,40 Myr. We shade this 3D-HST age-bin light-red, since it is artificially large and encompasses many galaxies that may have younger ages than 40\,Myr. In contrast to the 3D-HST study that only uses the \citet{Calzetti2000} dust-law, we fit all of our galaxies to SEDs attenuated by the \citet{Calzetti2000}, Milky Way (MW), Small Magellanic Cloud (SMC), Large Magellanic Cloud (LMC), and the average dust-law of \citet{Kriek2013} ($A_{1500}/A_V$\,=\,2.55, 2.66, 4.37, 2.79, and 2.91, respectively). Since several of the AGN templates used in the SED fitting included a dusty torus component, we do not apply a secondary dust-attenuation law to the AGN template. The respective best-fitting dust-attenuation law of the BC03 SEDs are color coded in Fig.~\ref{paramhists} in each $A_V$ bin. We also fit SEDs with solar ($Z$=0.02), subsolar ($Z$=0.004, 0.008), and supersolar ($Z$=0.05) metallicities. Despite the constraints in SED parameter space and increase in degrees of freedom in our fits, the profile of our parameter histograms are very similar to those from the 3D-HST SEDs. 

We also compare the redshift $z$ vs. $A_V$ and $A_V$ vs. log(mass) of our galaxies to the 3D-HST sample in Fig.~\ref{paramplots}. This allows us to compare the distributions of these parameters to the 3D-HST sample, and how the parameters correlate with one another in the two samples. Our SED parameters are seen to reside in the densest regions of the 3D-HST parameter space. The few outliers in our sample are seen to be consistent with the less dense regions in the 3D-HST parameter space as well. From these plots, we conclude that our sample resembles the larger 3D-HST sample and is approximately representative of galaxies at these redshifts. 

The rightmost panel in Fig.~\ref{paramplots} compares the $\chi^2$ value of the SED fits using only a \citet{Calzetti2000} dust-law to the $\chi^2$ of SED fits using the MW, SMC, LMC, and \citet{Kriek2013} dust-laws. We plot the \citet{Calzetti2000} $\chi^2$ vs the difference in the best-fitting SED dust-law and the \citet{Calzetti2000} dust-law, normalized by the \citet{Calzetti2000} dust-law $\chi^2$. We find mostly marginal improvements in $\chi^2$ for most cases. However, we find a subset of galaxies with substantial improvements when the best-fitting SED used a SMC dust-law. We indicate these larger improvements using a dashed horizontal line. This suggests that simply using a \citet{Calzetti2000} dust-law may not result in the most accurate SED fits for this subsample. 

\subsection{AGN Variability} \label{sec:qvar}
In order to maximize the accuracy of our SED fits and \fesc\ analyses that use them for modeling intrinsic LyC, we searched for galaxies with AGN that display obvious signs of variability and corrected the photometry to be consistent with flux measured closest to the time the WFC3/UVIS LyC data were taken. For a galaxy with variable flux, the photometry measured from non-coeval data in different bands may not fit well to a SED based on a static, physically-based stellar or AGN model. The resulting fit may therefore not accurately predict flux in bands not used in the fitting, e.g. bands used for LyC photometry. In order to improve the SED fit to a variable AGN, we must scale non-coeval flux to be the most consistent with photometry across all observed bands. 

Our first indication of AGN variability in $J$033208.3-274153.6 was the resulting poor SED-fit ($\chi^2$=30.9; see Fig.~\ref{variability}). The two-component SED fitting of FAST has an added degree of freedom compared to a single-component BC03 template, and the majority of AGN SED $\chi^2$-values were comparatively much lower, as seen in Fig.~\ref{agnseds}. The optical WFC/ACS data, indicated by red-open circles in the left and right panels in Fig.~\ref{variability}, were taken by the GOODS and Type Ia SNe surveys \citep{Riess2007} and collected from July~2002--Feb.~2003 in cycle 11 and from Apr.~2004--Feb.~2005 in cycles 12--13. The WFC3/IR and ACS/WFC F814W data, indicated in those panels by red-filled circles, were collected in Aug.~2010--Feb.~2012 by CANDELS. The wavelength captured by the CANDELS F814W photometry lies between the GOODS F775W and F850LP photometric data, and is seen to differ substantially from these earlier data points. The only other possibility to explain this offset might be a very bright emission line. However, no lines exist in this spectral region. $J$033208.3-274153.6 was also found to be variable by \citet{Sarajedini2011} at a high significance. This AGN was one of 85 to be classified as varying, with their population statistics showing a 74\% chance of variability in broad-line AGN and 15\% in narrow-line AGN, and a 2\% chance among all galaxies they surveyed. This roughly reflects our population as well, which was 1/17$\sim$6\% variable for our AGN sample and 1/111$\sim$1\% in our total sample of galaxies. 

In the middle panel of Fig.~\ref{variability}, we show how much the non-coeval data differs from one another. Here, we interpolated the later epoch WFC3/IR CANDELS photometry (blue circles) using a spline function (blue curve), then determined the average multiplicative difference between this curve and the observed early epoch WFC/ACS data (red circles). We find that the optical ACS/WFC flux decreased by a factor of $\sim$2.1 over the timespan between two epochs of $\sim$5--6 years. The \citet{Sarajedini2011} study found variability in $J$033208.3-274153.6 over a span of $\sim$6 months, however, our findings may indicate variability of this object over a timespan of years as well. We scale the flux down by this factor and perform the same fitting on this object with the modified photometry to compare the two fits. We see a substantial improvement in the $\chi^2$ value by a factor of $\sim$60 down to $\chi^2$=0.51 as shown in the right panel of Fig.~\ref{variability}, which is consistent with our 16 other two-component SED fits listed in Table~\ref{objtab} and Fig.~\ref{agnseds}. Since this AGN clearly displays variability in its photometry, we include $J$033208.3-274153.6 only after correcting the photometry for variability and re-fitting the SED to the corrected photometry. Subsequent analyses on samples with AGN that include $J$033208.3-274153.6 use this corrected photometry and the SED fit shown in the right panel of Fig.~\ref{variability}.

\begin{figure*}[th!]\centerline{
\includegraphics[width=.342\txw]{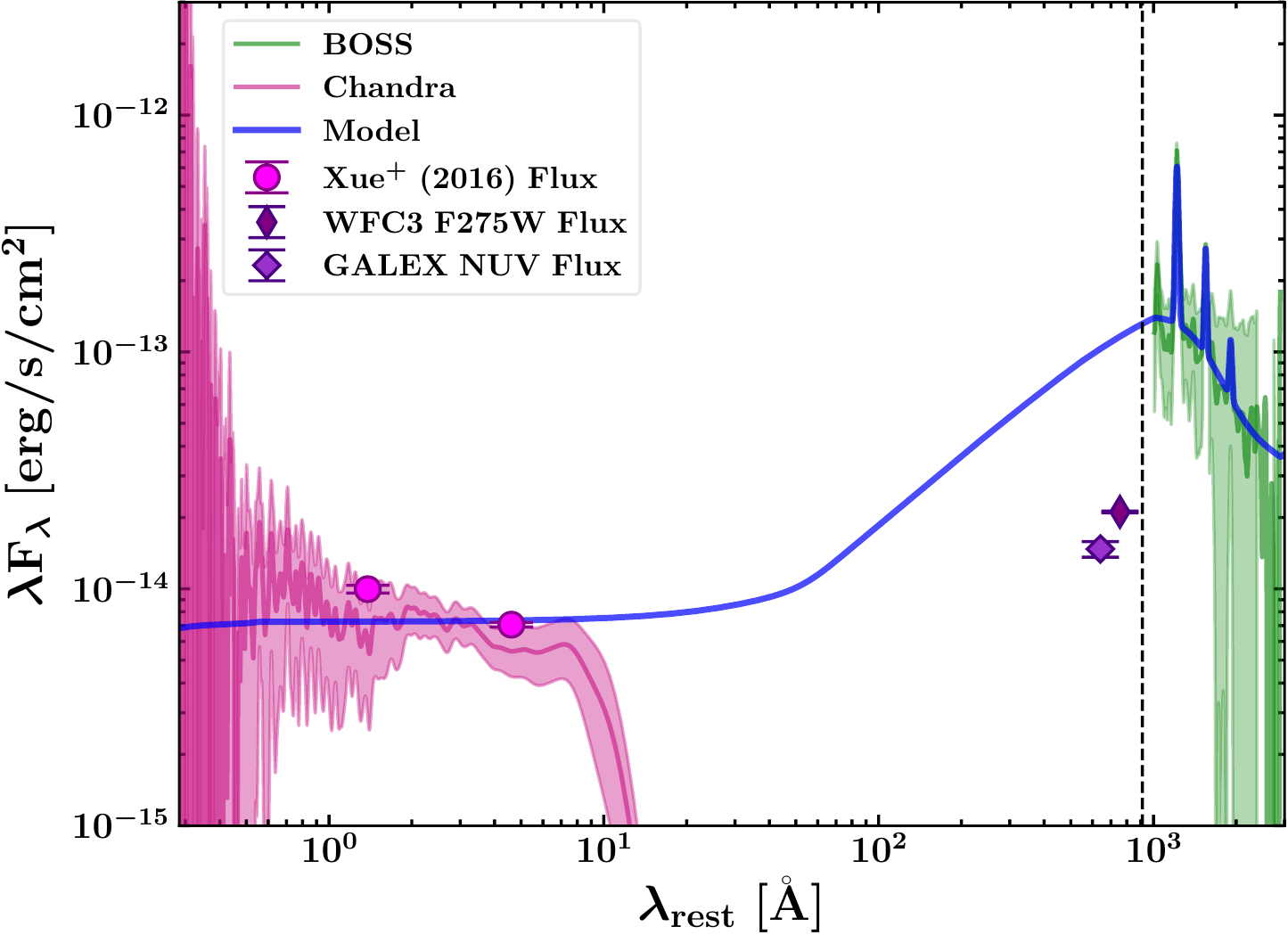}\hspace{5pt}
\includegraphics[width=.338\txw]{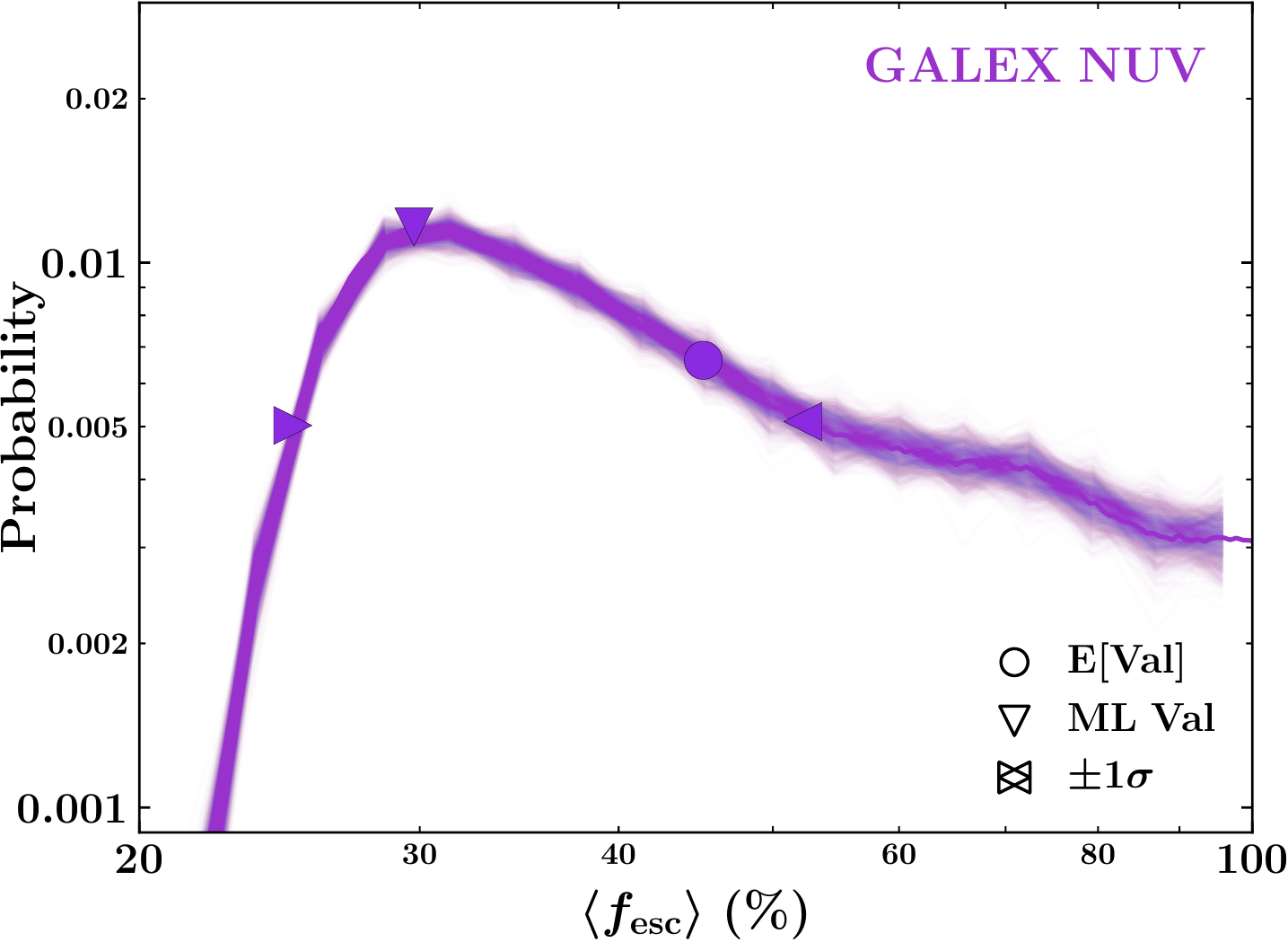}
\includegraphics[width=.309\txw]{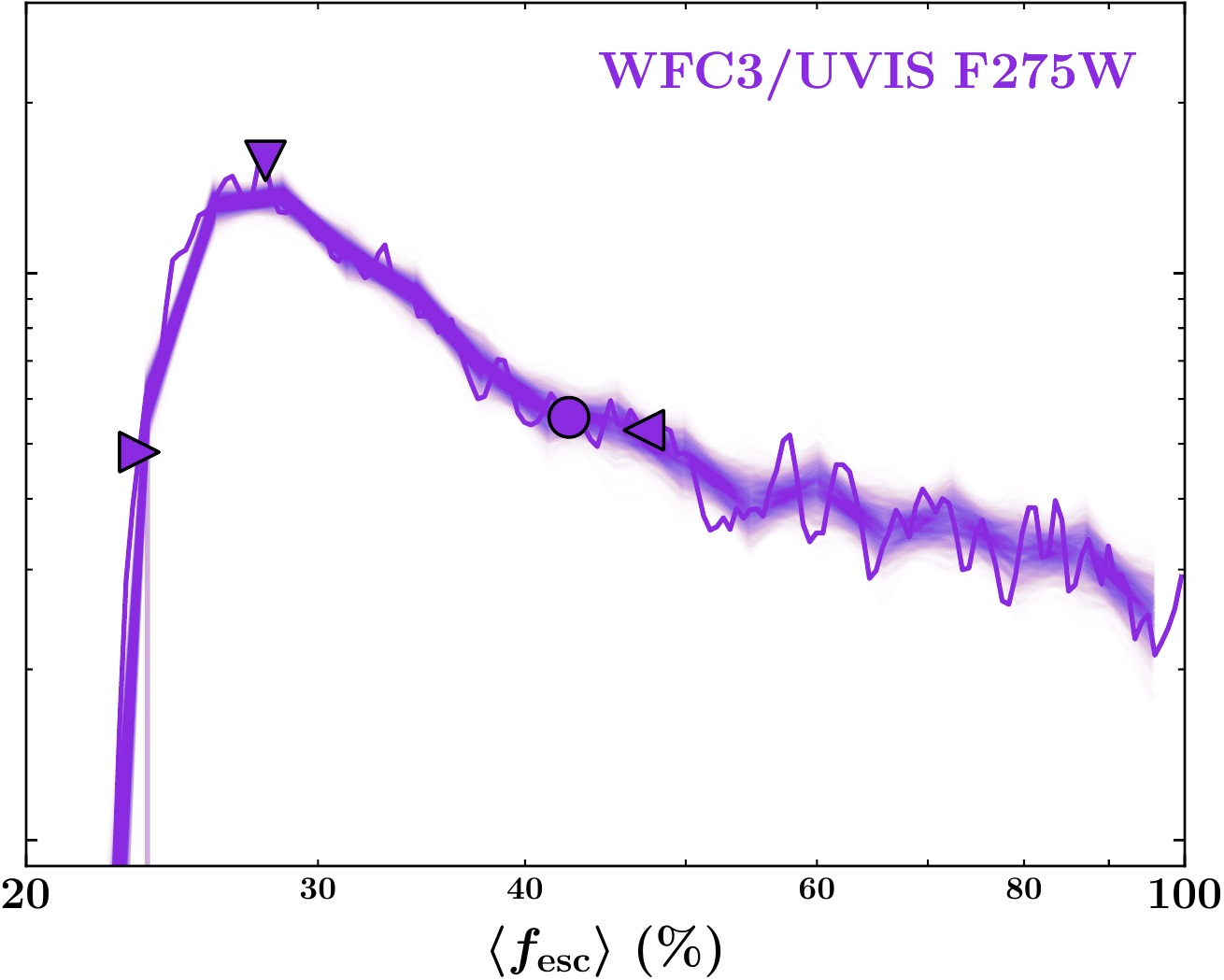}}
\caption{[\textbf{Left}]: SDSS BOSS \citep{Dawson2013} spectrum of QSO\,$J$123622.9+621526.7 and $\pm$1$\sigma$ uncertainty (green) together with its \Chandra\ ACIS Spectrum (magenta). The OPTXAGNF \citep{Done2012} model was fit to the \Chandra\ and BOSS spectra simultaneously using the XSpec software \citep[blue;][]{Arnaud1996}. The \emph{observed} \GALEX\ NUV and WFC3/UVIS F275W fluxes are plotted as light and dark violet diamonds, respectively, and the \Chandra\ soft (0.5--2.0\,keV) and hard (2--7\,keV) band fluxes as measured by \citet{Xue2016} are shown as magenta circles. [\textbf{Middle}]: The probability mass function (PMF) of escaping LyC flux derived from the \GALEX\ NUV flux. The shaded region represents the 1000 \fesc\ MC simulations, and the dark violet line is the combination of all 1000 simulations. The estimated LyC escape fraction for \GALEX\ NUV is estimated to be $\fesc^{\mathrm{NUV}}\!\simeq\!30^{+22}_{-5}$\%. [\textbf{Right}]: The PMF of escaping LyC flux observed in the WFC3/UVIS F275W filter. The estimated LyC escape fraction here is estimated to be $\fesc^{\mathrm{F275W}}\!\simeq\!28^{+20}_{-4}$\%. \label{brightqso}}
\end{figure*}

\section{Quasar LyC Detections and Escape Fractions} \label{sec:qphotfesc}
Our sample contains a single object (QSO\,$J$123622.9+621526.7) with a highly significant individual detection of LyC at \mAB=23.19$\pm$0.01\,mag in WFC3/UVIS F275W (S/N\,$\simeq$\,133) and with \GALEX\ NUV at \mAB=23.77$\pm$0.08\,mag (S/N\,$\simeq$\,13). This object is one of two bright QSOs as shown in Fig.~\ref{maghist}, each with \MAB\,$\leq$\,--24.5, and the other bright QSO is located in GOODS South at $z$=2.8280. \emph{Only QSO\,$J$123622.9+621526.7 displays a LyC detection from WFC3/UVIS F275W imaging and GALEX NUV imaging and is also highly significant in S/N, while QSO\,$J$033209.4-274806.8 at $z$\,=\,2.8280 in GOODS South is not detected in LyC.} This dichotomy in LyC detection between two similarly continuum-bright sources warrants further investigation and possible explanations for this emission. We therefore study QSO\,$J$123622.9+621526.7 in more detail in an attempt to infer why this AGN has such a bright LyC signal while other AGN in our sample show no significant LyC flux individually. 

\subsection{Quasar SED Fitting}\label{sec:qsosed} 
To characterize this LyC bright QSO, we first determine a physically-based model that fits well to all the available observations, from the \Chandra\ X-ray to the optical \HST\ data. We selected the OPTXAGNF model \citep{Done2012} since it incorporates the accretion disk black-body emission \emph{and} the optically thin and optically thick Comptonization components of the inner disk and SMBH corona. The Comptonization component of the OPTXAGNF model accounts for AGN SED flux from $\lambda$$\simeq$1--900\,\AA, which is important for modeling the intrinsic AGN LyC. We use the XSpec software \citep{Arnaud1996} to fit our observed data to this model, a method also used by \citet{Lusso2015} to characterize their stack of \zmean=2.4 AGN \HST\ WFC3/UVIS grism spectra. There are several input parameters in this model corresponding to physical properties of the AGN\footnote{See \url{https://heasarc.gsfc.nasa.gov/xanadu/xspec/manual/} for a full list and description of all parameters in the model.}, some of which we determined from available archival and published data and were held fixed during fitting. The first parameter we determined was the SMBH mass using the observed \ion{C}{4} line from the SDSS BOSS spectrum \citep{Dawson2013} released in DR14. We used the method from \citet{Coatman2016} to estimate this mass by first fitting the continuum-subtracted \ion{C}{4} line to a 6th order Gauss-Hermite polynomial \citep{Marel1993,Cappellari2002}, which allows for more robust estimations of the FWHM of the line. We then estimated the line's blueshift to correct the FWHM. This corrected \ion{C}{4} FWHM corresponded to a SMBH mass of log($\mathrm{\frac{M_{SMBH}}{M_{\odot}}}$)\,$\simeq$\,8.37, a mass similar to the SMBH mass of the Andromeda galaxy \citep{Bender2005}. 

Using this mass, we calculated the Eddington luminosity of this AGN to be $\mathrm{L_{Edd}}$$\simeq$2.9$\times$10$^{46}$\,erg/s. We then computed the bolometric luminosity to be $\mathrm{L_{bol}}$$\simeq$1.5$\times$10$^{46}$\,erg/s using the methodology of \citet{Runnoe2012} at $\lambda_{\rm rest}$\,=\,1450\,\AA. These parameters result in an Eddington ratio of $\mathrm{\lambda_{Edd}}$$\simeq$0.5, which, along with the SMBH mass, is typical of X-ray selected AGN \citep[see, e.g.,][]{Lusso2012}. We infer an accretion rate of $\dot{M}$$\simeq$3.4\,$\mathrm{M_{\odot}/yr}$ from $\mathrm{L_{bol}}$, assuming a matter-radiation conversion efficiency of $\epsilon$=8\% \citep{Marconi2004}. If this accretion rate represents the average accretion, the SMBH would have been accreting for $\sim$7.3$\times$$10^7$ years. We use the X-ray spectral index $\Gamma$$\simeq$1.687 from the \citet{Xue2016} catalog as input. The measured soft (0.5--2.0\,keV) and hard (2--7\,keV) X-ray fluxes from \citet{Xue2016} are plotted as magenta filled circles in the left panel of Fig.~\ref{brightqso} for reference. 

Our full XSpec model used a Tuebingen-Boulder ISM absorption model \citep{Wilms2000} to account for the X-ray absorption by the MW ISM, and five additional Gaussian profiles for fitting bright emission lines in the spectra. After inputting the GOODS North hydrogen column density of 1.6$\times$10$^{20}$\,cm$^{-2}$ \citep{Stark1992}, our calculated mass, comoving distance, and log($\mathrm{L_{Edd}}$), we simultaneously fit the SDSS BOSS spectrum and the \Chandra\ spectrum plotted in the left panel of Fig.~\ref{brightqso}. Before fitting, we first corrected the BOSS spectrum for aperture losses using the measured \HST\ ACS flux shown in Fig.~\ref{brightqso}. The \Chandra\ X-ray spectrum, background spectrum, and response curve were then extracted from the Chandra Deep Field North \citep[CDFN;][]{Brandt2001} data taken with the Advanced CCD Imaging Spectrometer \citep[ACIS;][]{Garmire2003} using the CIAO software \citep{Fruscione2006} tool \texttt{specextract}\footnote{\url{http://cxc.harvard.edu/ciao/threads/pointlike/}}. Because the CDFN data\footnote{For specific datasets, see: \url{http://cxc.harvard.edu/cda/DefSet/CDFN1.html} and \url{http://cxc.harvard.edu/cda/DefSet/CDFN2.html}} was taken at two different roll angles in Feb.~2000--Feb.~2002, we reprojected the CDFN event data onto a common tangent-plane using CIAO tool \texttt{reproject\_events}. We then fit these spectra using the Levenberg-Marquardt method \citep{Levenberg1944,Marquardt1963} to simultaneously minimize a combination of the $\chi^2$ and ``cstat'' \citep{Cash1979} statistics for the BOSS optical and \Chandra\ X-ray spectra, respectively. 

The resulting model is shown as the blue curve in Fig.~\ref{brightqso}, which was scaled by the response curve extracted from the combined CDFN ACIS data. The model also returned values for the dimensionless blackhole spin parameter $a_{\star}$\,=\,0.57, coronal radius $r_{cor}$\,=\,5.3\,$r_{s}$ where $r_s$ is the Schwarzschild radius, and electron temperature $T_e$\,$\simeq$\,1.4$\times$10$^{5}$\,K. We use this best-fit XSpec model to compute the \fesc\ values from the measured WFC3/UVIS F275W and \GALEX\ NUV flux as described in \S\ref{sec:qfesc}. 
\begin{figure*}[thp!]\centerline{
\includegraphics[width=.2951\txw]{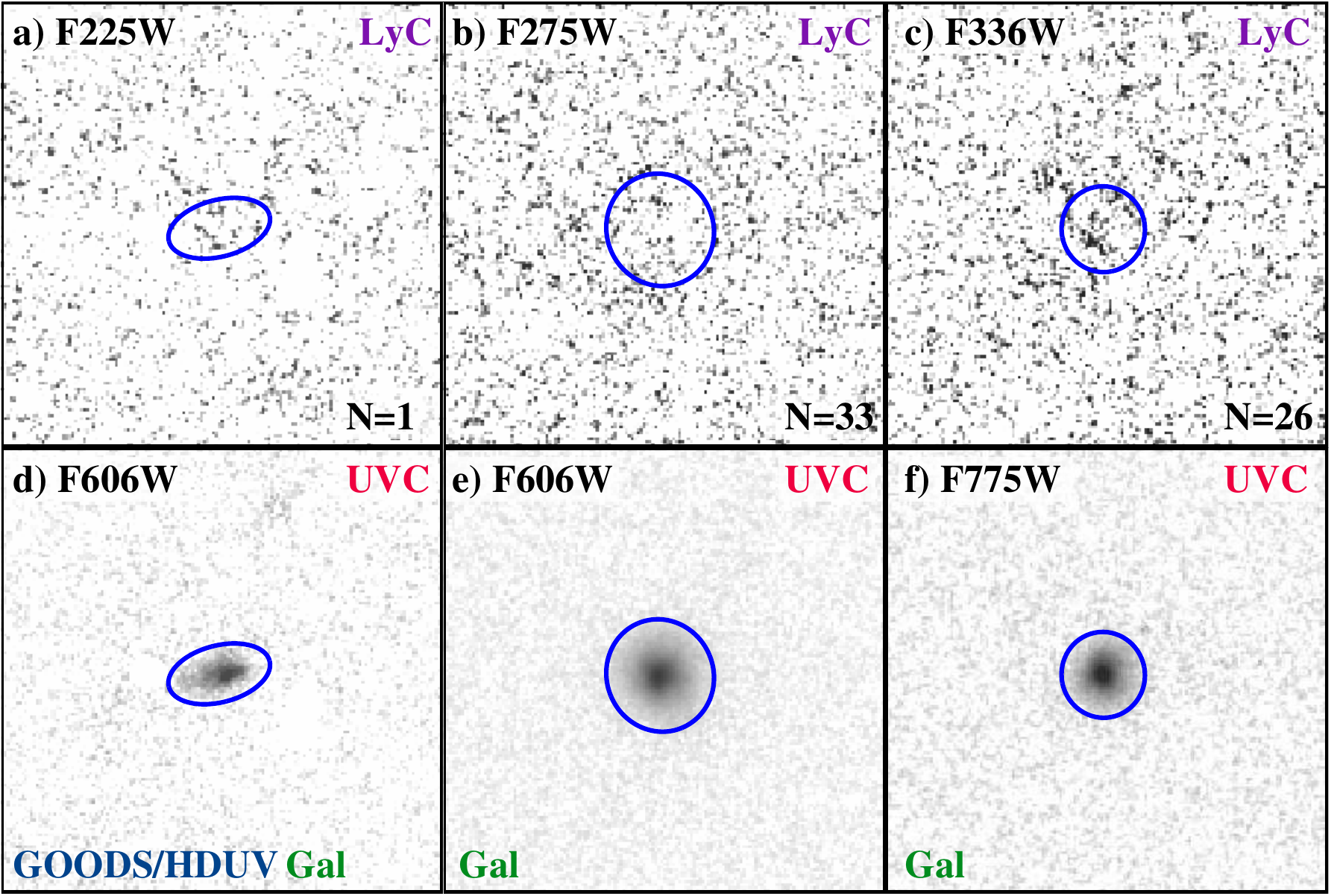}
\includegraphics[width=.2951\txw]{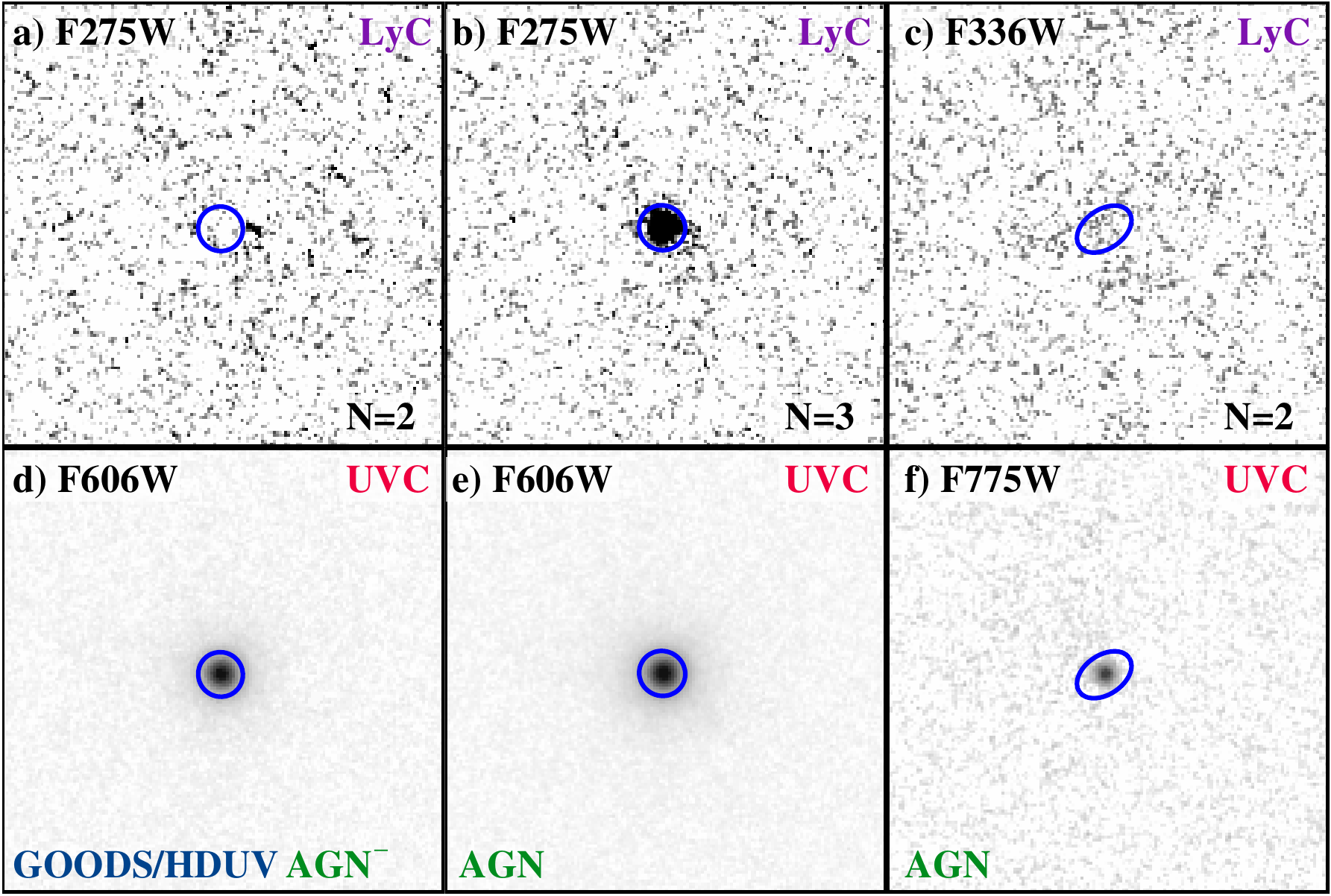}
\includegraphics[width=.4\txw]{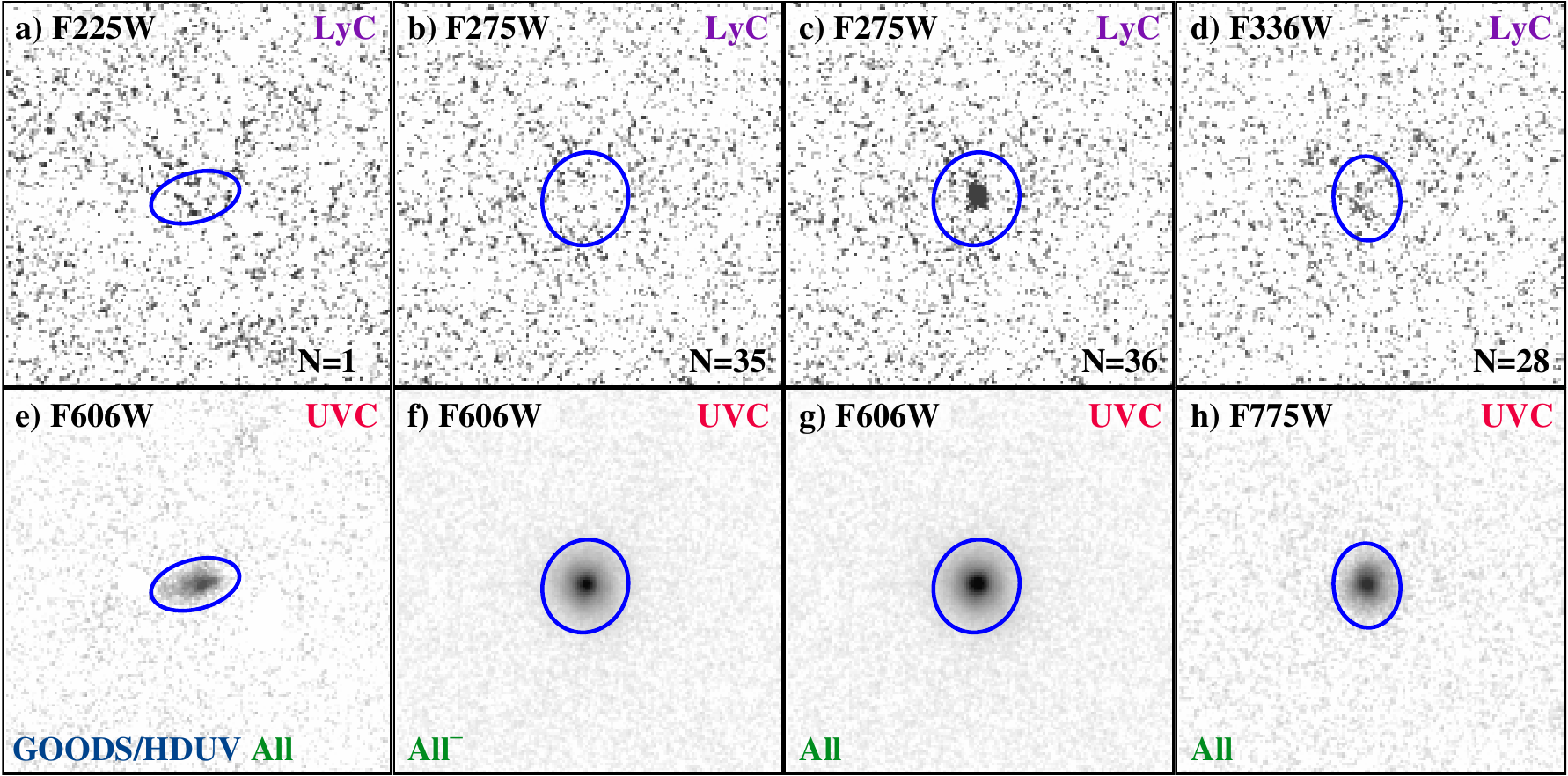}}
\caption{Sub-image stacks for the three different redshift bins in our sample selected from the GOODS/HDUV field for the galaxies \emph{without} AGN that have reliable spectroscopic redshifts (indicated by Gal in green), galaxies \emph{with} AGN (AGN in green), and \emph{all} galaxies (All in green). These stacks sample LyC emission in F225W, F275W, and F336W and the corresponding F606W and F775W stacks sample the UVC ($\sim$1400\,$\lesssim$\,$\lambda_{\mathrm{rest}}$\,$\lesssim$\,1800\AA) emission. The AGN$^{-}$ label indicates the exclusion of the LyC-bright QSO\,$J$123622.9+621526.7 from the stack. Blue ellipses indicate the \SExtractor\ \texttt{MAG\_AUTO} UVC detected matched apertures. All sub-images are 151$\times$151 pixels (4\farcs53$\times$4\farcs53) in size. \label{GOODSstacks}}
\end{figure*}
\begin{figure*}[thp!]\centerline{
\includegraphics[width=.33\txw]{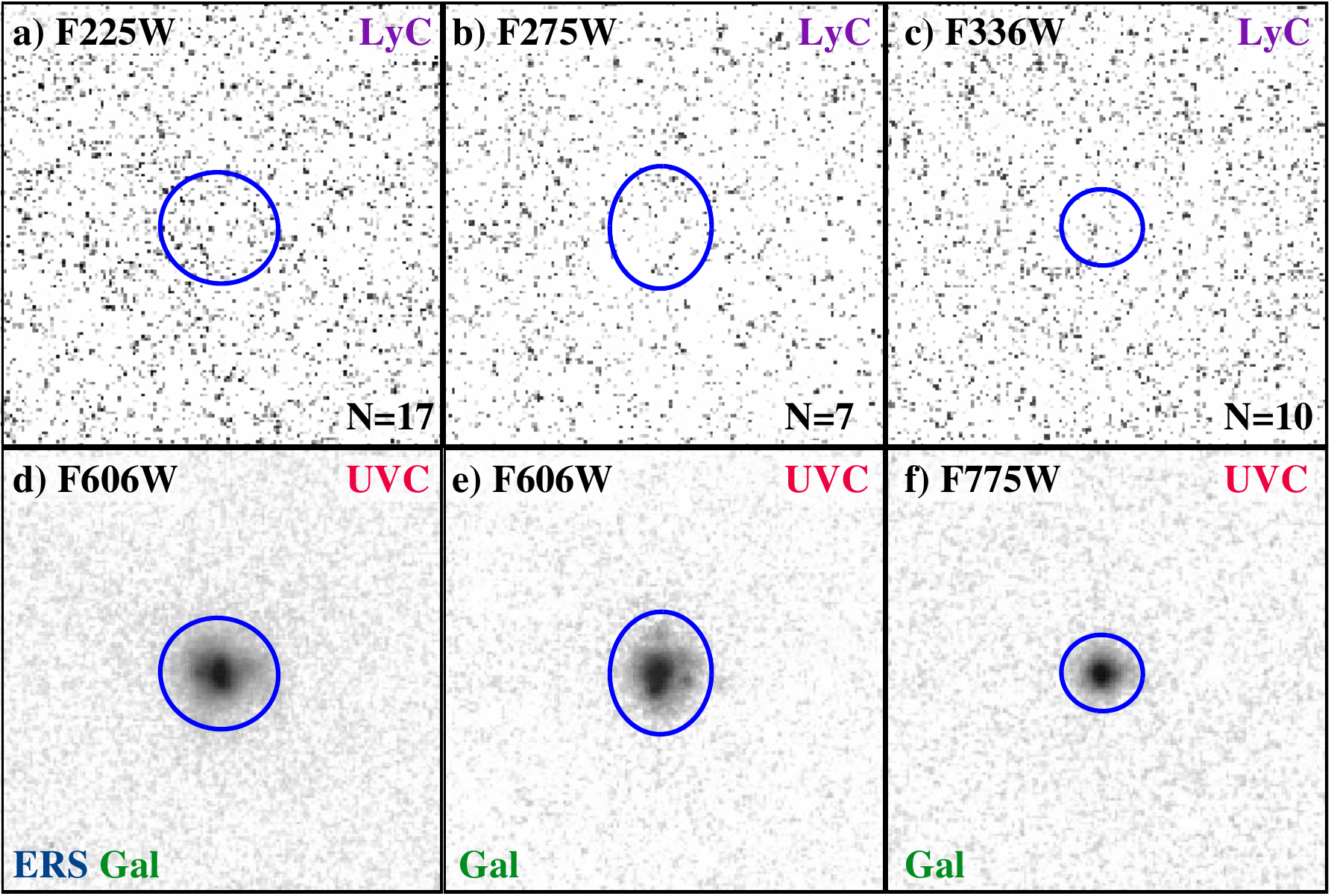}
\includegraphics[width=.33\txw]{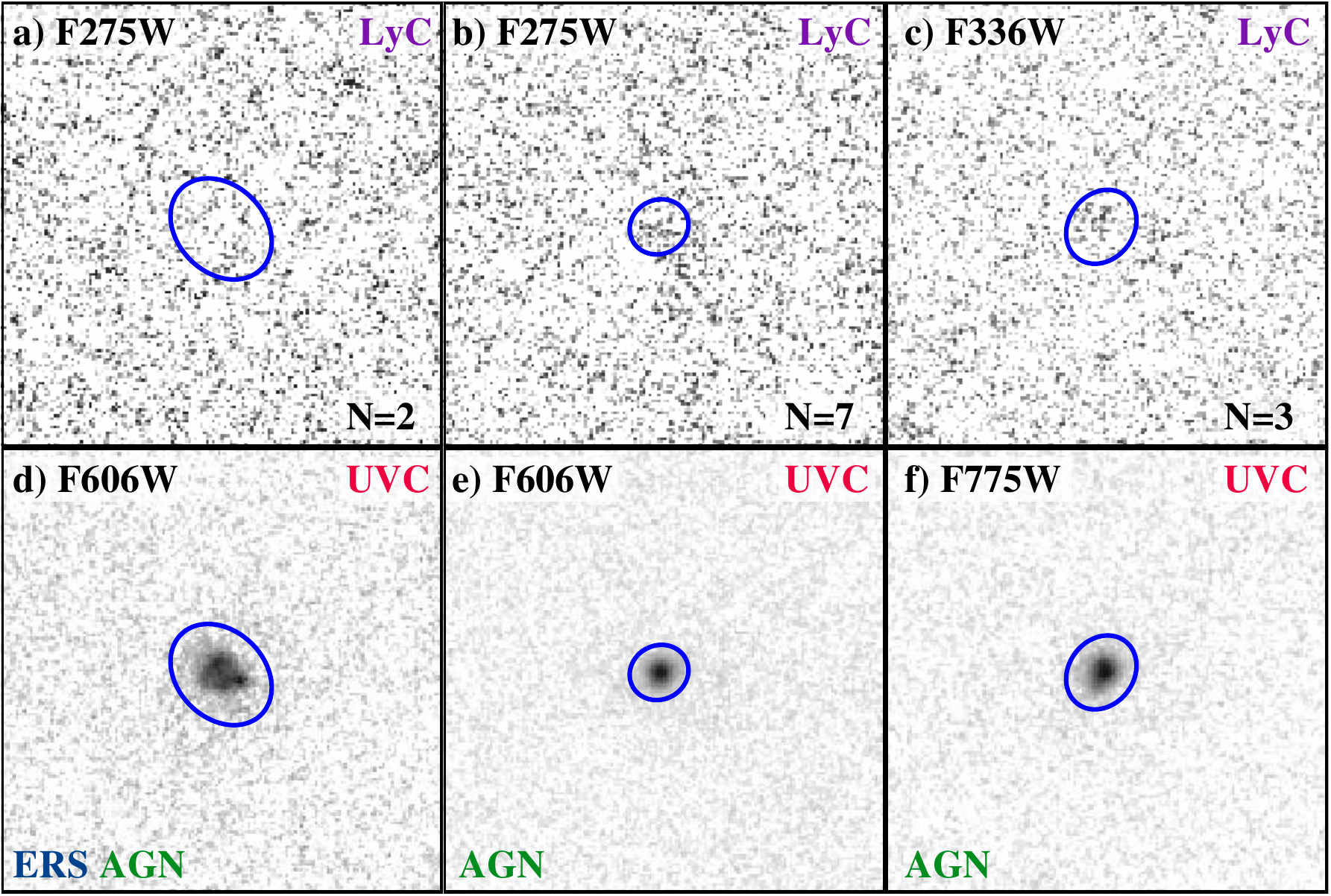}
\includegraphics[width=.33\txw]{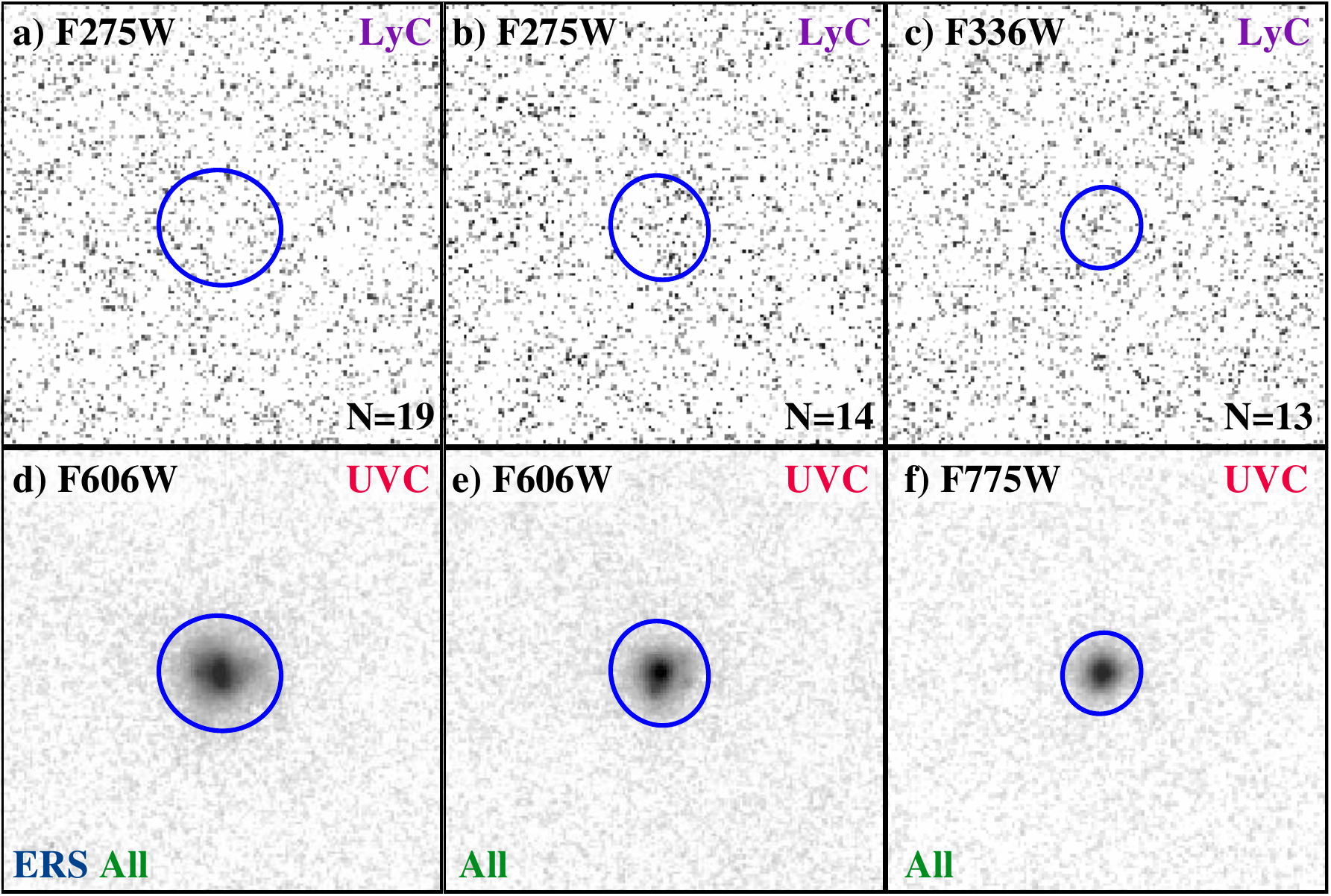}}
\caption{Same as Fig.~\ref{GOODSstacks}, but for galaxies selected from the ERS field. \label{ERSstacks}}
\end{figure*}
\begin{figure*}[thp!]\centerline{
\includegraphics[width=.267\txw]{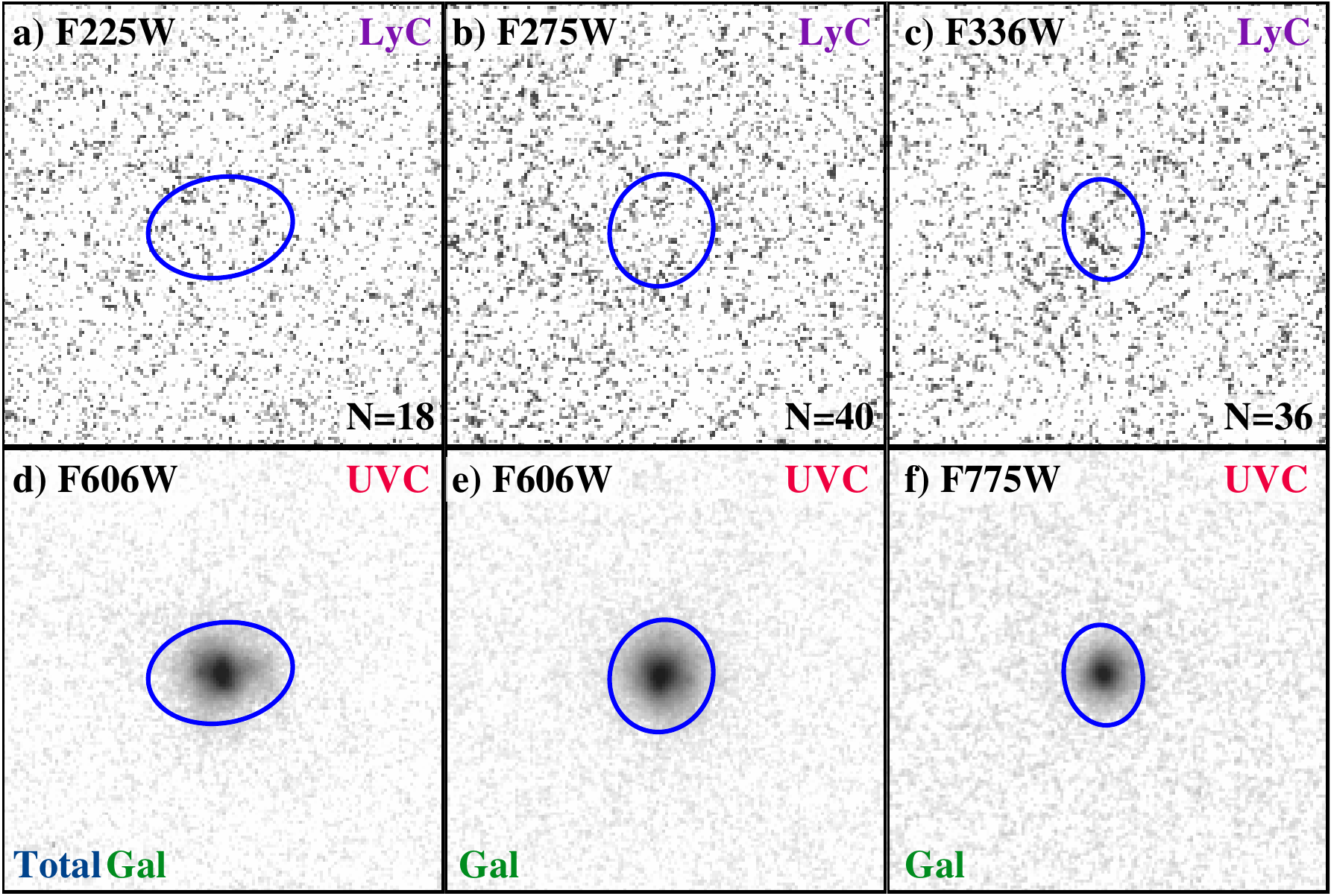}
\includegraphics[width=.363\txw]{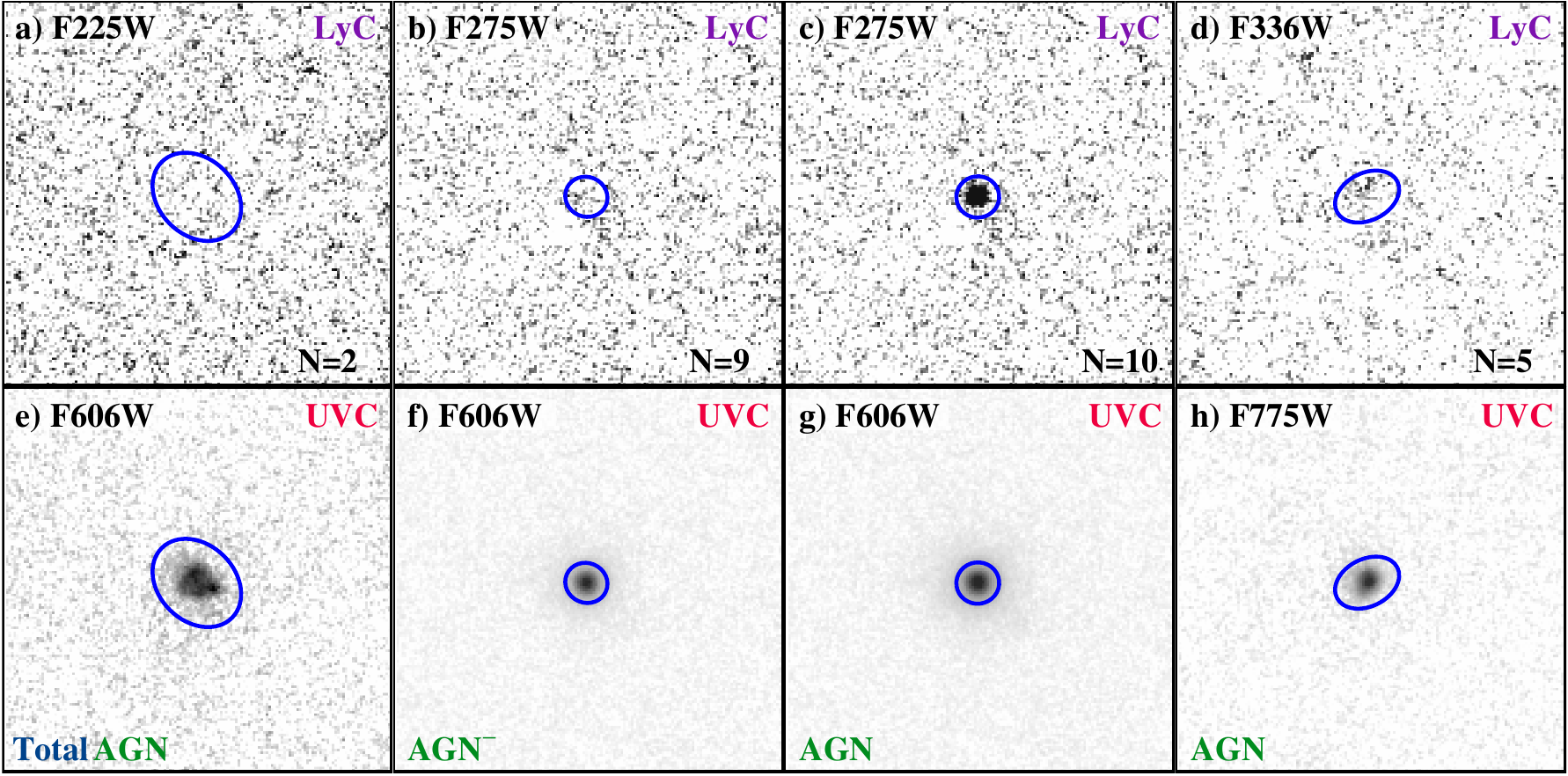}
\includegraphics[width=.363\txw]{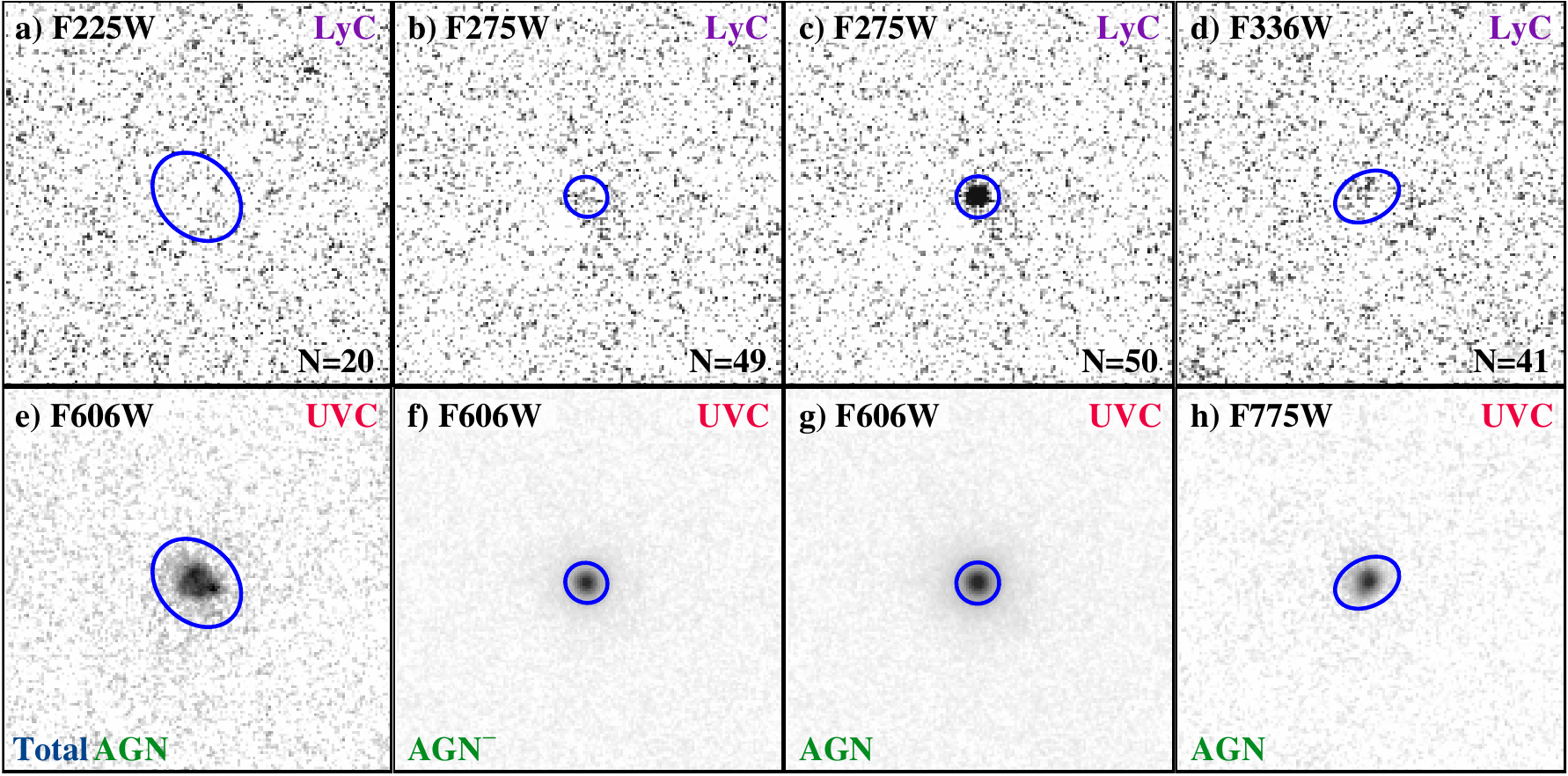}}
\caption{Same as Fig.~\ref{GOODSstacks}, but for galaxies from the ERS and GOODS/HDUV fields combined, or the ``Total'' sample. \label{GOODSERSstacks}}
\end{figure*}

\subsection{Quasar LyC Escape Fractions from \GALEX\ and WFC3/UVIS}\label{sec:qfesc}
We estimate the LyC escape fraction for all QSO\,$J$123622.9+621526.7 LyC measurements using the method from \citetalias{Smith2018}. In summary, we modeled the intrinsic LyC flux from QSO\,$J$123622.9+621526.7 by taking our best-fitting \emph{dust-free} SED from XSpec described in \S\ref{sec:qsosed} and attenuating it with the simulated line-of-sight intergalactic medium (IGM) transmission curves of \citet{Inoue2014}, then taking the inner-product of the attenuated SED and its respective LyC filter curve and \HST\ optical throughput. This calculation gives us the modeled, intrinsic LyC flux in this band, i.e., 
\begin{equation}
\mathrm{f_{\lyc,mod}}\!=\!\frac{\int F_{\rm SED}(\nu)\mathcal{T}_{\igm}(\nu,z)T^{\!\lyc}_{\rm obs}(\nu)\frac{d\nu}{\nu}}{\int T^{\!\lyc}_{\rm obs}(\nu)\frac{d\nu}{\nu}}\end{equation}
where $F_{\rm SED}(\nu)$ is the dust-free SED, $\mathcal{T}_{\igm}(\nu,z)$ is the IGM transmission at redshift $z$, and $T^{\!\lyc}_{\rm obs}(\nu)$ is the filter transmission+optical throughput used for the LyC observation. Components of the \emph{total} system throughput used in this model include the reflectivity of the telescope mirrors, quantum efficiency of the detectors, obscuration by the secondary mirrors, and transmission of the filter used for observation \citep{Morrissey2007,Kalirai2009}. The integrated product of the dust-free SED, IGM transmission, and filter curve+optical throughput simulates the effective intrinsically produced LyC from stellar and AGN components, and does not account for ISM effects captured by the \fesc\ parameter.

We perform $10^3$ MC simulations of the LyC \fesc\ parameter using $10^4$ line-of-sight IGM attenuation curves from the IGM absorber distribution-based \citet{Inoue2014} code. These attenuation curves were used to generate distributions of $10^4$ \emph{modeled} intrinsic LyC flux values as described above. Using the LyC flux measured with \SExtractor\ \citep{Bertin1996} from our GOODS North F275W mosaic described in \S\ref{sec:data} and the \GALEX\ NUV flux from \citet{Bianchi2017}, we modeled the \emph{observed} LyC fluxes ($\mathrm{f_{\lyc,obs}}$) as Gaussian Random Variables (GRV) in our MC runs with the mean of the Gaussian $\mu$ representing the measured flux values and the standard deviation of the Gaussian $\sigma$ representing the uncertainty of the measurements. We generated $10^4$ random values in these Gaussian distributions, which we used to estimate the \fesc\ values shown in Fig.~\ref{brightqso}. To generate our LyC \fesc\ distribution, we simply calculate the ratio of the observed LyC flux distribution to the modeled LyC flux distribution, i.e. $\fesc\!=\!f_{\lyc,obs}/f_{\lyc,mod}$, where $f_{\lyc,obs}$ and $f_{\lyc,mod}$ are the flux values in arbitrary linear units. We perform the \fesc\ calculation $10^3$ times on the generated $f_{\lyc,obs}$ and $f_{\lyc,mod}$ distributions in a MC fashion to get more robust statistics of the probability distributions of the \fesc\ values. The $10^3$ \fesc\ distributions for the measured \HST\ WFC3/UVIS F275W and \GALEX\ NUV fluxes are shown in Fig.~\ref{brightqso} as the group of transparent light and dark violet curves, respectively.

\begin{deluxetable*}{lcccrcccccr}
\centering\tablecaption{Lyman Continuum Stack Photometry\label{phottab}}
\tablewidth{\txw}
\tabletypesize{\scriptsize}
\tablehead{
\colhead{Filter} & \colhead{$z$-range} & \colhead{\zmean} & \colhead{$N_{obj}$} & \colhead{m$_{\!\lyc}$} & \colhead{ABerr$_{\!\lyc}$} & \colhead{S/N$_{\!\lyc}$} & \colhead{A$_{\uvc}$} & \colhead{m$_{\uvc}$} & \colhead{ABerr$_{\uvc}$} & \colhead{S/N$_{\uvc}$} \\
\multicolumn{1}{c}{(1)} & (2) & (3) & (4) & (5) & (6) & (7) & (8) & (9) & (10) & \multicolumn{1}{c}{(11)}}
\startdata
\multicolumn{10}{l}{\sc \underline{GOODS/HDUV}}\\
\multicolumn{10}{l}{\sc Galaxies without AGN:}\\
F225W & 2.4680-2.4680 & 2.4680 & 1 & $>$25.63 & \nodata & $(1.0)^{\ddagger}$ & 0.49 & 25.006 & 0.095 & 11 \\
F275W & 2.4845-3.0604 & 2.7263 & 33 & $>$26.94 & \nodata & $(1.0)^{\ddagger}$ & 1.00 & 24.674 & 0.010 & 110 \\
F336W & 3.1673-4.2830 & 3.6093 & 26 & $>$27.51 & \nodata & $(1.0)^{\ddagger}$ & 0.59 & 25.218 & 0.025 & 43 \\
\multicolumn{10}{l}{\sc Galaxies with AGN:}\\
F225W & \nodata & \nodata & \nodata & \nodata & \nodata & \nodata & \nodata & \nodata & \nodata & \nodata \\
F275W$^{-}$ & 2.5760-2.8280 & 2.7020 & 2 & $>$26.36 & \nodata & $(1.0)^{\ddagger}$ & 0.16 & 21.509 & 0.002 & 690 \\
F275W & 2.5760-2.8280 & 2.6653 & 3 & 24.66 & 0.17 & 6.23 & 0.17 & 21.118 & 0.001 & 1179 \\
F336W & 3.1930-3.6609 & 3.4270 & 2 & $>$26.89 & \nodata & $(1.0)^{\ddagger}$ & 0.20 & 25.508 & 0.078 & 14 \\
\multicolumn{10}{l}{\sc All Galaxies:}\\
F225W & 2.4680-2.4680 & 2.4680 & 1 & $>$25.63 & \nodata & $(1.0)^{\ddagger}$ & 0.96 & 25.006 & 0.094 & 12 \\
F275W$^{-}$ & 2.4845-3.0604 & 2.7249 & 35 & $>$27.45 & \nodata & $(1.0)^{\ddagger}$ & 0.51 & 23.922 & 0.004 & 261 \\
F275W & 2.4845-3.0604 & 2.7212 & 36 & 27.61 & 0.97 & 1.12 & 0.87 & 23.422 & 0.003 & 423 \\
F336W & 3.1673-4.2830 & 3.5962 & 28 & $>$27.45 & \nodata & $(1.0)^{\ddagger}$ & 0.62 & 25.224 & 0.024 & 46 \\
\midrule
\multicolumn{10}{l}{\sc \underline{ERS}}\\
\multicolumn{10}{l}{\sc Galaxies without AGN:}\\
F225W & 2.2760-2.4490 & 2.3496 & 17 & $>$27.69 & \nodata & $(1.0)^{\ddagger}$ & 1.08 & 24.430 & 0.012 & 92 \\
F275W & 2.5658-3.0762 & 2.7516 & 7 & $>$27.87 & \nodata & $(1.0)^{\ddagger}$ & 1.03 & 24.402 & 0.013 & 81 \\
F336W & 3.1320-4.1486 & 3.6029 & 10 & $>$32.56 & \nodata & $(1.0)^{\ddagger}$ & 0.51 & 24.843 & 0.023 & 47 \\
\multicolumn{10}{l}{\sc Galaxies with AGN:}\\
F225W & 2.2980-2.4500 & 2.3740 & 2 & $>$27.07 & \nodata & $(1.0)^{\ddagger}$ & 0.81 & 25.215 & 0.049 & 22 \\
F275W & 2.4700-2.7260 & 2.6184 & 7 & 28.32 & 0.83 & 1.30 & 0.27 & 25.102 & 0.014 & 80 \\
F336W & 3.2171-3.4739 & 3.3157 & 3 & 27.58 & 0.64 & 1.69 & 0.42 & 24.494 & 0.023 & 46 \\
\multicolumn{10}{l}{\sc All Galaxies:}\\
F225W & 2.2760-2.4500 & 2.3522 & 19 & $>$27.73 & \nodata & $(1.0)^{\ddagger}$ & 1.16 & 24.481 & 0.012 & 90 \\
F275W & 2.4700-3.0762 & 2.6850 & 14 & $>$27.70 & \nodata & $(1.0)^{\ddagger}$ & 0.84 & 24.681 & 0.013 & 87 \\
F336W & 3.1320-4.1486 & 3.5366 & 13 & $>$28.83 & \nodata & $(1.0)^{\ddagger}$ & 0.52 & 24.708 & 0.018 & 59 \\
\midrule
\multicolumn{10}{l}{\sc \underline{Total}}\\
\multicolumn{10}{l}{\sc Galaxies without AGN:}\\
F225W & 2.2760-2.4680 & 2.3562 & 18 & $>$27.54 & \nodata & $(1.0)^{\ddagger}$ & 1.20 & 24.442 & 0.012 & 92 \\
F275W & 2.4845-3.0762 & 2.7307 & 40 & $>$28.47 & \nodata & $(1.0)^{\ddagger}$ & 0.96 & 24.614 & 0.009 & 123 \\
F336W & 3.1320-4.2830 & 3.6075 & 36 & $>$28.60 & \nodata & $(1.0)^{\ddagger}$ & 0.65 & 25.078 & 0.018 & 62 \\
\multicolumn{10}{l}{\sc Galaxies with AGN:}\\
F225W & 2.2980-2.4500 & 2.3740 & 2 & $>$27.04 & \nodata & $(1.0)^{\ddagger}$ & 0.83 & 25.215 & 0.050 & 22 \\
F275W$^{-}$ & 2.4700-2.8280 & 2.6370 & 9 & 28.66 & 0.87 & 1.25 & 0.19 & 22.909 & 0.002 & 534 \\
F275W & 2.4700-2.8280 & 2.6325 & 10 & 26.23 & 0.08 & 13.00 & 0.19 & 22.251 & 0.001 & 874 \\
F336W & 3.1930-3.6609 & 3.3602 & 5 & 27.73 & 0.61 & 1.79 & 0.35 & 24.723 & 0.023 & 48 \\
\multicolumn{10}{l}{\sc All Galaxies:}\\
F225W & 2.2760-2.4680 & 2.3580 & 20 & $>$27.56 & \nodata & $(1.0)^{\ddagger}$ & 1.22 & 24.490 & 0.012 & 90 \\
F275W$^{-}$ & 2.4700-3.0762 & 2.7135 & 49 & $>$28.58 & \nodata & $(1.0)^{\ddagger}$ & 0.57 & 24.071 & 0.004 & 276 \\
F275W & 2.4700-3.0762 & 2.7111 & 50 & 27.84 & 0.26 & 4.11 & 0.89 & 23.630 & 0.003 & 354 \\
F336W & 3.1320-4.2830 & 3.5773 & 41 & 29.00 & 0.75 & 1.45 & 0.65 & 25.007 & 0.016 & 69 \\
\enddata
\vspace{5pt}
\begin{minipage}{.735\txw}{\footnotesize \textbf{Table columns:} (1) Observed WFC3/UVIS filter (\boldmath$^{-}$\unboldmath\! indicates the exclusion of the LyC-bright QSO\,$J$123622.9+621526.7); (2) Redshift range of galaxies included in LyC/UVC stacks; (3) Average redshift of all galaxies in each stack; (4) Number of galaxies with reliable spectroscopic redshifts included in each stack; (5) Observed total AB magnitude of LyC emission from stack (\SExtractor\ \texttt{MAG\_AUTO}) aperture matched to UVC, indicated by the blue ellipses in Figs.~\ref{GOODSstacks}--\ref{GOODSERSstacks}; (6) 1$\sigma$ uncertainty in LyC AB-mag; (7) Measured S/N of the LyC stack flux (\boldmath$^{\dagger}$\unboldmath\! indicates a 1$\sigma$ upper limit); (8) Area (in arcsec$^2$) of the UVC aperture; (9) Observed total AB magnitude of the UVC stack; (10) 1$\sigma$ uncertainties of UVC AB-mag. Listed uncertainties do not include systematic filter zeropoint uncertainty; (11) Measured S/N of the UVC stack flux. } 
\end{minipage}
\end{deluxetable*}
We then merged the $10^3$ \fesc\ distributions into one, and binned the data using equally-spaced logarithmic bins to optimize the resolution of \fesc\ in each decade between $10^{-4}$ and 1. We normalized the merged distribution by the sum of the bins to generate the probability mass function (PMF) of \fesc. We allowed the \fesc-values to vary unconstrained, resulting in some simulations going beyond 100\% due to the very low \emph{modeled} LyC flux. The middle and right panels of Fig.~\ref{brightqso} show the resulting distributions. The lighter shaded regions are individual MC realizations of \fesc\ and the darker lines are the full distribution of the merged simulated data. We extracted our statistics from these curves, taking the peak of the curve as the most-likely (ML) value of highest probability, the $\pm$1$\sigma$ values as the two points on the curve that have equal probability \emph{and} where the integrated area under the ML value down to these points is equal to 84\%. The expected value of \fesc\ (E[Val]), or the probability-weighted average \fesc, i.e., $\mathrm{E}[\fesc]\!=\!\sum_i p_i$\fesc$_{,i}$ is shown as well. For QSO\,$J$123622.9+621526.7, the \GALEX\ NUV data, which captures LyC at $\mathrm{\lambda_{rest}}$\,$\simeq$\,490--780\AA, we find the ML \fesc\ value to be \fesc$^{\rm \!\!\!\!\!\!NUV}$$\simeq$30$^{+22}_{-5}$\% and E[\fesc$^{\rm \!\!\!\!\!\!NUV}$]$\simeq$45\%. For the WFC3/UVIS F275W data, effectively covering LyC from $\mathrm{\lambda_{rest}}$$\sim$680--850\AA, we find the ML \fesc\ value to be $\fesc^{\rm F275W}$$\simeq$28$^{+20}_{-4}$\% and E[$\fesc^{\rm F275W}$]$\simeq$43\%. These values show a consistent escape fraction of LyC for this AGN to within their errors. These results are discussed in more detail in \S\ref{sec:agnlycdet}. 

\section{Results} \label{sec:results}
\subsection{LyC Image Stacks and Photometry} \label{sec:stackphot}
Since LyC measurements have been historically very faint or resulted in non-detections (see references in \S\ref{sec:intro}), we perform a weighted-sum based ``stacking'' algorithm used in \citetalias{Smith2018}, taking ``subimage'' cutouts of galaxies from the WFC3/UVIS mosaic observed in the filter that corresponds to the galaxy's LyC, then co-adding them all onto the same 151$\times$151\,pix (4\farcs53$\times$4\farcs53) grid. Before stacking, we first created $\chi^2$ images \citep{Szalay1999} of each galaxy in our sample, which were comprised of the \HST\ WFC3/UVIS F225W, F275W, F336W, WFC/ACS F435W, F606W, F814W, F850LP, WFC3/IR F098M, F105W, F125W, F140W, and F160W images when available. We then ran \SExtractor\ on the $\chi^2$ images to detect any faint objects in the \HST\ images that may potentially add contaminating, non-ionizing flux to our stacked image. Using the resulting segmentation map, we masked all neighboring and foreground objects in each subimage, except the galaxy from our sample in the center of the subimage. We improved our \SExtractor\ object detection and deblending parameters from \citetalias{Smith2018} to minimize the unintentional masking of image noise during stacking, and accounts for the differences between the photometry tabulated here and what is listed in table~2 of \citetalias{Smith2018}. We also subtracted the mean sky-level in each LyC subimage during stacking, which was determined by binning all the sky-pixels using the Freedman-Diaconis rule \citep[excluding the object pixels in the image;][]{Freedman1981}, and fitting a Gaussian function to this histogram to determine the mean and dispersion values of the sky-background. All subimages were then weighted by their corresponding Astrodrizzle weight-map, excised from the same RA-Dec region in the weight-image, then summed. 

We only stacked galaxy subimages excised from the same WFC3/UVIS filter mosaic for our analyses. The stacks therefore contain galaxies with redshift ranges listed in table~1 of \citetalias{Smith2018}, corresponding to 2.26\,$\leq$\,$z$\,$<$\,2.47 for F225W, 2.47\,$\leq$\,$z$\,$<$\,3.08 for F275W, and 3.08\,$\leq$\,$z$\,$<$\,4.35 for F336W. These redshift bins reduced the inclusion of non-ionizing flux into our LyC photometry down to $\sim$0.3\% of the total flux within the filter, based on our average SED. We created stacks for each subsample as described in \S\ref{sec:sample}, which can be seen in figs.~\ref{GOODSstacks}--\ref{GOODSERSstacks}. We created these stacks for the GOODS/HDUV, ERS, and the GOODS/HDUV+ERS (Total) fields. This corresponds to the 30 stacks shown in figs.~\ref{GOODSstacks}--\ref{GOODSERSstacks}, along with their corresponding UVC stacks that were created in the same way as the LyC stacks, but using the corresponding rest-frame non-ionizing UVC images ($\mathrm{\lambda_{rest}}$\,$\gtrsim$\,1400\,\AA) indicated in those figures. The number of galaxies in each stack is also indicated. 

\begin{figure*}[ht!]\centerline{
\includegraphics[width=.7\txw]{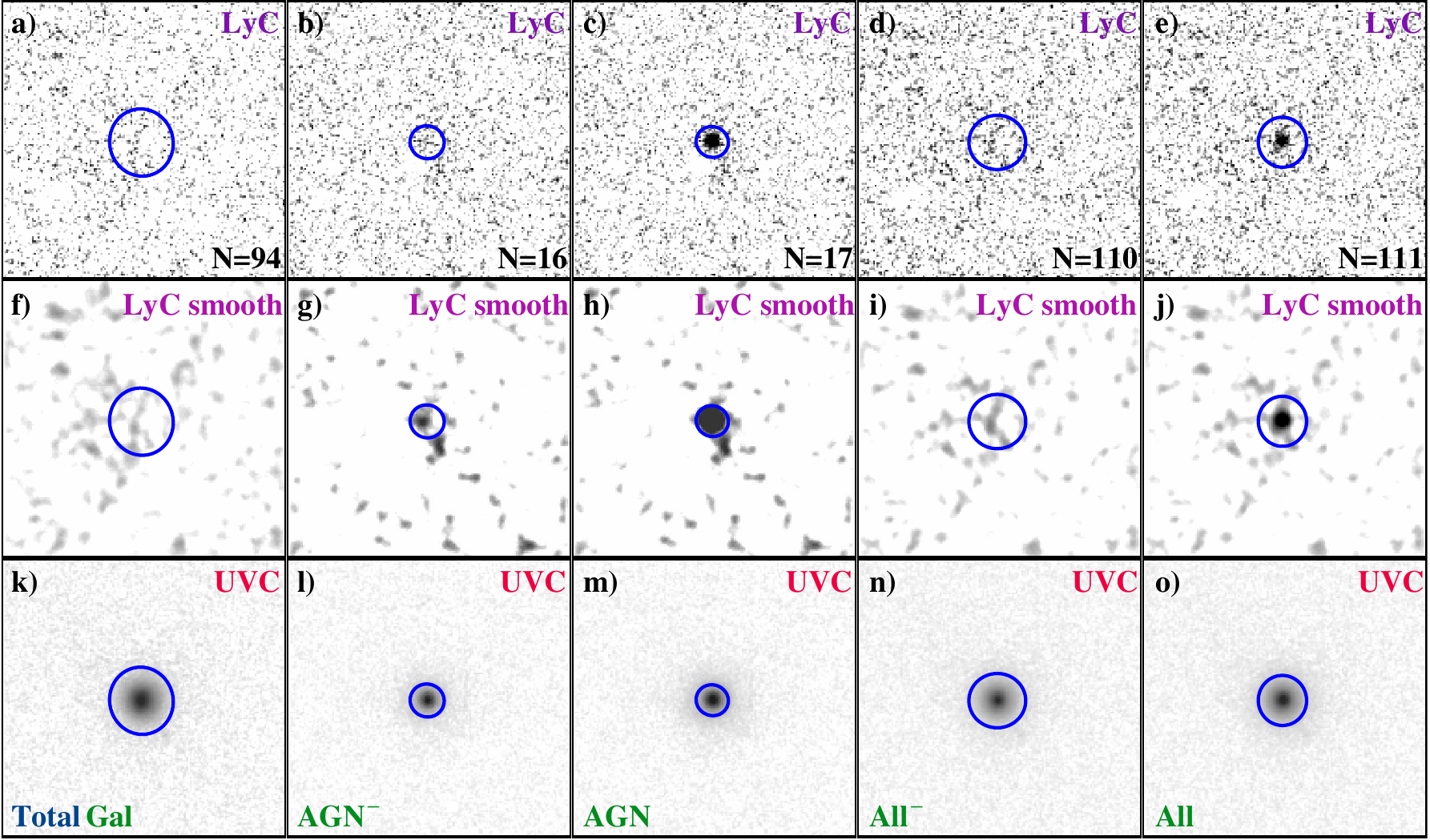}}
\caption{[\textbf{Top Row}]: Composite LyC stacks of all galaxies in our sample; [\textbf{Middle Row}]: The same as the top row but convolved with a $\sigma$=1\,pixel Gaussian kernel. [\textbf{Bottom Row}] The UVC counterparts of the top row; [\textbf{Left column of panels}]: Composite stacks of all galaxies without AGN in our sample (indicated by Gal in green) observed in the F225W, F275W, and F336W filters; [\textbf{2nd column}]: Composite stacks of all galaxies hosting AGN, excluding QSO\,$J$123622.9+621526.7 (indicated by AGN$^{-}$ in green); [3rd column:] Composite stacks of all galaxies hosting AGN; [\textbf{4th column}]: Composite stacks of 110 galaxies with \emph{and} without AGN in our sample, excluding QSO\,$J$123622.9+621526.7; [\textbf{5th column}]: Stack of all 111 galaxies \emph{with} and \emph{without} AGN in our sample. These stacks represent the average \emph{observed} LyC $F_{\nu}$ from all objects integrated from 2.28\,$\lesssim$\,$z$\,$\lesssim$\,4.28, scaled to a common F225W zeropoint magnitude. The blue ellipses were fit to the UVC stacks (bottom row). \label{SoS}}
\end{figure*}

The results of our photometric analyses on the stacks shown in figs.~\ref{GOODSstacks}--\ref{GOODSERSstacks} are listed in Table~\ref{phottab}. We performed our photometry in the same manner as in \citetalias{Smith2018} using a Gaussian additive noise model. In summary, we created flux datacubes for each galaxy based on the LyC subimage, where two dimensions corresponded to the pixel values in the LyC subimage, and the third dimension was based on the dispersion of the sky-background and the pixel's RMS value, calculated from the Astrodrizzle weight image. The voxels along each slice in the datacube had a mean value equal to the original pixel value of the LyC subimage and a variance equal to the sum of the total sky variance in the LyC subimage and the inverse of the corresponding pixel value in the weight map; i.e., the flux value of the voxel at coordinate x,\,y,\,z in the data cube is generated by $\mathbf{f}_{\rm x,y,z}\!=\!f_{\rm x,y}\!+\!\sigma_{\rm z}\!+\!\textsc{RMS}_{\rm z}$ where $f_{\rm x,y}$ is the pixel value of the LyC subimage at location x, y in the 151$\times$151 pixel grid, $\sigma_{\rm z}$ is the randomly selected sky-background value from a GRV defined by $\mathcal{N}(0,\sigma^2_{\rm sky})$ (where $\sigma_{\rm sky}$ is the sky-background dispersion), and $\mathrm{\textsc{RMS}}_{\rm z}$ is the randomly selected detector noise value from the GRV $\mathcal{N}(0,\sigma^2_{\textsc{RMS}})$ (where $\sigma_{\textsc{RMS}}\!=\!\sqrt{\frac{1}{W_{x,y}}}$ and $W_{x,y}$ is the pixel value in the Astrodrizzle weight map). More detail on inputs to the Astrodrizzle weight maps and how they affect RMS can be found in the Appendix of \citet{Casertano2000}.

We performed matched-aperture photometry with \SExtractor\ on our LyC stacks shown in figs.~\ref{GOODSstacks}--\ref{GOODSERSstacks} using the corresponding UVC stack as the detection image in all cases. To generate a representative distribution of the LyC flux, we iterated through all $10^4$ 2-dimensional slices along the z-dimension of our datacube and measured the flux within the UVC aperture. This  provided $10^4$ possible flux values that were based on the WFC3/UVIS LyC subimage, the sky variance in the LyC subimage, and the RMS in the pixels from detector noise. 

To generate the values listed in Table~\ref{phottab}, we took the mean of our flux distribution and the 16$^{\mathrm{th}}$ and 84$^{\mathrm{th}}$ percentile for the -1$\sigma$ and +1$\sigma$ uncertainty bounds, respectively. The ratio of the mean and the uncertainty was used to calculate the S/N ratios. When this S/N was less than one, we list the 84th percentile from the distribution as the 1$\sigma$ upper limit to the LyC flux, and denote the S/N by $(1.0)^{\dagger}$ in Table~\ref{phottab}. 

From the Total sample stacks, we find that the galaxies \emph{with} AGN at $z$=2.573--2.828 have the \emph{only} $>$5$\sigma$ detection, while other AGN samples have LyC S/N measurements around, or below, 2$\sigma$. All stacks of galaxies \emph{without} AGN show only upper limits, with S/N$<$1. Although there is visible flux in the GOODS/HDUV and Total F336W stacks for galaxies \emph{without} AGN, the significance is only $\sim$1$\sigma$. Thus, we cannot yet rule out the possibility this signal is spurious noise rather than real LyC flux. 

\subsection{Composite Stacks of the Total Sample}
To visualize the LyC flux from the various types of galaxies in our sample, we stack all LyC subimages from the various fields and WFC3/UVIS filters onto the same grid using the methodology outlined in \S\ref{sec:stackphot}. The resulting stacks are shown in Fig.~\ref{SoS} and the number of galaxy subimages they contain are indicated. While stacking the subimages, we also scaled the pixel values in all subimages in the stack such that all images had a common AB-zeropoint magnitude equal to that of the F275W filter (ZP\,=\,24.04\,mag) for the LyC stacks, and the UVC subimages were scaled to match the F606W zeropoint (ZP\,=\,26.51\,mag). 

We performed a similar photometric analysis on these LyC stacks described in \S\ref{sec:stackphot} to assess the S/N of the central flux measured within the blue UVC-detected apertures of Fig.~\ref{SoS}. For the Total sample of galaxies \emph{without} AGN, we find low levels of LyC flux (likely due to dilution from multiple stacked non-detections) with S/N$\sim$1. From all galaxies \emph{with} AGN, excluding QSO\,$J$123622.9+621526.7, we find a S/N\,$\sim$\,1.8, and for the Total sample of all galaxies \emph{with} AGN we measure a S/N\,$\simeq$\,10.3, which is dominated by the bright LyC flux from QSO\,$J$123622.9+621526.7. In the AGN stacks, we observe possible indications of LyC flux extending outside of the central UVC aperture, though the S/N cannot distinguish this fluctuation from sky noise. Based on the galaxy counts in \citet{Windhorst2011}, to the depth of the $\chi^2$ images used for object masking of AB$\lesssim$27.5\,mag, there are $\lesssim$5$\times$10$^5$ galaxies deg$^{-2}$. This results in a probability of $\simeq$3\% that this extended flux within 0\farcs5 of the central source is an interloper. Combining the LyC flux from all galaxies (aside from QSO\,$J$123622.9+621526.7), we find a S/N\,$\simeq$\,1.3 (again likely diluted from the non-detections in the galaxies \emph{without} AGN), while the LyC flux from all 111 galaxies from our sample amounts to a S/N\,$\simeq$\,3.1. 
\begin{deluxetable}{lcccrccr}
\centering\tablecaption{Composite Stack LyC Photometry\label{sosphot}}
\tablewidth{\txw}
\tabletypesize{\scriptsize}
\tablehead{
\colhead{Stack} & \colhead{$z$-range} & \colhead{\zmean} & \colhead{$N_{obj}$} & \colhead{m$_{\!\lyc}$} & \colhead{ABerr$_{\!\lyc}$} & \colhead{S/N$_{\!\lyc}$} & \colhead{A$_{\uvc}$}\\\vspace{-13pt}\\
\multicolumn{1}{c}{(1)} & (2) & (3) & (4) & (5) & (6) & (7) & \multicolumn{1}{c}{(8)}}
\startdata
Gal & 2.2760--4.2830 & 2.9948 & 94 & $>$28.3 & \nodata & $(1.0)^{\dagger}$ & 0.89 \\
AGN$^{-}$ & 2.2980--3.6609 & 2.8541 & 16 & $>$27.8 & \nodata & $(1.0)^{\dagger}$ & 0.23 \\
AGN & 2.2980--3.6609 & 2.8377 & 17 & 26.5 & 0.1 & 10.3 & 0.21 \\
All$^{-}$ & 2.2760--4.2830 & 2.9754 & 110 & $>$28.4 & \nodata & $(1.0)^{\dagger}$ & 0.64 \\
All & 2.2760--4.2830 & 2.9719 & 111 & 28.3 & 0.4 & 3.1 & 0.50 \\
\enddata
\vspace{1pt}
\begin{minipage}{.515\txw}{\footnotesize \textbf{Table columns}: (1) Galaxy type subsample (\boldmath$^{-}$\unboldmath\! indicates the exclusion of the LyC-bright QSO\,$J$123622.9+621526.7); (2) Redshift range of galaxies included in LyC composite stacks; (3) Average redshift of all galaxies in each stack; (4) Number of galaxies with reliable spectroscopic redshifts included in each stack; (5) Observed total AB magnitude of LyC emission from stack (\SExtractor\ \texttt{MAG\_AUTO}) aperture matched to UVC, indicated by the blue ellipses in Fig.~\ref{SoS}; (6) 1$\sigma$ uncertainty in LyC AB-mag (7) Measured S/N of the LyC stack flux (\boldmath$^{\dagger}$\unboldmath\! indicates a 1$\sigma$ upper limit); (8) Area (in arcsec$^2$) of the UVC aperture.} 
\end{minipage}
\end{deluxetable}

On average, we find that the flux from AGN outshines the galaxies \emph{without} AGN by a factor of $\sim$10, and excluding QSO\,$J$123622.9+621526.7 the AGN still outshine galaxies \emph{without} AGN by $\sim$2 times. The low S/N of these measurements makes these ratios highly uncertain, and only larger samples of spectroscopically verified high-redshift galaxies can reduce these uncertainties. Larger samples will also increase the chances of observing sources of brighter LyC flux, e.g., the bright QSO\,$J$123622.9+621526.7 found among the 17 AGN in our sample. Increasing the sample size may also increase the chance of including more rare sources of LyC emission, e.g., lower mass, star-bursting galaxies with extreme [\ion{O}{3}] emission \citep[e.g.,][]{Fletcher2019}. 

\subsection{Stacked LyC Escape Fractions} \label{sec:fesc}
\begin{figure*}[thp!]\centerline{
\includegraphics[width=.364\txw]{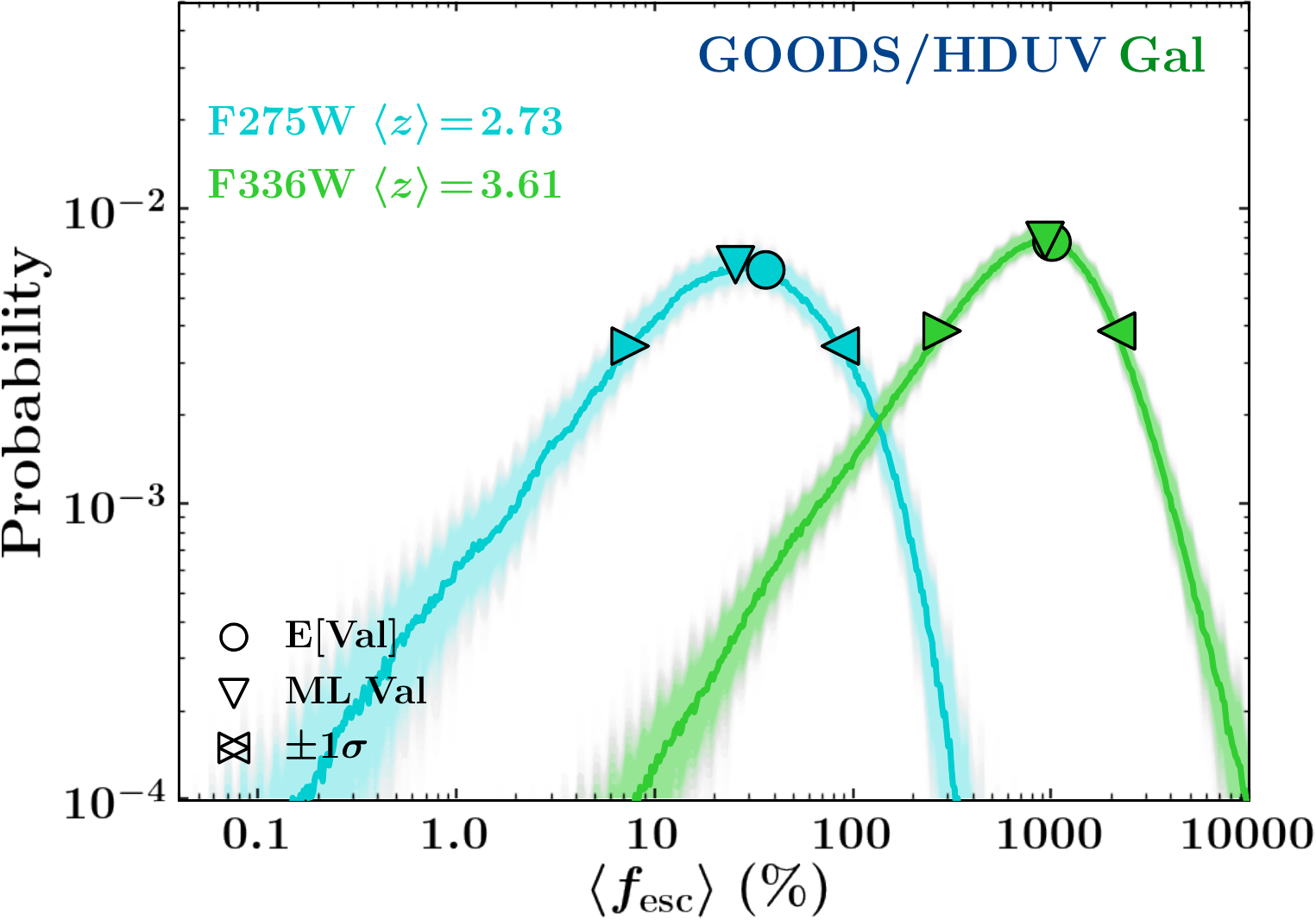}
\includegraphics[width=.316\txw]{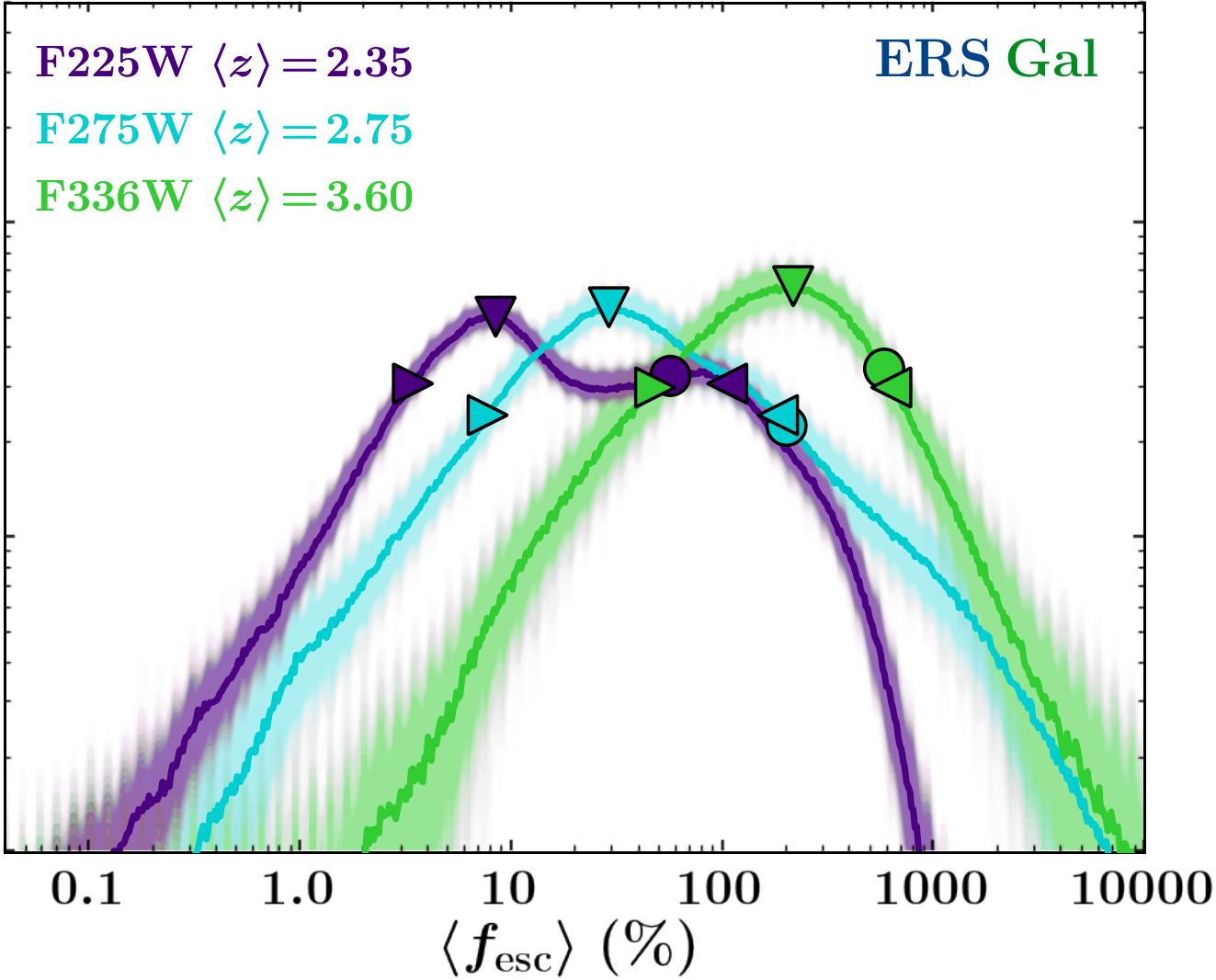}
\includegraphics[width=.316\txw]{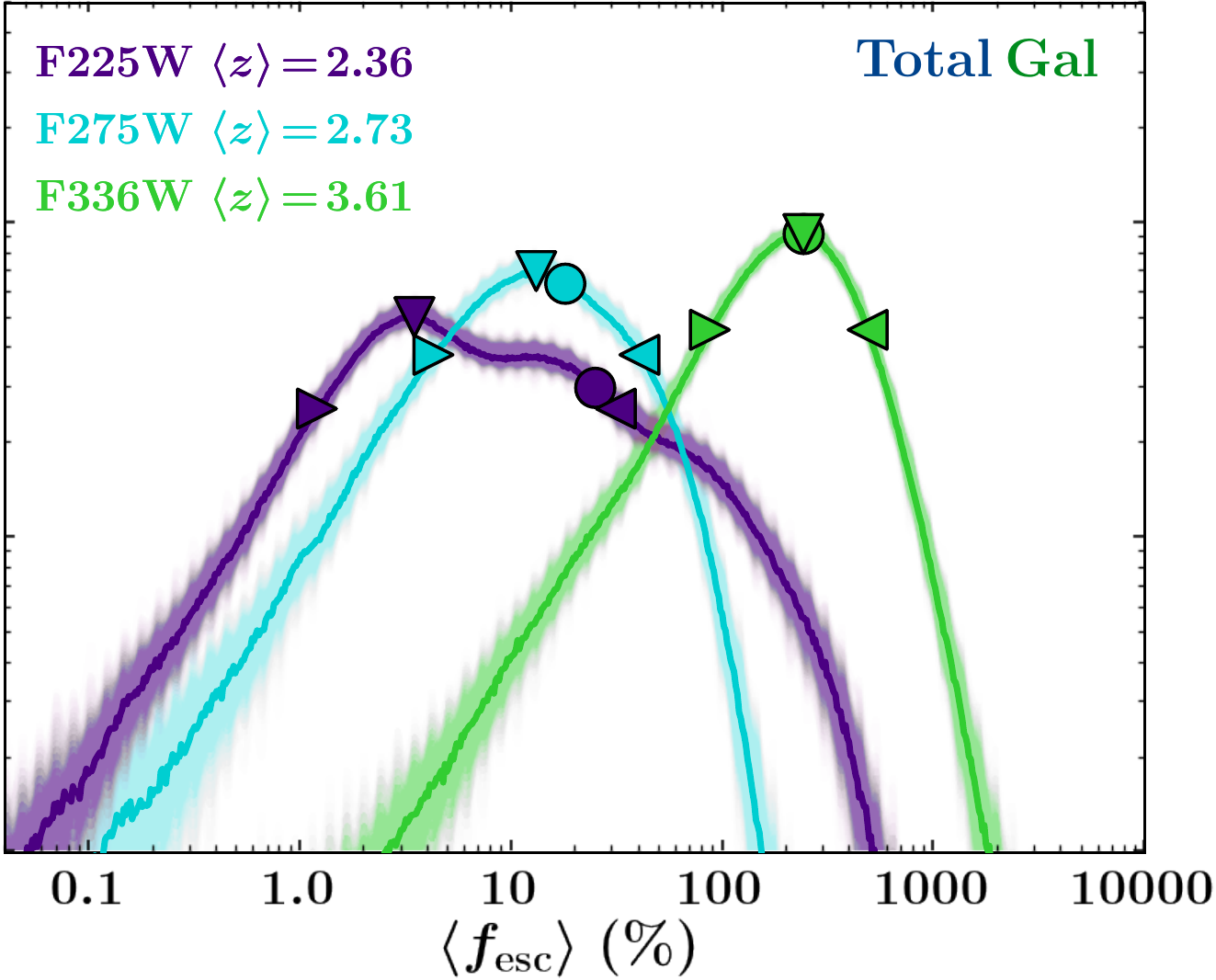}}
\caption{PMFs of the sample-averaged \fesc\ values of galaxies without AGN from the MC simulations described in \citepalias{Smith2018}, plotted against their relative probability. The respective sample is indicated in each top right corner. This analysis was performed 10$^3$ times using the measured and modeled intrinsic stacked \emph{apparent} LyC flux and their $\pm$1$\sigma$ ranges. We apply the IGM attenuation models of \citet{Inoue2014} to our modeled LyC fluxes. Downwards triangles and circles indicate the resulting ML and expected values of \fesc\ in each probability distribution function, respectively, while the left/right facing triangles indicate the $\pm$1$\sigma$ range around the ML value. \label{fescgal}}
\end{figure*}

\begin{figure*}[thp!]\centerline{
\includegraphics[width=.364\txw]{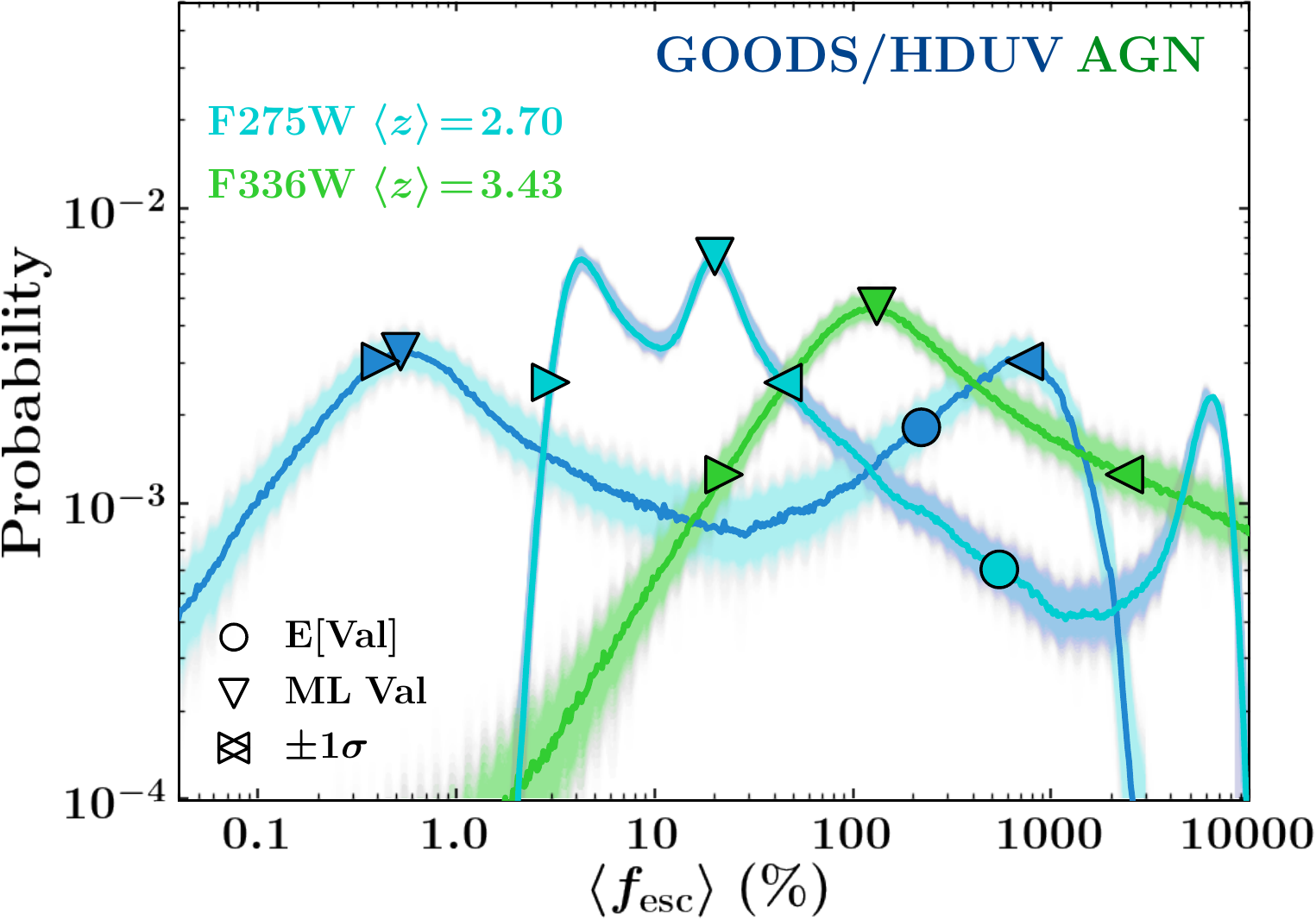}
\includegraphics[width=.316\txw]{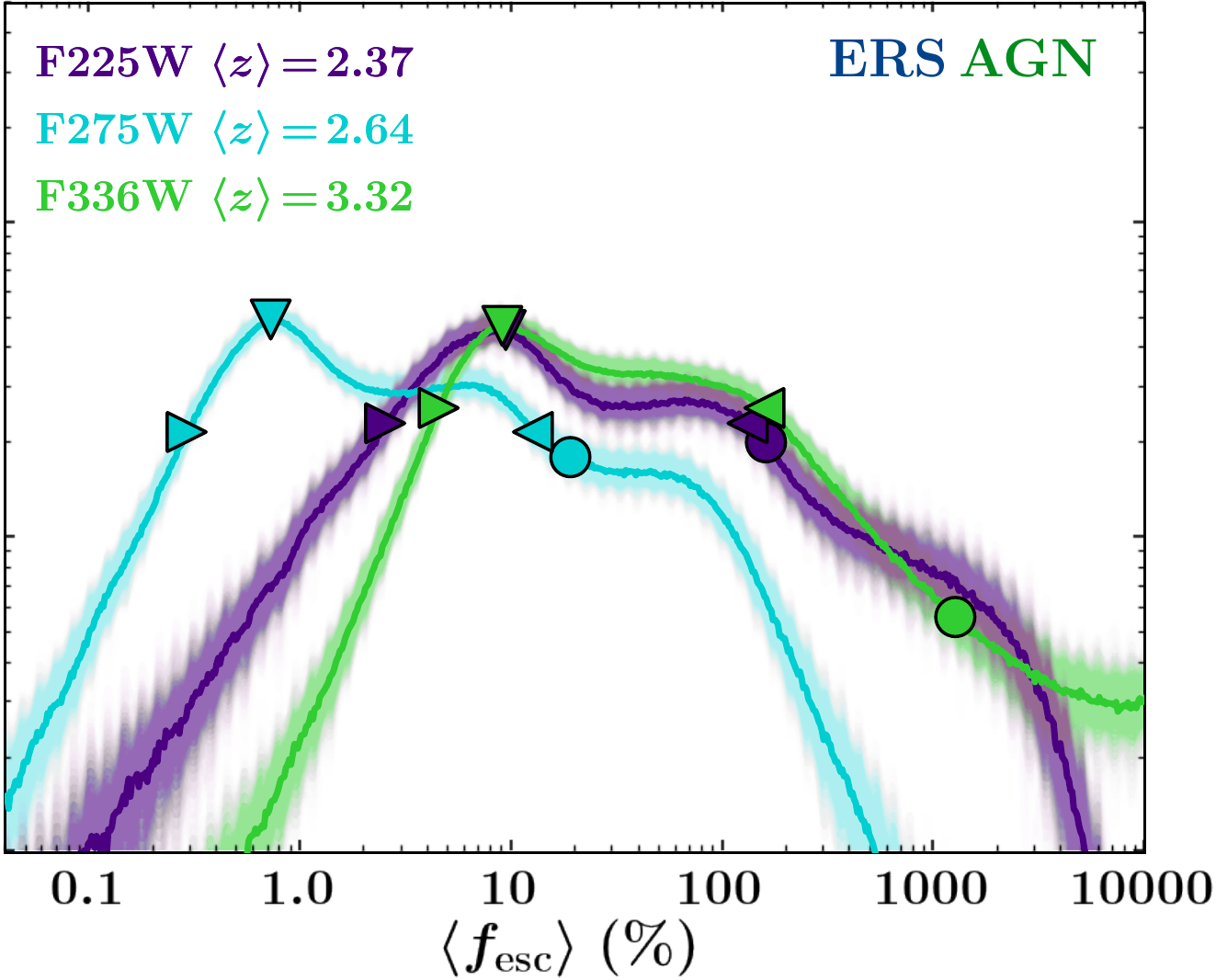}
\includegraphics[width=.316\txw]{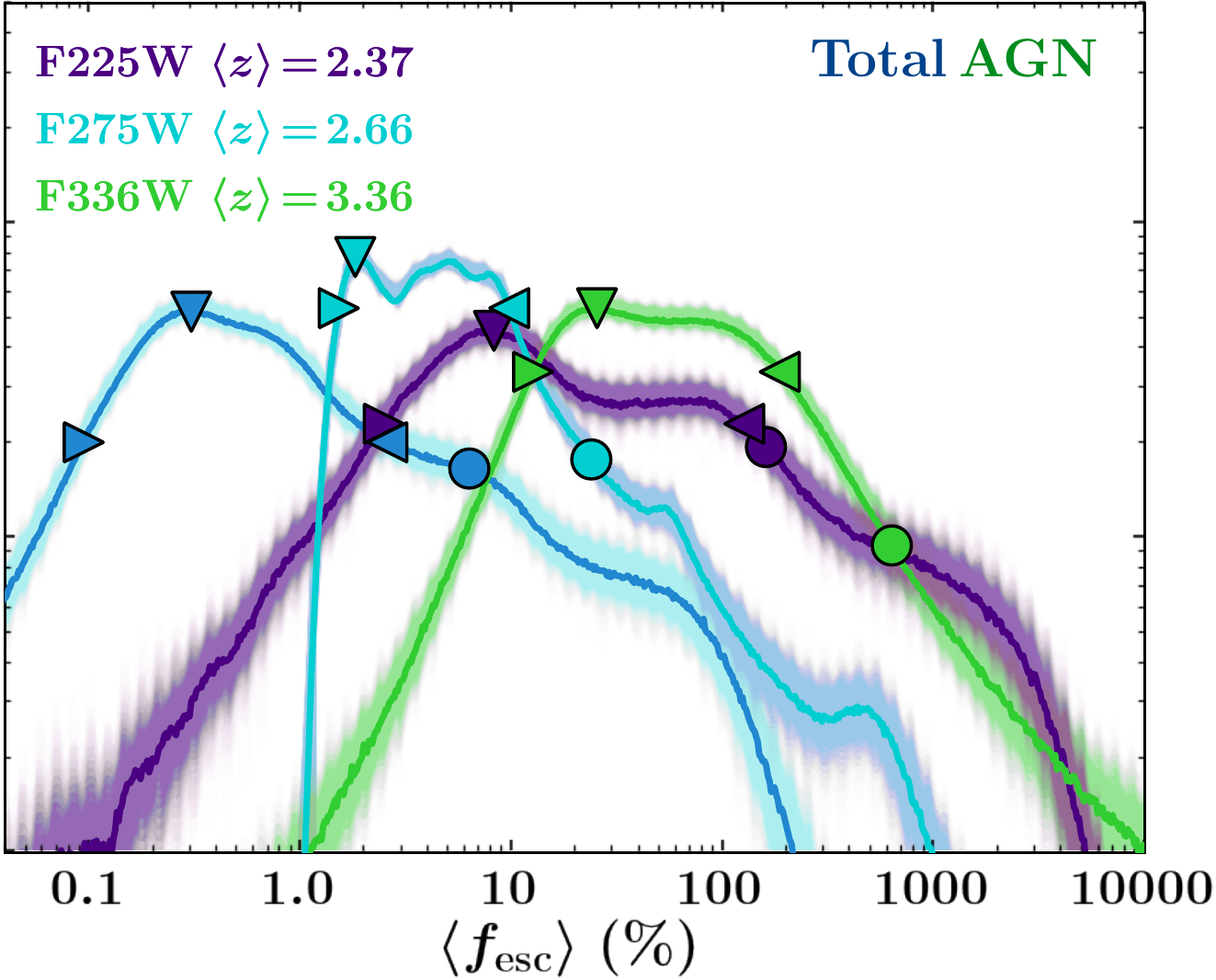}}
\caption{PMFs of the sample-averaged \fesc\ values of galaxies \emph{with} AGN. This analysis was performed in the same manner as in Fig.\ref{fescgal}, and the symbols are equivalent in meaning as well. The darker blue curve is the same analysis as the lighter blue curve, except the LyC-bright QSO\,$J$123622.9+621526.7 is excluded. \label{fescagn}}
\end{figure*}

\begin{figure*}[thp!]\centerline{
\includegraphics[width=.364\txw]{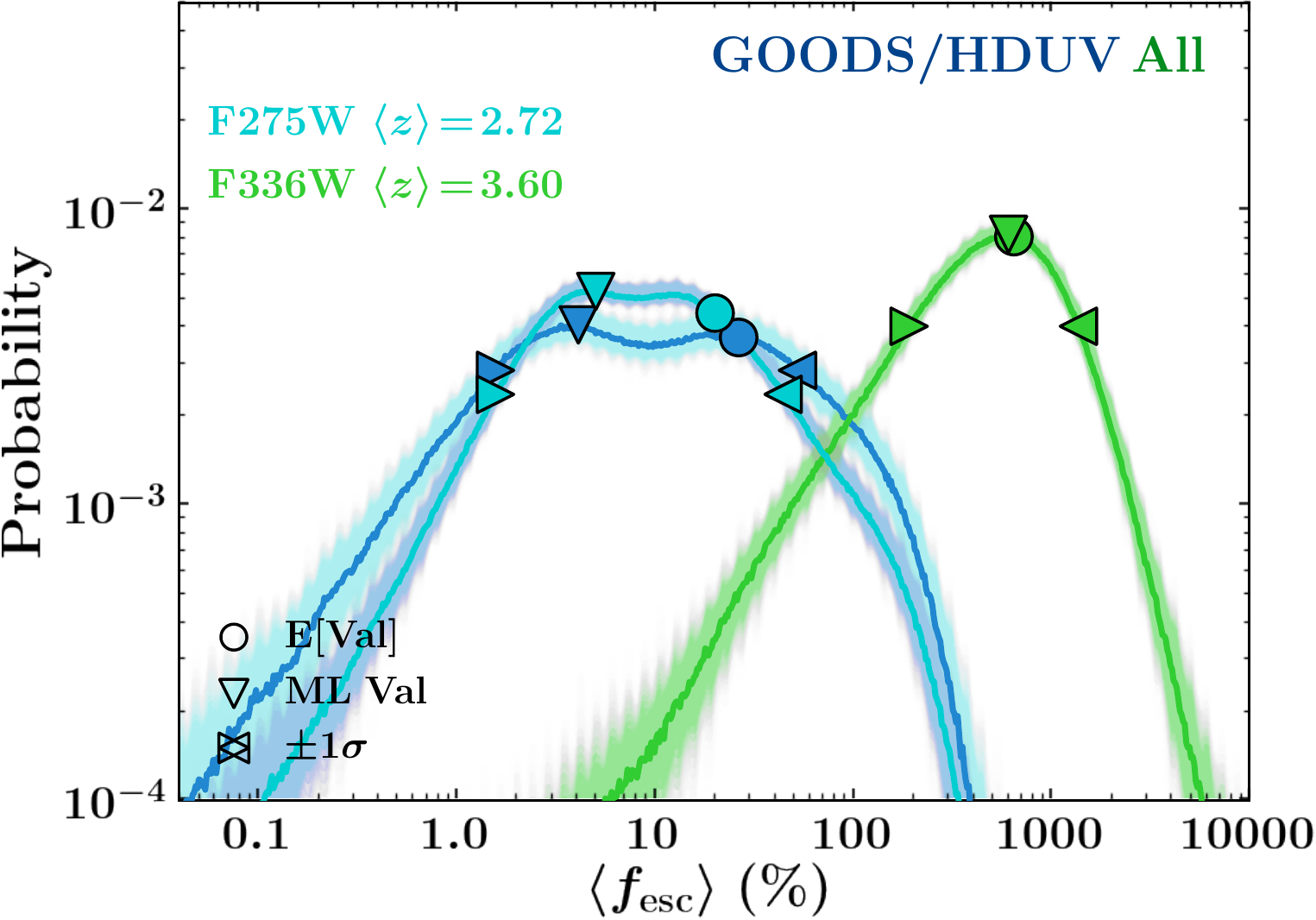}
\includegraphics[width=.316\txw]{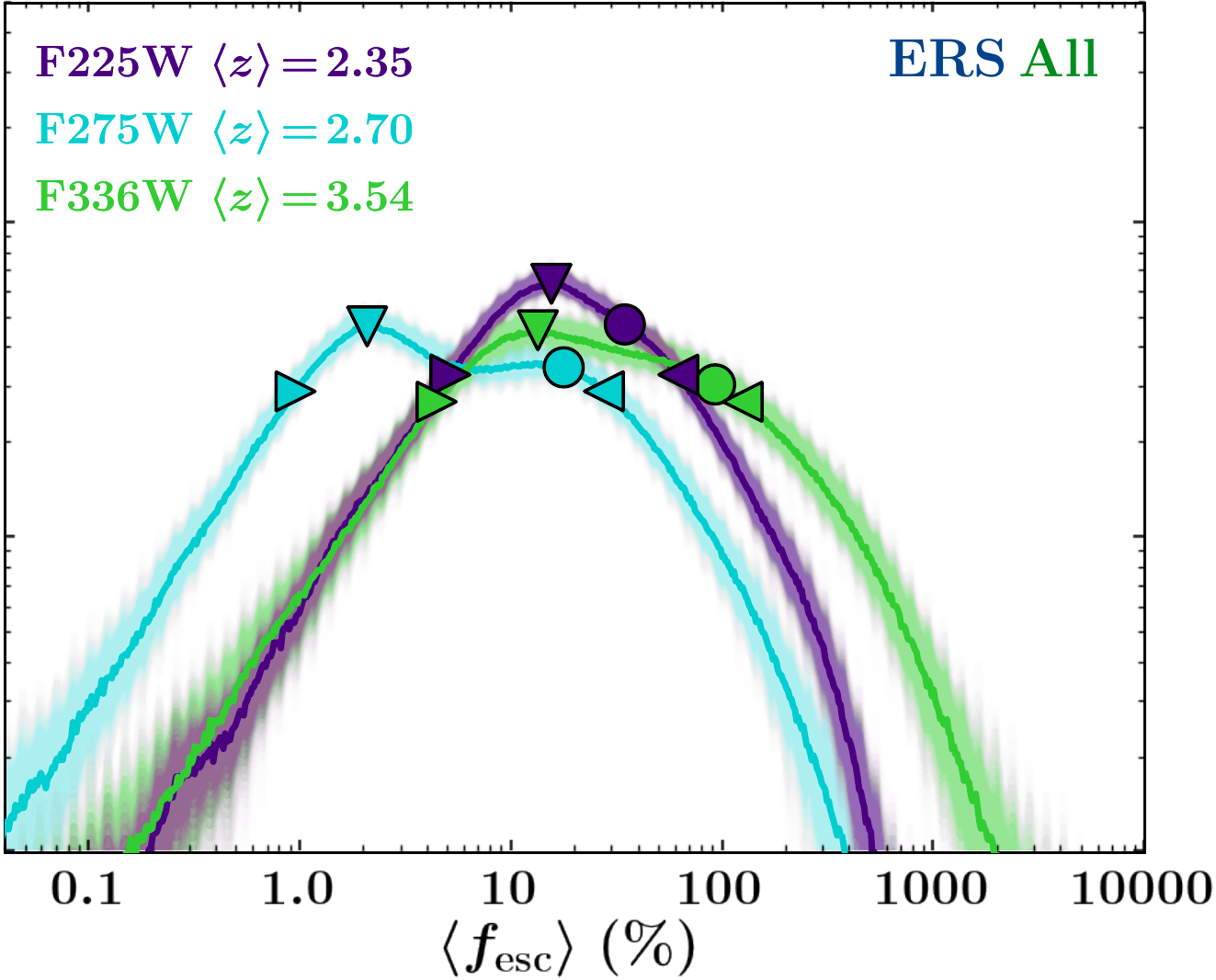}
\includegraphics[width=.316\txw]{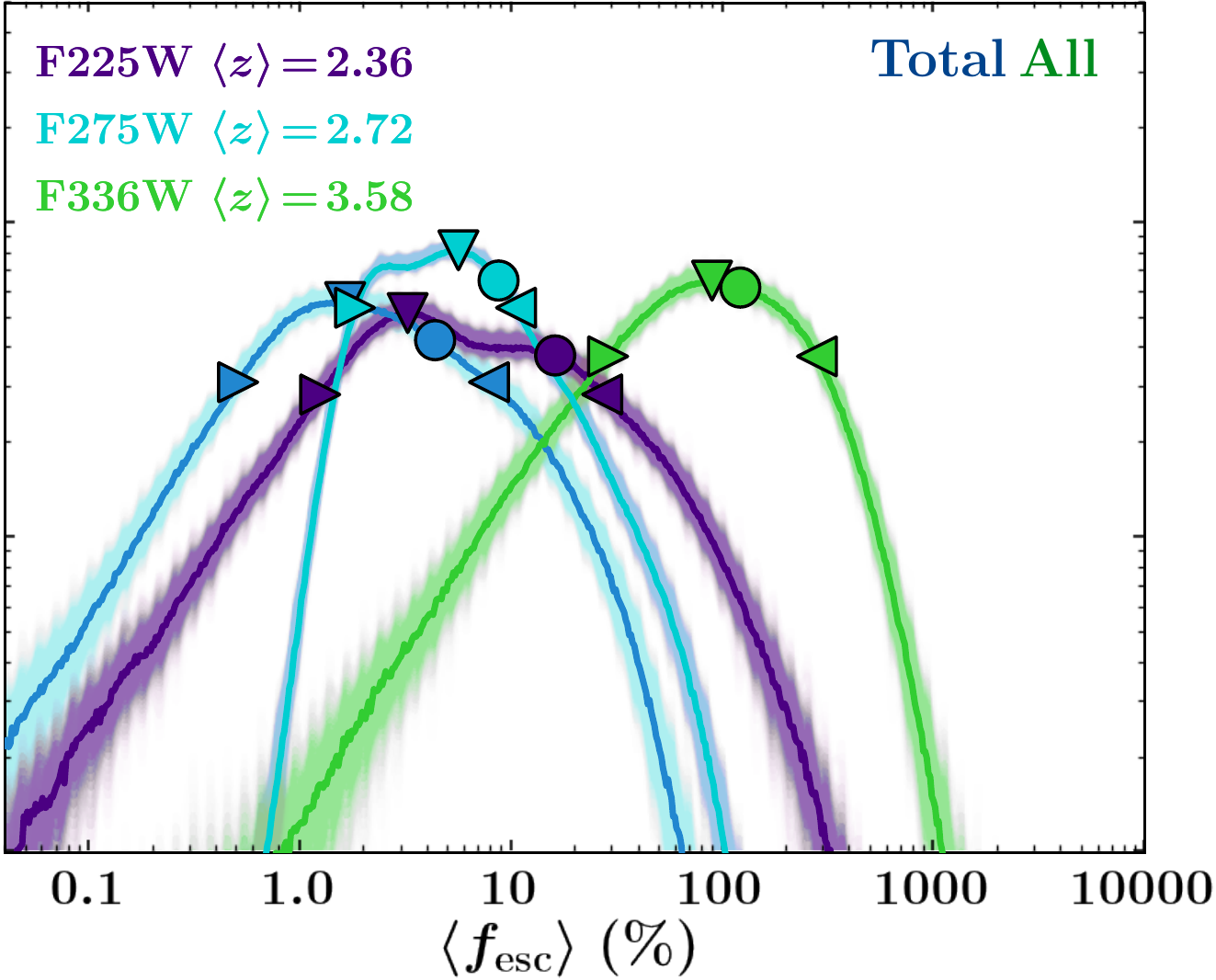}}
\caption{PMFs of the sample-averaged \fesc\ values of \emph{all} galaxies. This analysis was performed in the same manner as in Fig.\ref{fescagn}, and the symbols are equivalent in meaning as well. \label{fescall}}
\end{figure*}

To estimate the LyC \fesc\ from our subsamples discussed in \S\ref{sec:sample}, we apply the same statistical methodology of \citetalias{Smith2018} described in \S\ref{sec:qfesc}. The \fesc\ distributions shown in figs.~\ref{fescgal}--\ref{fescall} are generated by taking the ratio of the photometric distributions measured in \S\ref{sec:stackphot} to the distribution of intrinsic LyC flux derived from the best-fitting SED, the IGM attenuation models at the galaxy's respective redshift, and the WFC3/UVIS throughput curve corresponding to the filter used for the LyC observation. These distributions are useful for \emph{inferring} the most likely, \emph{sample-averaged} \fesc\ values of a given photometric dataset, even if the photometry is highly uncertain, or shows a non-detection of LyC. A much simpler approach would be to adopt constant values for each factor in the \fesc\ calculation, while using the measured UVC-to-LyC flux ratio, i.e. $\langle f_{\uvc}/f_{\!\lyc}\rangle_{obs}$. This method would reduce to \begin{equation}\label{fescsimp}
\fesc^{\mathrm{\mathrm{\,simp}}}\!=\!\frac{\langle f_{\uvc}/f_{\!\lyc}\rangle_{\mathrm{int}}}{\langle f_{\uvc}/f_{\!\lyc}\rangle_{\mathrm{obs}}}e^{\langle\tau_{_{\igm}}\rangle} 10^{-0.4\langle A_{\uvc}\rangle}
\end{equation}
where $\langle\tau_{_{\igm}}\rangle$ is the average IGM opacity at the average redshift of the galaxies in the stack, and $\langle A_{\uvc}\rangle$ is the average extinction from dust in the sample averaged SED near $\lambda\!\simeq\!1500$\AA in magnitudes. Typical values adopted for $f_{\uvc}/f_{\!\lyc}$$_{\mathrm{int}}$ usually range from 2--7 \citep[e.g.,][]{Steidel2001,Shapley2006,Siana2007}. This quantity is useful for comparison with other studies that use a similar technique, though this method does not account for the variations in intrinsic UVC-to-LyC flux ratios and amounts of dust extinction between galaxies, nor the variations in IGM opacity for different sight-lines and redshifts. We therefore calculate this $\fesc^{\mathrm{\,simp}}$ quantity for comparison with other literature, and use our MC method to constrain the \fesc\ values more representatively for our galaxy samples.

We calculate the intrinsic LyC flux distributions in the same way as in \S\ref{sec:stackphot} for each galaxy in a stack, then we perform a weighted average of all these distributions using the average weight map value used in the image stacks. This was to ensure that the same proportions of flux from the sets of galaxies included in the image stack used for photometry matched the modeled intrinsic LyC flux. We generated $10^3$ of these distributions (shaded regions in figs.~\ref{fescgal}--\ref{fescall}) and merged them into one (solid lines in figs.~\ref{fescgal}--\ref{fescall}) after constraining each individual \fesc\ distribution to physical values between 0--100\%. The \fesc\ values produced by the simulations were binned into histograms using equally-spaced logarithmic bins and normalized by the sum of the bins to produce the \fesc\ PMF. The y-axis of these PMFs thus represents the relative probabilities of the \fesc\ values in the x-axis. We extracted the ML, $\pm$1$\sigma$ uncertainties, and the E[\fesc] statistics from the merged distributions and plotted each in figs.~\ref{fescgal}--\ref{fescall}. We quote the ML and its $\pm$1$\sigma$ values in Table~\ref{fesctab} as the estimated \fesc, since it has the highest probability of representing the global-average $\langle\fesc\rangle$ of galaxies at their average redshifts. Each subsample and its redshift-range is color-coded and indicated in the figure, along with the type of galaxy and the field the analysis was performed on. 

Several of these distributions display large asymmetries and some show bimodalities. These asymmetries are caused mostly by the variations in IGM transmission along different lines-of-sight, and bimodalities result from some sources in the stacks dominating the modeled intrinsic LyC flux. This creates a peak of higher $f_{\lyc,mod}$ values amongst the (on average) fainter modeled LyC flux as in, e.g., the light-blue curve of the left panel of Fig.~\ref{fescagn}. However, with more sources added to the stacks, the distributions begin to become more Gaussian-like (see, e.g., the right panel of Fig.~\ref{fescall}). 

Due to the uncertain flux estimation in the GOODS/HDUV F336W stack, whether spurious or not, our \fesc\ peak is located near 100\%, indicating that either the current IGM+SED models do not properly account for flux this bright at \zmean$\simeq$3.5. If this flux is indeed real LyC emission, these high \fesc\ values may be a result of anisotropic LyC escape mechanisms \citep{Nakajima2014, Paardekooper2015} or stochastic periods of star-formation \citep{Kimm2017, Trebitsch2017} in these galaxies, allowing for higher \fesc\ than average during the life-time of these galaxies. Starbursts composed of star-formation with multiple waves of SNe relatively close in time can sustain the higher pressure in the ISM needed to drive galactic winds \citep{Veilleux2005}.

By virtue of the random nature in selecting IGM attenuation sight-lines for our \fesc\ simulations, another possibility is that the IGM models underestimate the frequency of regions of lower IGM \ion{H}{1} column densities. A realistic, highly inhomogeneous large-scale structure may provide more clear sight-lines in the IGM on scales smaller than a galactic halo \citep[e.g.,][]{Daloisio2015,Keating2018,Bosman2018}. Our MC analysis redraws simulated \fesc\ values above 100\% in an attempt to reject these over-estimated column densities, which causes the observed pile up of \fesc\ values near 100\% as a result of this constraint. These \fesc\ values are listed in column 11 of Table~\ref{fesctab} as upper limits, and the LyC fluxes they are based on are listed in column 5 of Table~\ref{phottab}. 


\begin{deluxetable*}{lllcrrrrrrrr}
\centering\tablecaption{Summary of \fesc\ Constraints \label{fesctab}}
\tablewidth{\txw}
\tabletypesize{\scriptsize}
\tablehead{
\colhead{Filter} & \colhead{\zmean} & \colhead{$N_{obj}$} & \colhead{$\langle f_{\uvc}/f_{\!\lyc}\rangle_{\rm obs}$} & \colhead{$\langle
f_{\uvc}/f_{\!\lyc}\rangle_{\rm int}$} & \colhead{Age} & \colhead{$\mathrm{A_V}$} & \colhead{$\mathrm{log(M_{\star}/M_{\odot})}$} & \colhead{$\mathrm{SFR}$} & \colhead{$\langle \tau_{_{\!\igm}} \rangle$} & \colhead{$\fesc^{\mathrm{\,simp}}$} & \colhead{$\langle\fesc\rangle$} \\[-7pt]
\null & \null & \null & \null & \null & \colhead{[$\mathrm{log(yr)}$]} & \colhead{[mag]} & \colhead{\null} & \colhead{[log($\mathrm{M_{\odot}/yr}$)]} & \null & \colhead{[\%]} & \colhead{[\%]} \\[-5pt]
\multicolumn{1}{c}{(1)} & (2) & (3) & (4) & \multicolumn{1}{c}{(5)} & \multicolumn{1}{c}{(6)} & \multicolumn{1}{c}{(7)} & \multicolumn{1}{c}{(8)} & \multicolumn{1}{c}{(9)} & \multicolumn{1}{c}{(10)} & \multicolumn{1}{c}{(11)} & \multicolumn{1}{c}{(12)}}
\startdata
\multicolumn{12}{l}{\sc \underline{GOODS/HDUV}}\\
\multicolumn{12}{l}{\sc Galaxies without AGN:}\\
F275W & 2.726 & 33 & $6.2^{+15.8}_{-6.2}$ & $4.567^{+0.009}_{-0.006}$ & $8.82^{+0.01}_{-1.82}$ & $0.10^{+0.50}_{-0.10}$ & $9.4^{+0.3}_{-1.1}$ & $0.8^{+0.7}_{-0.3}$ & $1.54^{+7.02}_{-0.94}$ & $<$100 & $30^{+57}_{-22}$ \\
F336W & 3.609 & 26 & $8.0^{+19.3}_{-8.0}$ & $6.99^{+0.03}_{-0.02}$ & $8.9^{+0.3}_{-0.7}$ & $0.03^{+0.08}_{-0.03}$ & $9.0^{+0.6}_{-0.6}$ & $0.8^{+0.2}_{-0.6}$ & $3.12^{+7.69}_{-0.85}$ & $39^{+331}_{-39}$ & $<$100 \\
\multicolumn{12}{l}{\sc Galaxies with AGN:}\\
F275W$^{-}$ & 2.702 & 2 & $>$208 & $2.666^{+0.002}_{-0.002}$ & $6.9^{+0.2}_{-0.2}$ & $0.75^{+0.17}_{-0.17}$ & $9.4^{+0.7}_{-0.7}$ & $2.2^{+1.2}_{-1.2}$ & $1.51^{+7.01}_{-0.92}$ & $0.06^{+3.57}_{-0.06}$ & $0.6^{+732}_{-0.2}$ \\
F275W & 2.665 & 3 & $23.8^{+7.2}_{-1.4}$ & $2.760^{+0.002}_{-0.002}$ & $6.9^{+0.2}_{-0.3}$ & $0.55^{+0.29}_{-0.32}$ & $9.4^{+0.7}_{-0.6}$ & $2.2^{+1.4}_{-1.1}$ & $1.48^{+6.95}_{-0.91}$ & $24^{+4}_{-4}$ & $20^{+24}_{-17}$ \\
F336W & 3.427 & 2 & $3.0^{+8.2}_{-3.0}$ & $3.92^{+0.02}_{-0.03}$ & $8.9^{+0.3}_{-0.3}$ & $0.40^{+0.20}_{-0.20}$ & $9.9^{+0.3}_{-0.3}$ & $<-4$ & $2.46^{+8.35}_{-0.98}$ & $<$100 & $<$100 \\
\multicolumn{12}{l}{\sc All Galaxies:}\\
F275W$^{-}$ & 2.725 & 35 & $16.9^{+51.5}_{-16.9}$ & $2.945^{+0.002}_{-0.002}$ & $8.82^{+0.01}_{-1.88}$ & $0.15^{+0.51}_{-0.15}$ & $9.4^{+0.3}_{-1.1}$ & $0.7^{+0.8}_{-0.2}$ & $1.53^{+7.02}_{-0.94}$ & $<$38 & $3^{+49}_{-2}$ \\
F275W & 2.721 & 36 & $25.8^{+67.4}_{-8.6}$ & $2.999^{+0.002}_{-0.001}$ & $8.82^{+0.03}_{-1.92}$ & $0.15^{+0.49}_{-0.15}$ & $9.4^{+0.3}_{-1.1}$ & $1.0^{+0.6}_{-0.4}$ & $1.53^{+7.02}_{-0.93}$ & $13^{+26}_{-11}$ & $5^{+39}_{-3}$ \\
F336W & 3.596 & 28 & $7.1^{+18.6}_{-7.1}$ & $6.58^{+0.02}_{-0.02}$ & $9.1^{+0.1}_{-0.9}$ & $0.01^{+0.16}_{-0.01}$ & $9.4^{+0.4}_{-1.0}$ & $0.8^{+0.2}_{-0.7}$ & $3.11^{+7.71}_{-0.85}$ & $<$100 & $<$100 \\
\midrule
\multicolumn{12}{l}{\sc \underline{ERS}}\\
\multicolumn{12}{l}{\sc Galaxies without AGN:}\\
F225W & 2.350 & 17 & $18.7^{+51.7}_{-18.7}$ & $3.83^{+0.02}_{-0.02}$ & $7.7^{+0.5}_{-0.3}$ & $0.75^{+0.01}_{-0.64}$ & $8.81^{+0.85}_{-0.03}$ & $0.1^{+0.6}_{-1.8}$ & $1.12^{+4.34}_{-0.78}$ & $6^{+14}_{-6}$ & $9^{+98}_{-4}$ \\
F275W & 2.752 & 7 & $>$51.1 & $6.80^{+0.03}_{-0.03}$ & $7.30^{+1.40}_{-0.03}$ & $0.33^{+0.31}_{-0.33}$ & $9.0^{+0.7}_{-0.2}$ & $0.6^{+0.7}_{-0.9}$ & $1.68^{+6.92}_{-0.98}$ & $<$31 & $30^{+154}_{-23}$ \\
F336W & 3.603 & 10 & $>$23.5 & $6.60^{+0.02}_{-0.02}$ & $8.99^{+0.01}_{-1.09}$ & $0.01^{+0.01}_{-0.01}$ & $9.2^{+0.3}_{-0.7}$ & $0.9^{+0.2}_{-0.3}$ & $3.11^{+7.70}_{-0.85}$ & $<$130 & $<$100 \\
\multicolumn{12}{l}{\sc Galaxies with AGN:}\\
F225W & 2.374 & 2 & $3.6^{+11.1}_{-3.6}$ & $2.70^{+0.02}_{-0.02}$ & $7.6^{+0.8}_{-0.8}$ & $1.10^{+0.27}_{-0.27}$ & $9.6^{+0.4}_{-0.4}$ & $2.0^{+0.6}_{-0.6}$ & $1.24^{+4.60}_{-0.84}$ & $3^{+20}_{-3}$ & $9^{+123}_{-7}$ \\
F275W & 2.618 & 7 & $10.9^{+27.3}_{-2.1}$ & $2.64^{+0.01}_{-0.02}$ & $7.9^{+1.3}_{-1.4}$ & $0.62^{+0.83}_{-0.53}$ & $9.0^{+1.1}_{-0.1}$ & $0.8^{+2.1}_{-0.6}$ & $1.31^{+6.72}_{-0.83}$ & $20^{+19}_{-14}$ & $0.7^{+12.0}_{-0.4}$ \\
F336W & 3.316 & 3 & $11.9^{+22.3}_{-1.8}$ & $7.61^{+0.07}_{-0.07}$ & $7.4^{+0.7}_{-1.6}$ & $1.50^{+0.55}_{-1.50}$ & $9.9^{+0.8}_{-1.8}$ & $2.75^{+0.04}_{-0.09}$ & $2.12^{+8.68}_{-0.95}$ & $24^{+16}_{-14}$ & $9^{+149}_{-5}$ \\
\multicolumn{12}{l}{\sc All Galaxies:}\\
F225W & 2.352 & 19 & $19.9^{+52.6}_{-19.9}$ & $3.61^{+0.01}_{-0.02}$ & $7.6^{+0.5}_{-0.5}$ & $0.75^{+0.05}_{-0.58}$ & $8.90^{+0.79}_{-0.04}$ & $0.1^{+1.0}_{-1.7}$ & $1.22^{+4.46}_{-0.83}$ & $3^{+17}_{-3}$ & $16^{+46}_{-10}$ \\
F275W & 2.685 & 14 & $13.8^{+46.8}_{-13.8}$ & $2.85^{+0.01}_{-0.01}$ & $7.1^{+1.9}_{-0.6}$ & $0.42^{+0.87}_{-0.42}$ & $9.0^{+1.0}_{-0.2}$ & $0.8^{+1.4}_{-0.9}$ & $1.50^{+6.98}_{-0.91}$ & $17^{+23}_{-17}$ & $2^{+26}_{-1}$ \\
F336W & 3.537 & 13 & $29.4^{+80.1}_{-29.4}$ & $7.51^{+0.07}_{-0.05}$ & $8.7^{+0.3}_{-2.5}$ & $0.15^{+0.02}_{-0.15}$ & $9.5^{+0.2}_{-1.2}$ & $0.8^{+1.9}_{-0.2}$ & $2.82^{+8.00}_{-0.94}$ & $0.6^{+2.7}_{-0.6}$ & $13^{+114}_{-9}$ \\
\midrule
\multicolumn{12}{l}{\sc \underline{Total}}\\
\multicolumn{12}{l}{\sc Galaxies without AGN:}\\
F225W & 2.356 & 18 & $16.2^{+49.4}_{-16.2}$ & $3.30^{+0.01}_{-0.01}$ & $7.7^{+0.5}_{-0.6}$ & $0.75^{+0.05}_{-0.61}$ & $9.6^{+0.1}_{-0.8}$ & $1.2^{+0.1}_{-2.8}$ & $1.23^{+4.49}_{-0.83}$ & $8^{+15}_{-8}$ & $3^{+28}_{-2}$ \\
F275W & 2.731 & 40 & $25.4^{+80.5}_{-25.4}$ & $4.884^{+0.009}_{-0.007}$ & $8.82^{+0.02}_{-1.82}$ & $0.15^{+0.45}_{-0.15}$ & $9.1^{+0.6}_{-0.7}$ & $0.9^{+0.5}_{-0.8}$ & $1.54^{+7.02}_{-0.94}$ & $6^{+22}_{-6}$ & $13^{+27}_{-9}$ \\
F336W & 3.607 & 36 & $23.1^{+76.1}_{-23.1}$ & $6.86^{+0.02}_{-0.02}$ & $9.0^{+0.2}_{-0.9}$ & $0.03^{+0.08}_{-0.03}$ & $9.3^{+0.4}_{-1.0}$ & $0.8^{+0.2}_{-0.6}$ & $3.12^{+7.69}_{-0.85}$ & $57^{+61}_{-57}$ & $<$100 \\
\multicolumn{12}{l}{\sc Galaxies with AGN:}\\
F225W & 2.374 & 2 & $5.0^{+9.4}_{-5.0}$ & $2.70^{+0.01}_{-0.02}$ & $7.6^{+0.8}_{-0.8}$ & $1.10^{+0.27}_{-0.27}$ & $9.6^{+0.4}_{-0.4}$ & $2.0^{+0.6}_{-0.6}$ & $1.24^{+4.60}_{-0.84}$ & $<$23 & $8^{+120}_{-5}$ \\
F275W$^{-}$ & 2.637 & 9 & $108^{+298}_{-21}$ & $2.656^{+0.003}_{-0.006}$ & $7.9^{+1.2}_{-1.4}$ & $0.42^{+0.90}_{-0.26}$ & $9.0^{+1.3}_{-0.4}$ & $0.9^{+2.5}_{-0.6}$ & $1.33^{+6.86}_{-0.85}$ & $2^{+3}_{-1}$ & $0.3^{+2.3}_{-0.2}$ \\
F275W & 2.632 & 10 & $38.9^{+3.3}_{-2.7}$ & $2.721^{+0.004}_{-0.005}$ & $7.9^{+1.2}_{-1.4}$ & $0.42^{+0.85}_{-0.32}$ & $9.0^{+1.3}_{-0.3}$ & $0.9^{+2.4}_{-0.6}$ & $1.33^{+6.83}_{-0.84}$ & $10.7^{+1.0}_{-0.6}$ & $1.8^{+8.0}_{-0.3}$ \\
F336W & 3.360 & 5 & $10.8^{+20.0}_{-1.2}$ & $7.48^{+0.08}_{-0.05}$ & $7.5^{+1.6}_{-1.6}$ & $0.38^{+1.16}_{-0.38}$ & $9.0^{+1.9}_{-0.7}$ & $<-4$ & $2.36^{+8.45}_{-0.97}$ & $79^{+50}_{-43}$ & $26^{+162}_{-13}$ \\
\multicolumn{12}{l}{\sc All Galaxies:}\\
F225W & 2.358 & 20 & $18.0^{+46.7}_{-18.0}$ & $3.24^{+0.01}_{-0.01}$ & $8.2^{+0.1}_{-1.1}$ & $0.53^{+0.28}_{-0.33}$ & $8.90^{+0.82}_{-0.01}$ & $1.18^{+0.22}_{-2.69}$ & $1.23^{+4.50}_{-0.83}$ & $2^{+8}_{-2}$ & $3^{+24}_{-2}$ \\
F275W$^{-}$ & 2.713 & 49 & $62.2^{+184}_{-62.2}$ & $2.915^{+0.005}_{-0.004}$ & $8.88^{+0.04}_{-2.02}$ & $0.16^{+0.62}_{-0.16}$ & $9.1^{+0.8}_{-0.7}$ & $1.0^{+0.6}_{-0.8}$ & $1.52^{+7.01}_{-0.93}$ & $3^{+12}_{-3}$ & $2^{+6}_{-1}$ \\
F275W & 2.711 & 50 & $42.0^{+21.8}_{-3.2}$ & $2.957^{+0.003}_{-0.005}$ & $8.88^{+0.03}_{-2.08}$ & $0.16^{+0.59}_{-0.16}$ & $9.1^{+0.8}_{-0.7}$ & $0.8^{+0.9}_{-0.6}$ & $1.52^{+7.01}_{-0.93}$ & $18^{+5}_{-4}$ & $6^{+5}_{-4}$ \\
F336W & 3.577 & 41 & $24.2^{+58.2}_{-3.7}$ & $7.36^{+0.04}_{-0.06}$ & $9.1^{+0.1}_{-1.3}$ & $0.15^{+0.02}_{-0.15}$ & $9.5^{+0.3}_{-1.3}$ & $1.0^{+0.1}_{-0.8}$ & $3.08^{+7.74}_{-0.86}$ & $4^{+4}_{-3}$ & $90^{+191}_{-61}$ \\
\enddata
\vspace{5pt}
\begin{minipage}{1.03\txw}{\footnotesize
\textbf{Table columns}: (1) Observed WFC3/UVIS filter (\boldmath$^{-}$\unboldmath\! indicates the exclusion of QSO\,$J$123622.9+621526.7);
(2) Mean redshift range of galaxies included in LyC/UVC stacks;
(3) Number of galaxies included in the stack;
(4): Mean \emph{observed} flux ratio $f_{\nu,\uvc}/f_{\nu,\!\lyc}$ and its $\pm$1$\sigma$ uncertainty, as measured from the LyC and UVC stacks in their respective apertures (see \S\ref{sec:stackphot} and Table~\ref{phottab}); 
(5): Mean \emph{intrinsic} flux ratio $f_{\nu,\uvc}/f_{\nu,\!\lyc}$ and its $\pm$1$\sigma$ uncertainty, as derived from the best-fit BC03 SED models and their respective WFC3/UVIS and WFC/ACS filter curves for each of our redshift bins;
(6): Peak age of the stellar populations distribution from the best-fit BC03 models and their $\pm$1$\sigma$ standard deviations in years;
(7): Peak dust extinction \AV\ and its $\pm$1$\sigma$ uncertainty of the best-fit BC03 SED model in AB magnitudes;
(8): Peak stellar mass and its $\pm$1$\sigma$ uncertainty of the best-fit BC03 SED model in solar masses;
(9): Peak star-formation rate and its $\pm$1$\sigma$ uncertainty of the best-fit BC03 SED model in solar masses/year;
(10): Average filter-weighted IGM opacity of all sight-lines and redshifts in the stacks and their $\pm$1$\sigma$ standard deviations, calculated from the \citet{Inoue2014} models;
(11): The simplified stacked \fesc\ and $\pm$1$\sigma$ uncertainties from Eq.~\ref{fescsimp} adopting a constant intrinsic ratio $f_{\uvc}$/$f_{\!\lyc}$=3.0, the observed  $f_{\uvc}$/$f_{\!\lyc}$ ratio in column 4, the average IGM opacity in column 10, and the $A_{\uvc}$ scaled from the $A_V$ listed in column 7.
(12): ML and $\pm$1$\sigma$ uncertainties for the sample-average \fesc\ inferred from the MC analysis described in \S\ref{sec:fesc}, \ie\ the escape fraction of LyC including effects from all components of the ISM and reddening by dust, corrected for IGM attenuation.}
\end{minipage}
\end{deluxetable*}

\section{Discussion} \label{sec:discussion}
\subsection{AGN LyC Detections} \label{sec:agnlycdet}
We surveyed an area of $\sim$\,175\,arcmin$^2$ from $z$=2.26--4.35, corresponding to a comoving volume of $\sim$1.3$\times$10$^6$\,Mpc$^3$, and uncovered a single LyC detection from QSO\,$J$123622.9+621526.7. This space density of $\simeq$\,8$\times$$10^{-7}$\,Mpc$^{-3}$ is consistent with the space densities of very luminous ($\mathrm{L_X}$\,=\,10$^{45}$--10$^{47}$\,erg\,s$^{-1}$) Compton-thin AGN at $z$=2.59 \citep{Ueda2014}. The X-ray luminosity of this AGN is $\mathrm{L_X}$\,=\,$10^{44.9}$\,erg\,s$^{-1}$ and has an intrinsic \NH=0.6$\pm$0.1$\times$$10^{23}$\,cm$^{-2}$ \citep{Laird2006}. Therefore, this AGN falls within the parameter space of the observed space density trends for AGN with similar properties. This may allude to the existence of more LyC-bright AGN that have so far not been detected. 

In \S\ref{sec:qsosed}, we describe the SED model fitting and resulting parameters of the best-fit, which were revealed to be typical of AGN at $z$\,$\sim$\,2.6. Since this object does not exhibit exotic or extreme AGN accretion parameters, possible causes of the bright LyC emission may be that the escape path of LyC is advantageously aligned to the line-of-sight of the observation, the neutral density of the IGM along this line-of-sight was especially lower than the average, and/or that the LyC production and subsequent escape from AGN is delayed compared to the optical, unobscured portions of the spectrum. 

This first hypothesis has support from the bright flux detected in all other \HST\ bands. The UVC band for this AGN (WFC/ACS F606W) is measured at \mAB\,$\simeq$\,20.42\,mag, and the WFC3/IR bands F125W, F140W, and F160W measure \mAB\,$\simeq$\,20.25\,mag. The \MAB$_{\rm 1500\mathring{A}}$ is $\sim$\,--24.44\,mag, which falls well outside of the luminosity function of galaxies at these redshifts. It is possible that the luminous nature of this AGN can be due to collimation of the AGN jet or outflow clearing a direct path for LyC escape into the line-of-sight of the observation alone. 

This single line-of-sight may also have much lower neutral column density than predicted by IGM simulations, which is also hinted at by very high (though highly uncertain) \fesc\ values found at $z$\,$\gtrsim$\,3. Other LyC studies in the protocluster SSA22 \citep{Micheva2017,Fletcher2019} have suggested a spatially varying \ion{H}{1} density in the IGM or CGM in the field in order to explain the LyC non-detections in their sample. Spatial variations of \ion{H}{1} in the IGM may also be present in the GOODS North field, allowing for higher \fesc\ values in under-dense IGM regions.

The possibility of a time-lag effect between lower energy continuum and high energy, ionizing photons has been observed in the production of hard X-rays, where multiple inverse Compton scattering (ICS) events of thermally produced optical and UV photons from the accretion disk gain energy by scattering off of relativistic electrons in AGN coronae \citep[e.g.,][]{Fabian2009,Kara2016}. Furthermore, LyC from AGN has also been found to be produced via this same ICS mechanism at wavelengths of $\mathrm{\lambda_{rest}}$\,$\lesssim$\,$1000$\,$\mathrm{\mathring{A}}$ \citep[e.g.,][]{Zheng1997,Telfer2002,Shull2012,Stevans2014}. Thus, the discrepancy between the bright LyC escape measured from this AGN and its ordinary SMBH/accretion parameters inferred from the unobscured continuum may be explained by a time lag between the production and subsequent escape of the LyC photons, where the time lag accumulates from multiple LyC-producing ICS events.

\begin{figure*}[th!]\centerline{
\includegraphics[width=.48\txw]{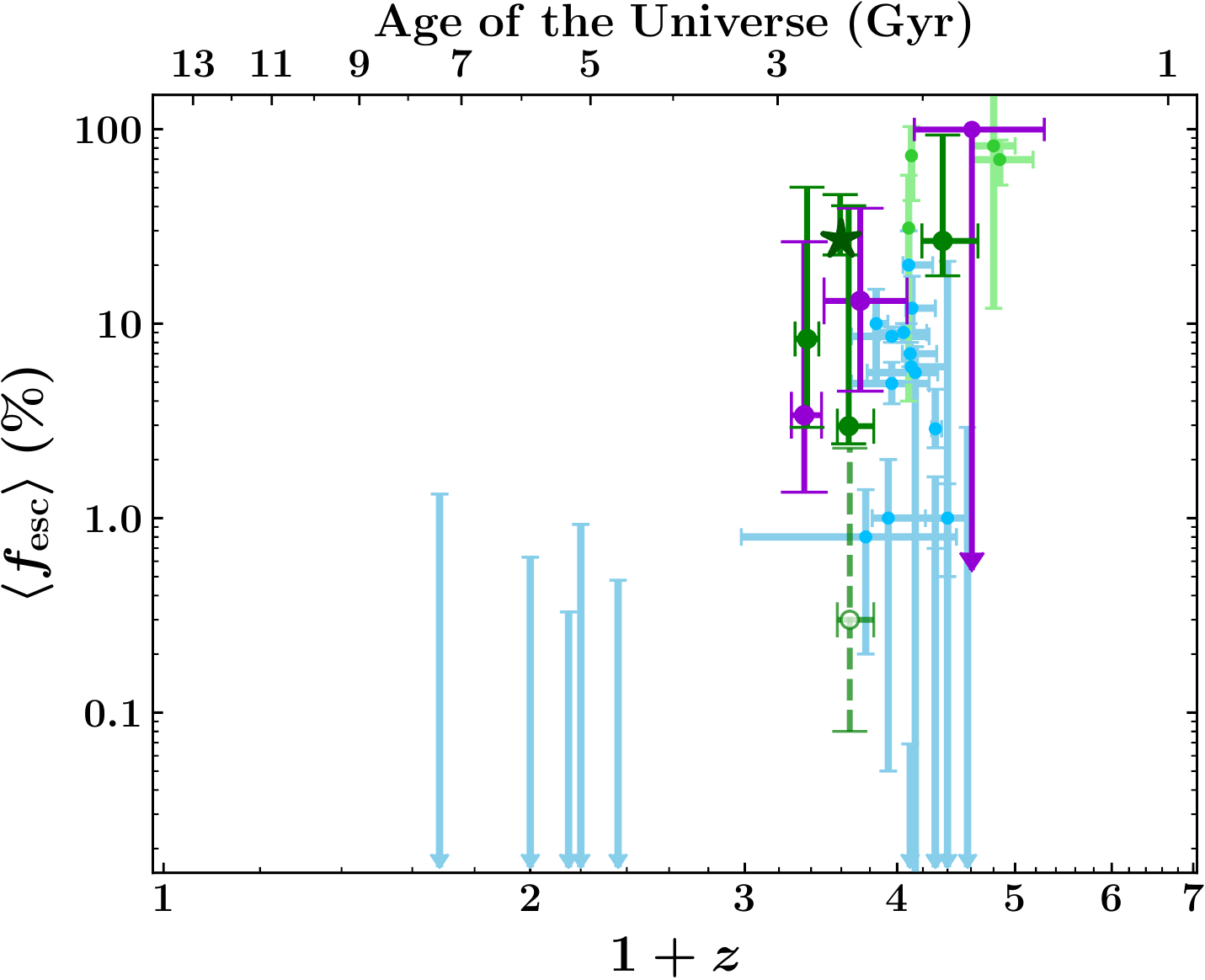}
\includegraphics[width=.2385\txw]{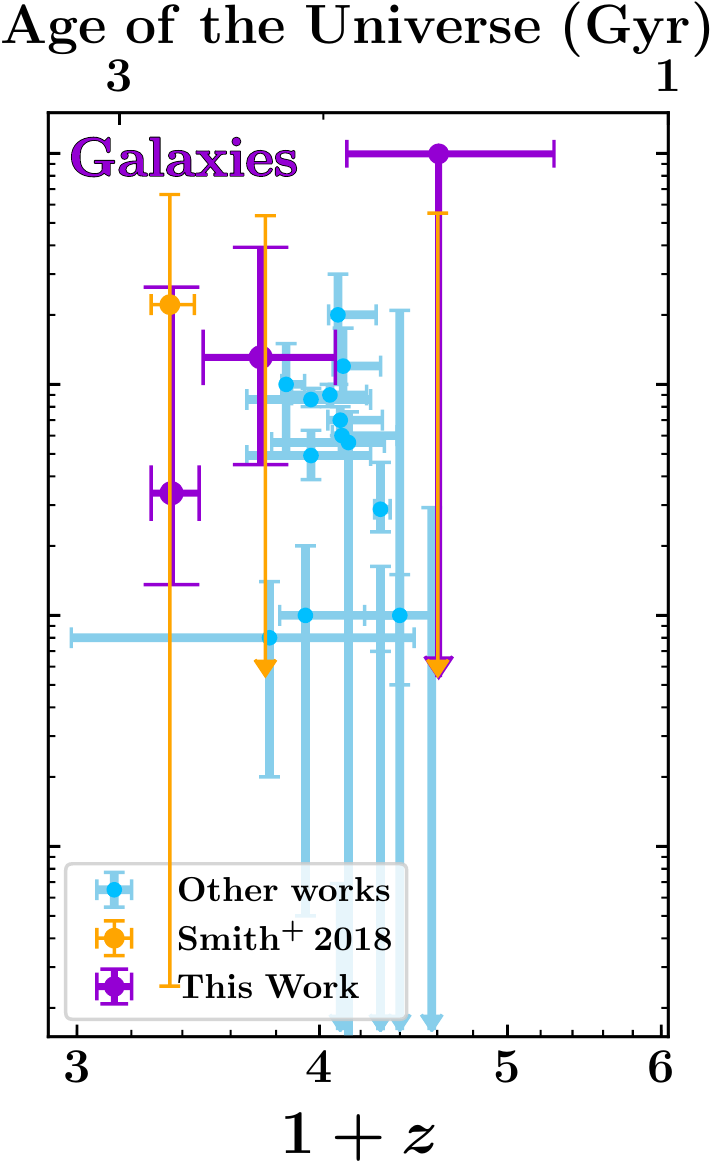}
\includegraphics[width=.255\txw]{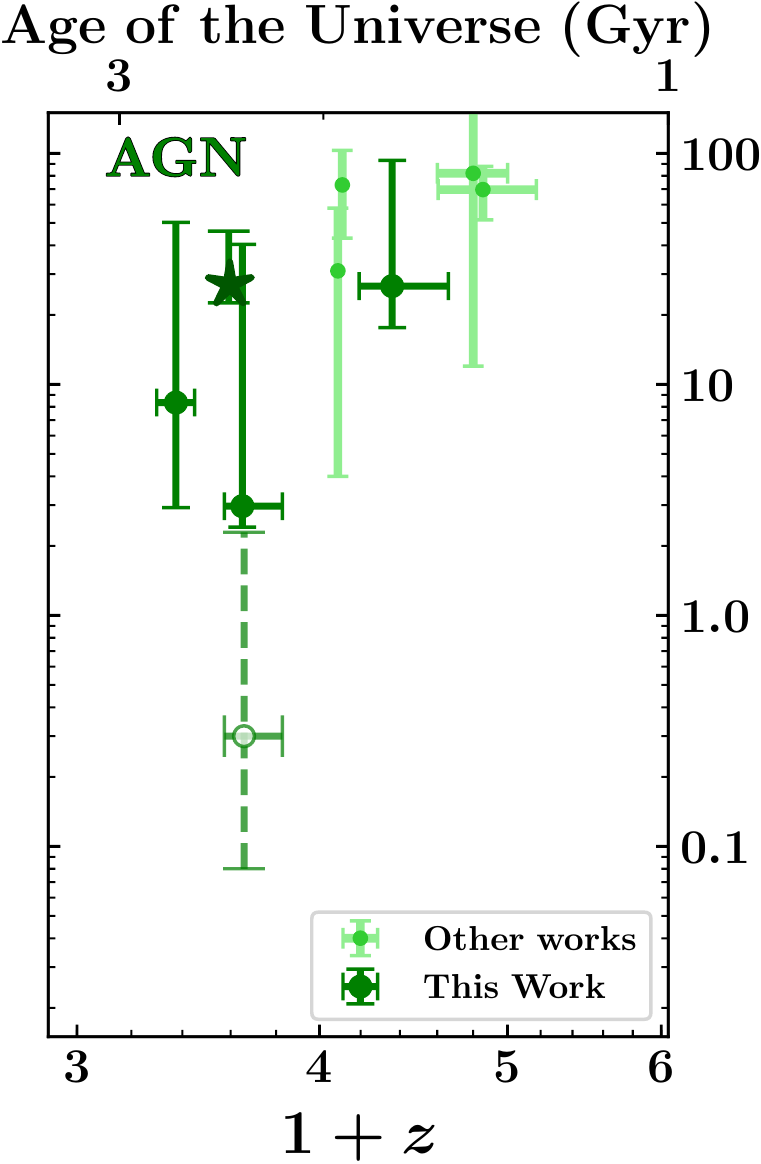}}
\caption{The inferred sample-averaged LyC escape fraction for various galaxy samples as a function of redshift. Plotted are our ML average \fesc\ value with their $\pm$1$\sigma$ range or 1$\sigma$ upper limits for our galaxies \emph{without} AGN (purple filled circles) and galaxies \emph{with} AGN (green filled circles; the open green circle indicates the exclusion of QSO\,$J$123622.9+621526.7) taken from the probability mass functions of Figs.~\ref{fescgal}--\ref{fescall}, generated from our MC simulations described in \S\ref{sec:fesc}. The green $\star$ symbol represents \fesc\ in F275W of QSO\,$J$123622.9+621526.7. Orange points are taken from \citet{Smith2018} for comparison to updated values in purple. The blue points indicate available published \emph{sample-averaged} \fesc\ values for galaxies \emph{without} AGN that have $\langle M_{\rm 1500\mathring{A}}\rangle$\,$\lesssim$\,-21. Some blue points were converted from the quoted \fescrel\ values using extinction values from the literature source. Light green points are available published sample-average \fesc\ values for AGN. References to these data are found in \S\ref{sec:intro} and \S1 of \citetalias{Smith2018}. All vertical error bars are the $\pm$1$\sigma$ uncertainty on the inferred \fesc\ values. Some errors were converted from the quoted 2--3$\sigma$ uncertainties in the literature. Upper limits are shown as downward triangles. Although the blue points represent galaxy samples with properties that differ from those of our samples, and the quoted errors were derived from uncertainties based on different methods of error assessment, the combined data is consistent with a trend of \fesc\ with redshift for both galaxies with \emph{and} without AGN based on their upper limits, given that observations are limited by the brightness of the galaxies. This compiled dataset does not rule out the possibility that massive galaxies may have had high enough LyC \fesc\ values to complete hydrogen reionization by $z$\,$\sim$\,6, if galaxies at $2\!\lesssim\! z\!\lesssim\!4$ can serve as analogs for those at $z\!\gtrsim\!6$. \label{fescvsz}}
\end{figure*}
\subsection{Galaxy \fesc\ Evolution}
The results of our \fesc\ simulations are tabulated in Table~\ref{fesctab}, and the \fesc\ results for all galaxies \emph{without} AGN in our Total sample are plotted in Fig.~\ref{fescvsz} as filled purple circles. We plot our results along with \fesc\ parameters inferred by other authors in the literature as light blue filled circles. Upper limits are shown as downward light blue arrows. 

Compared to \citetalias{Smith2018}, the observed trend in \fesc\ is more pronounced, mainly due to the higher implied \fesc\ upper limit at \zmean\,$\simeq$\,3.6. These high \fesc\ upper limits at $z$\,$\gtrsim$\,3 are dominated by two uncertainties in the data used to construct the PMF for the \fesc\ ranges. The first is large uncertainty in the LyC photometry of the \zmean\,$\simeq$\,3.6 F336W stack. Table~\ref{phottab} shows that galaxies \emph{without} AGN at \zmean\,$\simeq$\,3.6 consistently have the faintest upper limits of all the stacks without AGN. The F336W Gal stack from the ERS field in Fig.~\ref{ERSstacks} shows no clear flux, though the same stack from the GOODS/HDUV field shows very faint but possibly spurious flux in fig~\ref{GOODSstacks}. We cannot rule out the possibility that this flux is a background fluctuation due to its low S/N ratio. The GOODS/HDUV F336W flux from galaxies is still visible in the Total Gal F336W stack, though diluted from the combination with the ERS non-detections. 

The addition of the ERS also improved the S/N by a factor of $\sim$3.7, caused by a reduction in the background noise by a factor of $\sim$4.3. The S/N of the Total stacked flux for galaxies \emph{without} AGN is $\sim$0.9, with a formal \mAB\,=29.4$\pm$1.2. The possibility that the marginal fluxes of galaxies \emph{without} AGN in F336W seen in figs.~\ref{GOODSstacks}--\ref{GOODSERSstacks} are real cannot yet be ruled out. Only adding more galaxy LyC subimages to the stack can improve the uncertainties in the photometry, which is shown to still increase the S/N of the photometry in Table~\ref{phottab}, i.e., we have not yet reached the noise floor in the stack. 

The second cause of the high \fesc\ values at \zmean\,$\simeq$\,3.6 is a result of the high frequency of sight-lines with low LyC transmission through the IGM, where the average transmission value of photons passing through the F336W filter is $\mathcal{T}$$\simeq$\,4.4\%. Furthermore, the percentage of the sight-lines with $\mathcal{T}$$\leq$1\% in the MC simulations of \citet{Inoue2014} is $\sim$68.7\%, and $\sim$\,99.9\% of the lines-of-sight have transmission values below $\mathcal{T}$\,$<$\,50\%. These low transmission values cause our inferred \fesc\ to increase, due to the inverse proportionality of the modeled IGM transmission values with \fesc. The combination of a large percentage of low IGM transmission, and therefore low modeled LyC flux, with low S/N LyC photometry that varies widely allows the implied \fesc\ values to grow very large, even well above 100\% if not constrained. 

If the fluxes measured through the F336W filter for the \zmean\,$\simeq$\,3.6 stacks are taken at face value, this may allude to an IGM that has more variation on smaller scales in their lines-of-sight than sampled here. As mentioned in \S\ref{sec:agnlycdet}, similar studies in over-dense regions find LyC detections in only a portion of their homogeneous samples with no obvious reason for the dichotomy. One reason suggested is that IGM and CGM \ion{H}{1} is spatially varying in the field, allowing for some lines-of-sight to have much lower LyC attenuation than the bulk in the field. More variation in the IGM transmission at smaller scales may be needed in future transmission models in order to explain the LyC detections and non-detections at $z$\,$>$\,3. Only adding more galaxies from spectroscopic samples to improve the S/N of LyC stacks will be able to address this question.

The current stacked \fesc\ upper limit at \zmean\,$\simeq$\,3.6 is not well constrained, and the highest likelihood value from the \fesc\ MC simulations result in extreme values of \fesc\,$\sim$\,100\%. Until these values can be better constrained, they only hint at \fesc\ values that may be increasing more significantly at $z$\,$>$\,3. Since the \fesc\ values for galaxies presented in Fig.~\ref{fescvsz} are based on upper limits to LyC flux, the exact trend of \fesc\ vs. redshift cannot be determined with the current data. We note that \fesc\ values from the literature show similar trends, with \fesc\ increasing at $z$\,$\gtrsim$3. This is consistent with the scenario expressed in \citetalias{Smith2018}, where galaxies undergo more accretion and mergers until the peak of the star-formation history at $z$\,$\simeq$2, which causes \fesc\ to decrease during this epoch due to the accumulated higher \ion{H}{1} densities. The feedback from AGN accretion and SNe heat reduces the SFR, thereby reducing the formation of OB stars that could clear channels for LyC escape after going supernova, which consequently reduces \fesc\ further at lower redshift. Declining AGN and star-formation feedback at $z$\,$\lesssim$\,2 can also reduce \fesc, as these mechanisms can also assist in carving out paths in the ISM for LyC to escape. 

\subsection{The \fesc\ of Galaxies with AGN}
It is now believed that AGN LyC \fesc\ values are expected to be less than 100\% \citep{Cristiani2016,Micheva2017,Grazian2018}, rather than $\sim$\,100\% as assumed in earlier models \citep{Giallongo2015,Madau2015}. The production of LyC in galaxies by AGN is mostly well understood from theoretical models and observations of AGN spectra \citep[e.g.,][]{Telfer2002,Done2012,Kubota2018,Petrucci2018}. In short, the accretion disk emits like a blackbody when the heat energy produced by accretion thermalizes \citep{Shakura1973}, with a disk temperature increasing radially towards the SMBH. The spectrum produced by such a model \citep[e.g.,][]{Mitsuda1984} does not reproduce the observed UV spectrum of AGN, which requires a broken power-law to fit \citep[e.g.,][]{Zheng1997,Davis2007,Lusso2015}. The same warm Comptonization component that can explain the soft X-ray excess seen in some AGN spectra \citep[e.g.,][]{Kaufman2017,Petrucci2018} could also extend across the \ion{H}{1} absorption gap and connect the UV broken power law to the soft X-ray component \citep[e.g.,][]{Mehdipour2011,Mehdipour2015}. In this model, a portion of the accretion energy is not thermalized in the disk, but rather is emitted from a warm ($kT_e$\,$\simeq$\,0.1--1\,keV), optically thick region \citep[$\tau$\,$\simeq$\,10--25; see, e.g.,][]{Petrucci2013,Middei2018,Porquet2018}. UV photons down to $\sim$\,1000\,\AA\ can also be produced by the same region \citep{Kubota2018}. However, the physical origin of this emission is still not well understood phenomenologically \citep{Crummy2006,Walton2013,Rozanska2015,Petrucci2018}. 

The determinants of LyC photon absorption or escape are more unclear. As mentioned in \S\ref{sec:intro}, it is likely that the well studied, nearby, LyC emitting galaxies Tol 1247-232 \citep{Leitherer2016}, Haro 11 \citep{Leitet2011}, and J0921+4509 \citep{Borthakur2014} have AGN exhibited by their X-ray detections (\citealt{Kaaret2017}, \citealt{Prestwich2015}, \citealt{Jia2011}, respectively). \citet{Grazian2018} suggest that these galaxies likely have measurable \fesc\ values due to their AGN component creating a mechanical force to drive away nearby ISM, thereby increasing \fesc\ from the AGN and surrounding stars. \citet{Giallongo2012} propose that AGN outflow shock-waves triggered by accelerating disk outflows can clear enough paths in the ISM surrounding the AGN to increase the \fesc\ parameter (see \citealt{Menci2008}, \citealt{Dashyan2018}, \citealt{Penny2018}, and \citealt{Menci2019} for models and observations supporting this scenario). Here, the LyC can escape along a narrow bi-polar cone unobscured, and paths outside this cone will have a reduced \fesc. In these highly ionized cones however, dust extinction becomes the dominant source of LyC absorption, which is later re-emitted as IR radiation \citep[e.g.,][]{Netzer2013}. 

Dust can survive in the environment of the inner accretion disk only in regions where the effective temperature of the disk is below the dust-sublimation temperature \citep[e.g.,][]{Czerny2011}. In these regions, pressure acting on the grains from the ambient radiation field can raise dusty clumps above the disk surface. As the clumps rise, they become exposed to the central radiation source and clumps get pushed in a radial direction into sublimation regions, and the dust can evaporate. 

Bipolar outflows in the narrow-line region (NLR) of AGN have been observed from UV and optical (integral field) spectroscopy \citep{MullerSanchez2011,Harrison2014,Karouzos2016,Woo2017}, which can potentially clear some emission line emitting clouds that absorb LyC. \citet{Liu2013} were able to detect dense, optically thick, dusty gas clouds embedded in hot, low-density winds in the NLR transitioning into optically thin clouds at a distance of $\simeq$\,7.0$\pm$2\,kpc away from the SMBH, caused by declining radiation pressure on the cloud. As the clouds flow outwards with the AGN wind, the binding external radiation pressure declines, allowing the clouds to expand from their own internal gas pressure. These dense, pressure bounded clouds in the NLR produce emission lines on a thin, outer shell. Their main source of ionization is likely from the AGN continuum, both thermal and non-thermal. The transition from optically thick to optically thin would have have a direct effect on the \fesc\ of AGN as well, and thus outflows may play a dominant role in determining the \fesc\ from their resulting dynamics. 

The \fesc\ values for AGN shown in Fig.~\ref{fescvsz} (dark and light green points) are relatively consistent across all redshift ranges, showing a possible slight downward trend from $z$\,$\simeq$\,4--2. However, the mean values of these data points are consistent with a constant \fesc\ to within their error bars. This may indicate that optically selected AGN with broad emission lines may modulate their \fesc\ with the same mechanisms. With AGN accretion rates on the rise at $z$\,$\sim$\,3--4 and peaking near $z$\,$\sim$2--3 \citep[e.g.,][]{Fanidakis2012, Ueda2015}, the \fesc\ parameter could show decline due to this effect at lower redshift ($z$\,$\lesssim$\,2), mainly due to the reduction in powerful shocks from outflows and winds. AGN disks are fed by cold-accretion from the nearby ISM and inflows onto dark-matter halos, as well as advection dominated hot-accretion. These accretion modes are generally associated with outflows, AGN winds, and/or jets \citep{Ho2002,Chatterjee2011,Yuan2014}, thus decreasing accretion rates should also accompany a reduction in outflow luminosity. This may provide a mechanism for an \fesc\ parameter for AGN that declines with cosmic time, as AGN accretion and outflow energy evolves with redshift. To determine the evolution \fesc\ of AGN with redshift more precisely, additional data at lower and higher redshifts are needed, and the intrinsic nature of the AGN must also be better understood through more detailed theoretical modeling. 

\section{Conclusions} \label{sec:conclusion}
We analyzed and quantified the LyC radiation escaping from a survey of 111 spectroscopically verified galaxies in the GOODS North, GOODS South, and the ERS fields in three WFC3/UVIS filters where LyC can be observed at $z$\,$\simeq$\,2.26--4.35. We independently drizzled the GOODS/HDUV data together with the available CANDELS and UVUDF WFC3/UVIS data and found good agreement with the \citet{Oesch2018} publicly released data, with modest improvement in sky background and depth. The ERS UV images are more shallow than the GOODS/HDUV mosaics. Nevertheless, since the ERS data were taken shortly after WFC3 was installed onto \HST, losses in sensitivity from CTE degradation are not a concern for this dataset.

We studied several subsamples of these galaxies based on their redshift, observed field, and spectroscopic evidence of (weak) AGN activity. We studied our single LyC QSO\,$J$123622.9+621526.7 in more detail alone, as well as including it in stacks, and combined the various subsamples to determine any biases from the imaging data and to study the LyC escaping from galaxies with and without AGN. 

We first stacked extracted sub-images centered on galaxies from the GOODS North and South and ERS mosaics in their appropriate LyC filters, and quantified the LyC in the stack using a MC approach. We removed all potential neighboring and foreground galaxies during the stacking process using $\chi^2$ images of all available \HST\ data for each galaxy. We performed SED fitting on all the galaxies and used these to estimate their intrinsic LyC flux, then estimated the \fesc\ parameter of the stacked galaxies, as well as for QSO\,$J$123622.9+621526.7, using the modeled intrinsic flux and the MC simulated IGM transmission curves of \citet{Inoue2014} for various lines-of-sight. We find the following main results:

\smallskip\noindent (1) Our quantitative analysis of the LyC flux from the stacks of galaxies at $z$\,$\simeq$\,2.3--4.3 is tabulated in
Table~\ref{phottab}. We find upper limits to the total LyC flux of \MAB$\simeq$$M^*$ galaxies \emph{without} AGN at \zmean\,$\simeq$\,2.36, 2.73, and 3.61 to be \mAB\,$>$\,27.5, $>$\,28.5, and $>$\,28.6\,mag, respectively. For galaxies \emph{with} (weak) AGN, we find fluxes of \mAB\,$\gtrsim$\,27.6, $\simeq$\,26.15, and $\simeq$\,27.73\,mag at \zmean\,$\simeq$\,2.37, 2.65, and 3.36, where the first flux is a 1$\sigma$ upper limit and the other two measurements have S/N\,$\simeq$\,13.1 and $\sim$\,1.8, respectively. 

\smallskip\noindent (2) Our only LyC detection was measured from the galaxy QSO\,$J$123622.9+621526.7, detected at \mAB=23.19$\pm$0.01\,mag in the WFC3/UVIS F275W filter and with \GALEX\ NUV at \mAB\,=\,23.77$\pm$0.08\,mag, with S/N\,$\simeq$\,133 and 13, respectively. We calculate an \fesc\ of the WFC3/UVIS F275W and \GALEX\ NUV flux using our best-fitting XSpec SED and IGM transmission models from \citet{Inoue2014}, corresponding to \fesc$^{\mathrm{F275W}}$\,$\simeq$\,28$^{+20}_{-4}$\% and \fesc$^{\mathrm{NUV}}$\,$\simeq$\,30$^{+20}_{-4}$\%. Using the simplified approach for calculating \fesc\ in Eq.~\ref{fescsimp}, and assuming a constant intrinsic ratio of UVC-to-LyC of $\langle f_{\uvc}/f_{\lyc}\rangle_{\mathrm{int}}$=2.0 and using the average, filter-weighted IGM opacity at $z$\,=\,2.59 of $\tau_{\igm}$\,$\simeq$\,1.29, we find \fesc$^{\mathrm{simp}}$\,$\simeq$\,62\% and \fesc$^{\mathrm{simp}}$\,$\simeq$\,36\% for the WFC3/UVIS F275W and \GALEX\ NUV fluxes, respectively. The differences in values for each \fesc\ calculation method can be attributed to the wavelength dependence in the attenuation of ionizing photons by the IGM, which must be incorporated for more accurate \fesc\ estimations. Our modeling suggest that this AGN is not especially extreme in its SMBH parameters, and neither in its accretion characteristics. This implies that the LyC escaping from this AGN may instead be advantageously directed toward the line-of-sight of observation, that the LyC production and subsequent escape took much longer than the time-scale of the peak of its accretion, and/or that the particular line-of-sight of this AGN had a very low \ion{H}{1} IGM column density.

\smallskip\noindent (3) The combined LyC emission averaged over the three WFC3/UVIS filters implies that the AGN dominate the LyC production in the epoch of \zmean$\simeq$2.3--4.3 by a factor of $\sim$10. The overall LyC flux distribution of AGN may also be non-centrally concentrated, though additional data are needed to make this feature more visible above the deeper image noise. If real, this could suggest a radial dependence of \fesc\ based on axial direction of the AGN LyC escape, ISM porosity, and/or scattering of the LyC photons from ionized regions in the galaxy. 

\smallskip\noindent (4) Our best-fit BC03 SED models fit to \HST\ continuum observations longwards of \Lya\ suggest that the observed LyC fluxes for galaxies \emph{without} AGN correspond to \emph{average} LyC escape fraction of \fesc\,$\simeq$3$^{+23}_{-2}$\% at \zmean$\simeq$2.4, \fesc\,$\simeq$13$^{+26}_{-9}$\% at \zmean$\simeq$2.7, and \fesc\,$\lesssim$99.5$^{+0.5}_{-2.1}$\% at \zmean$\simeq$3.6. This large \fesc\ at \zmean$\simeq$3.6 is due to a combination of two effects. The first is caused by the low S/N LyC flux in the image stack, which cannot be ruled out as spurious, and the second is implied by the majority of very low IGM line-of-sight transmission in simulations, with 91\% of the transmission values lying at $\mathcal{T}$\,$<$\,20\%, and 68\% having transmission values $\mathcal{T}$\,$<$\,1\%. This effect is seen to be mitigated in the AGN \fesc\ simulations, which had higher LyC S/N values than galaxies \emph{without} AGN. We measure \emph{average} LyC \fesc\ values of all galaxies \emph{with} AGN to be \fesc\,$\simeq$8$^{+42}_{-5}$\% at \zmean$\simeq$2.4, \fesc\,$\simeq$3$^{+37}_{-0.6}$\% at \zmean$\simeq$2.7, and \fesc\,$\simeq$27$^{+67}_{-9}$\% at \zmean$\simeq$3.4. 

\smallskip\noindent (5) Our uncertainty ranges on \fesc\ for galaxies \emph{without} AGN remain large, though they are generally consistent and improved from \citetalias{Smith2018}. This data is still consistent with the observed increasing trend of \fesc\ with redshift from \citetalias{Smith2018}, though this trend cannot be well constrained since the current \fesc\ data is based on upper limits to the LyC flux. For our galaxies, the steepest decline in \fesc\ appears to occur near $z$\,$\simeq$\,2 from \fesc$<$26\% to $\lesssim$2\%, which correlates with the peak of the cosmic star-formation history within an interval of $\pm$1\,Gyr \citep{Madau2014}. 

For galaxies \emph{with} AGN, their stacked \fesc\ appears to remain roughly constant with redshift within their error ranges, though shows hints of decline within their 1$\sigma$ limits, dropping from $<$94\% to $<$26\% from $z$\,$\simeq$\,3.6 to $z$\,$\simeq$\,2.4, respectively. The AGN \fesc\ trend can also be compared to trends in AGN luminosity functions and space density, which steadily peak near $z$\,$\simeq$\,2 and decrease at $z$\,$<$\,2 \citep{Ueda2015}. The evolution of these parameters may be linked by the decrease in AGN fuel from major mergers and/or accretion \citep{Fabian2012} and star-formation feedback through cosmic time. The evolution in these parameters may be correlated to changing dynamics of galaxies, where infall/merger driven star-formation at 2\,$\lesssim$\,$z$\,$\lesssim$\,6 transitions to a more passively evolving universe by giant galaxies at $z$\,$\lesssim$\,1--2. This may result in gas and dust rapidly accumulating in the disks and nuclei of forming galaxies, combined with a SNe rate that has progressively less impact on clearing gas/dust in galaxies that are steadily growing in mass with cosmic time. The accumulating \ion{H}{1} gas and decreasing SFR may have caused \fesc\ to decrease over a relatively narrow interval of cosmic time from 2$<$\,$z$\,$<$3 ($\sim$1\,Gyr), as feedback effects inhibit the formation of new massive stars that could clear LyC escape paths. When AGN outflows began to increase after the peak in the cosmic star-formation history at $z$\,$\simeq$\,2, their outflows may have cleared enough paths in the ISM of host galaxies to enhance the fraction of escaping LyC radiation produced by massive stars and from the accretion disk, resulting in AGN beginning to dominate the ionizing background at $z$\,$\lesssim$\,2--3. 

\smallskip\noindent (6) If the trend in \fesc\ for galaxies \emph{without} AGN, based on their upper limits to the LyC flux, continues beyond $z$$\sim$5, these galaxies may have had a sufficient LyC escape fraction to reionize the IGM by $z$\,$\gtrsim$\,6. Since AGN outshine galactic-stellar LyC by a factor of $\sim$10, combined with their consistently larger \fesc\ seen in this work and in the literature, AGN likely contributed a significant portion of the ionizing photons needed to finish \emph{and} maintain cosmic reionization at $z$\,$\lesssim$\,3. Due to the downsizing of SMBHs with decreasing redshift \citep{Ueda2003,Hasinger2005}, AGN of similar masses to QSO\,$J$123622.9+621526.7 may exist in higher abundance during reionization and could provide a significant portion to the ionizing background. Resolving the X-ray background above $\sim$10\,keV may reveal more heavily obscured, Compton-thick AGN that would increase the observed space density of all AGN at $z$\,$\gtrsim$\,2 \citep{Alexander2008}. Improvements in the accuracy of AGN space densities and their luminosity functions may provide enough statistics to prove the effectiveness of AGN in assisting SFGs significantly with reionization. 

More data on LyC \fesc\ are essential to reducing the uncertainties in these trends. The current sample of deep, high quality spectra are still very small, and larger spectroscopic samples taken with the \JWST\ FGS/NIRISS grisms and with NIRSpec \citep{Gardner2006} could improve uncertainties in LyC stacks. Additional deep imaging of wider \HST\ fields in the UV would also supplement these studies, e.g., in the COSMOS and EGS fields where large spectroscopic samples at high redshift already exist. The future release of the Ultraviolet Imaging of the Cosmic Assembly Near-infrared Deep Extragalactic Legacy Survey Fields (UVCANDELS; PI: \citeauthor{Teplitz2018}) data will provide additional LyC sources larger than our current sample size, and further increase the sensitivity to faint LyC in the GOODS fields. Additional theoretical and observational work is needed to improve the statistics of IGM line-of-sight transmission curves, in order to explain the observed larger \fesc\ values at $z$\,$\simeq$\,3 in this study and others in the literature. 

\acknowledgments
This paper is based in part on Early Release Science observations made by the WFC3 Scientific Oversight Committee. We are grateful to the Director of the Space Telescope Science Institute, Dr. Matt Mountain, for generously awarding Director's Discretionary time for this program. Finally, we are deeply indebted to the crew of STS-125 for refurbishing and repairing \HST. We thank Prof. Charles Steidel and Prof. Naveen Reddy who provided invaluable Keck spectra used for our sample selection. We thank Dr. Keith Arnaud for his instrumental guidance on running Xspec. Finally, we thank Gil Speyer and Jason Yalim from ASU Research Computing for providing helpful advice with running our MC simulation code on the Agave cluster. Support for \HST\ programs GO-11359, AR-13877, and AR-14591 was provided by NASA through grants from the STScI, which is operated by the Association of Universities for Research Inc., under NASA contract NAS 5-26555. R.A.W. acknowledges support from NASA JWST Interdisciplinary Scientist grants NAG5-12460, NNX14AN10G, and 80NSSC18K0200 from GSFC.

\facilities{\HST(ACS, WFC3/UVIS, WFC3/IR), VLT(FORS, FORS2, VIMOS, and MUSE), Keck(LRIS)}

\software{astropy \citep{Astropy2018},
		  DrizzlePac \citep{Gonzaga2012},
		  CIAO tools \citep{Fruscione2006},
		  FAST \citep{Kriek2009,Aird2018},
		  FAST++ \citep{Schreiber2018},
          SExtractor \citep{Bertin1996},
		  Xspec v12 \citep{Arnaud1996}
          }
\clearpage


\bibliography{ms}

\appendix

\section{Best-fitting BC03 SEDs for galaxies without AGN} \label{app:sed}
The remaining SED fits for galaxies \emph{without} AGN continued from Fig.~\ref{galseds} are shown here, and BC03, dust extinction, and AGN parameters of our sample are shown in Table~\ref{objtab}.
\noindent\begin{figure*}[th!]\includegraphics[width=\txw]{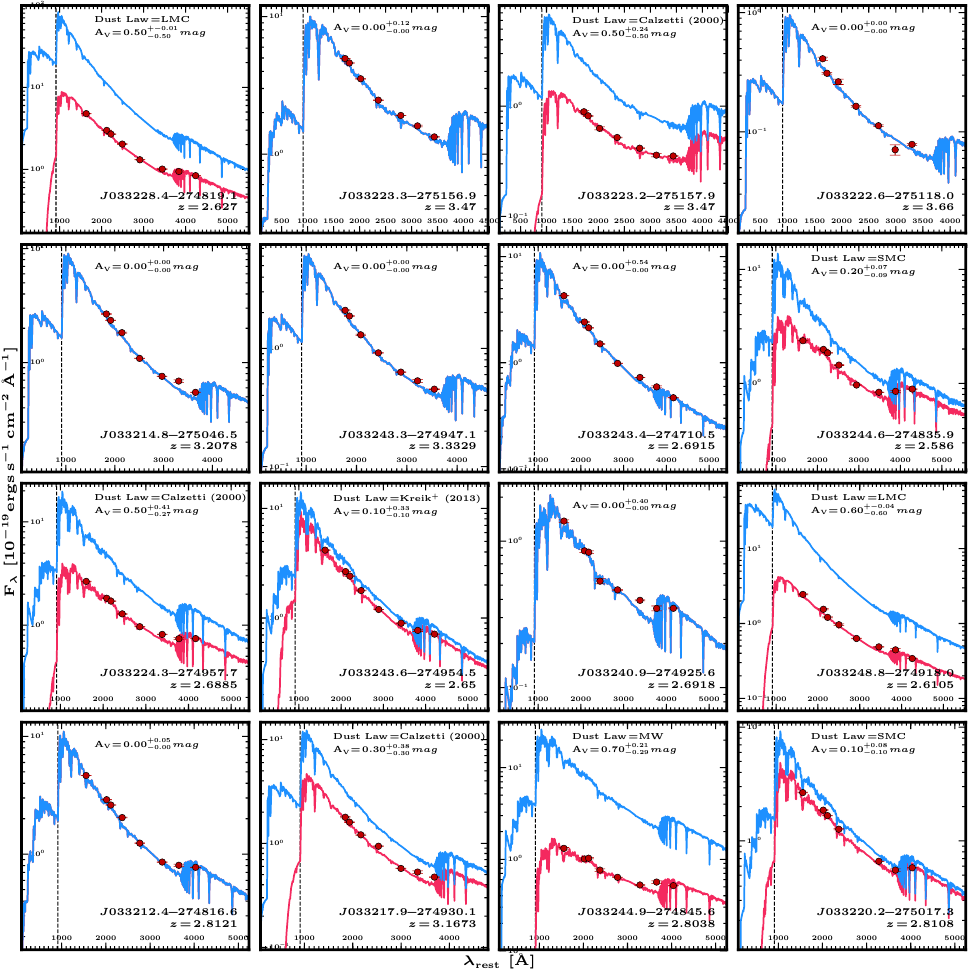}
\caption{The remainder of the BCO3 SEDs in our sample, continued from Fig.~\ref{galseds}.}
\end{figure*}
\setcounter{figure}{16}
\noindent\begin{figure*}[th!]\includegraphics[width=\txw]{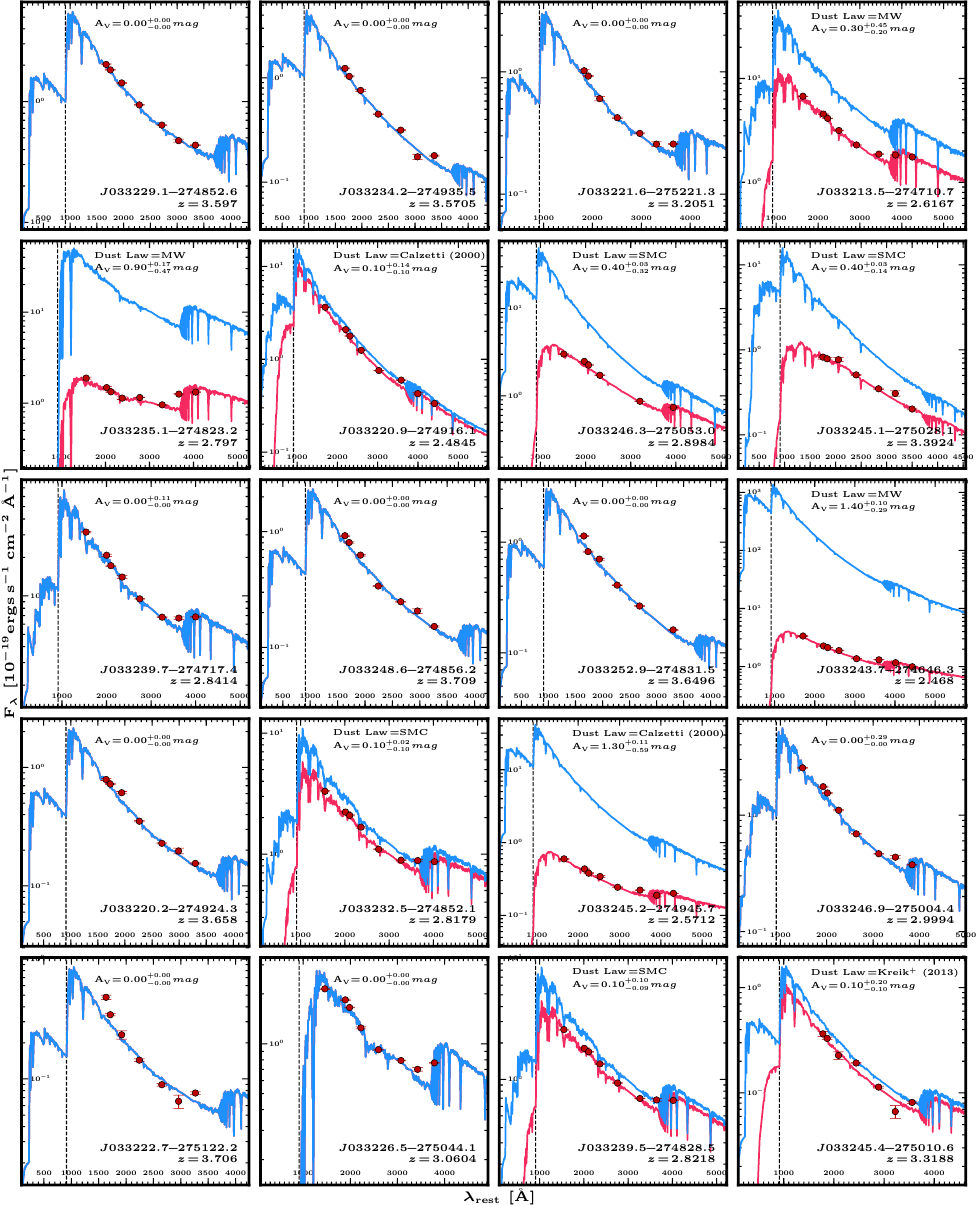}\vspace*{-6pt}
\caption{Continued}
\end{figure*}
\setcounter{figure}{16}
\noindent\begin{figure*}[th!]\includegraphics[width=\txw]{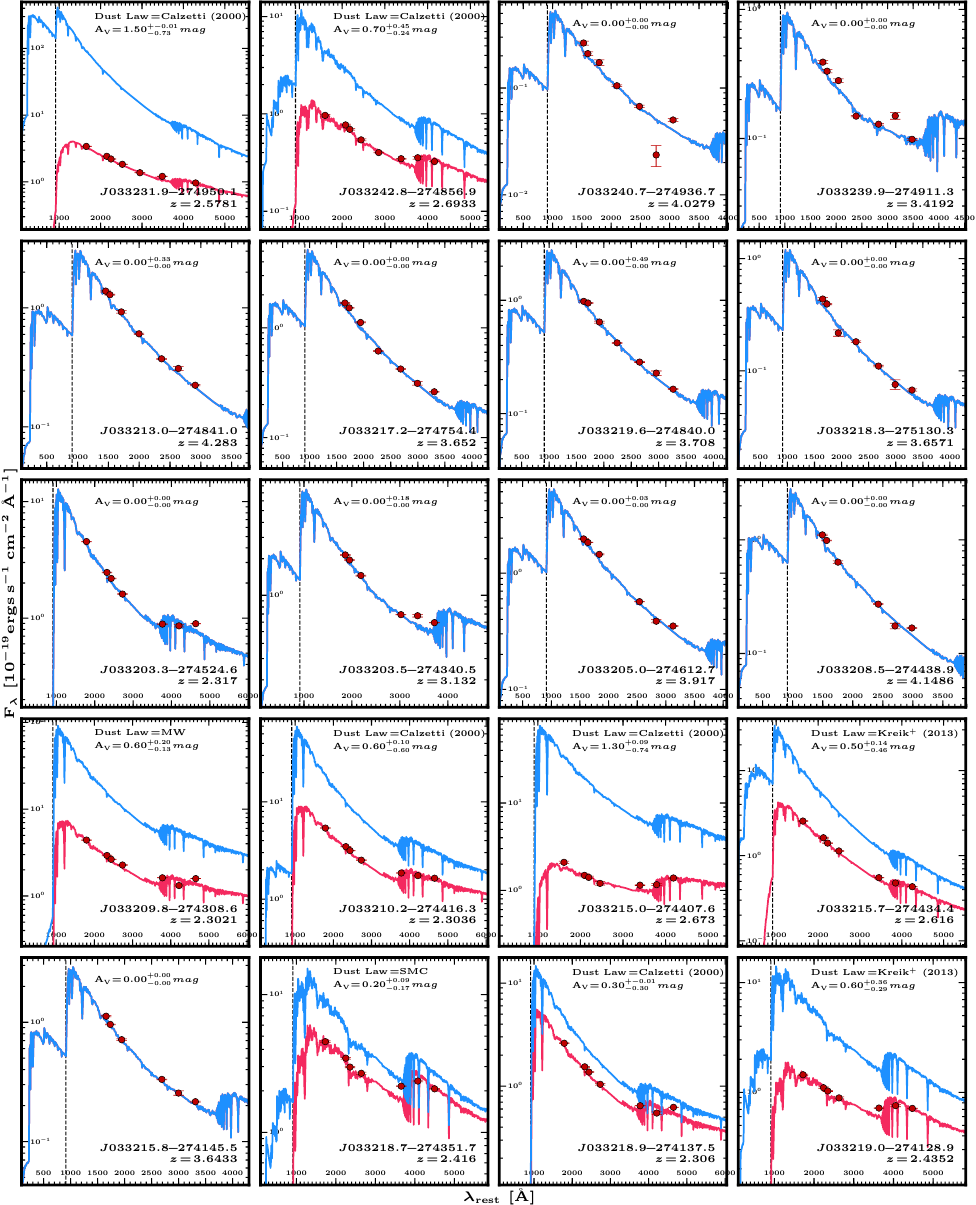}\vspace*{-6pt}
\caption{Continued}
\end{figure*}
\setcounter{figure}{16}
\noindent\begin{figure*}[th!]\includegraphics[width=\txw]{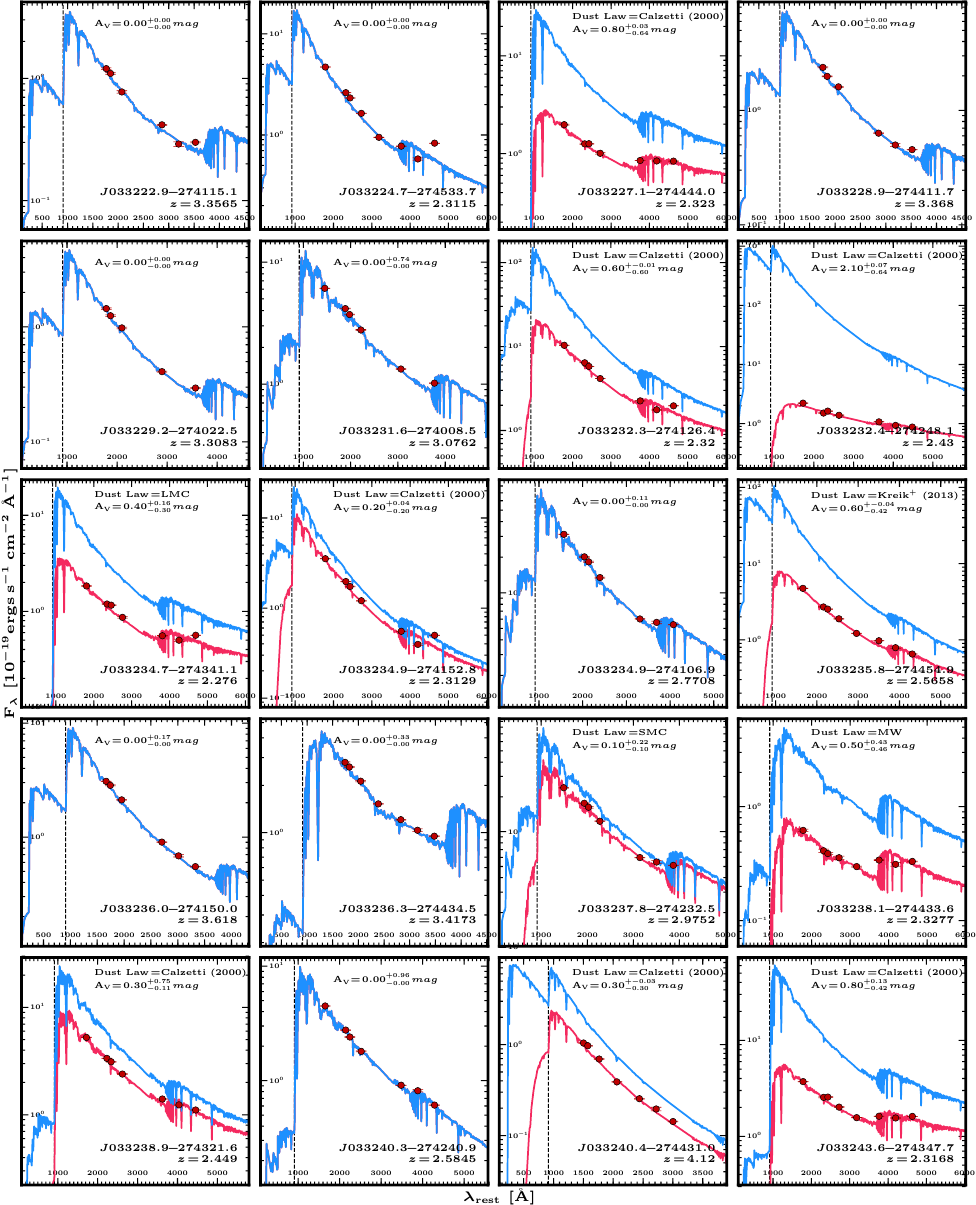}\vspace*{-6pt}
\caption{Continued}
\end{figure*}
\setcounter{figure}{16}
\noindent\begin{figure*}[th!]\includegraphics[width=0.5\txw]{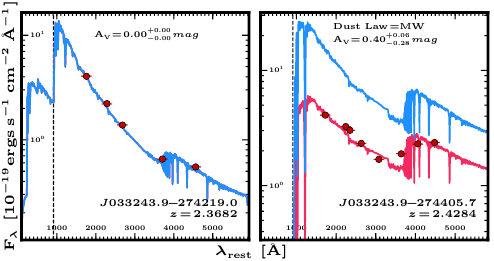}\vspace*{-6pt}
\caption{Continued}
\end{figure*}

\clearpage
{\setlength{\tabcolsep}{1pt}
\begin{longtable}[c]{L{50pt}C{50pt}C{29pt}C{17pt}C{1pt}C{24pt}cC{85pt}ccccC{25pt}R{15pt}}
\caption{List of Galaxies and SED parameters \label{objtab}}\\
\toprule \\ [-19pt]
\colhead{RA (J2000)} & \colhead{Dec (J2000)}  & \colhead{$z_{\mathrm{spec}}$} & \colhead{$m_{J,{\rm AB}}$} & \colhead{$M_{\!1500\mathrm{\mathring{A}}}^{\mathrm{AB}}$} & \colhead{($V\!-\!I$)} & \colhead{$\mathrm{f_{AGN}}$} & \colhead{Dust\ Law} & \colhead{$\mathrm{A_V}$} & \colhead{Age} & \colhead{log($\mathrm{M_{\star}/M_{\odot}}$)} & \colhead{$\mathrm{SFR}$} & $Z$ & $\chi^2$ \\ [-5pt]
\colhead{[hh:mm:ss]}  & \colhead{[dd:mm:ss]} & \colhead{\null} & 
 \colhead{[mag]} & \colhead{[mag]} & \colhead{[mag]} & \colhead{[\%]} & \colhead{\null} & \colhead{[mag]} & \colhead{[$\mathrm{log(yr)}$]} & \colhead{\null} & \colhead{[log($M_{\odot}/yr$)]} & \colhead{\null} & \colhead{\null} \\ [2pt]
\midrule \\ [-19pt]
03:32:01.580 & $-$27:43:27.01 & 2.7212 & 24.06 & $-$20.98 & \phantom{$-$}0.07 & 38 & Calzetti (2000) & 1.1 & 6.5 & 9.09 & \phantom{$-$}2.89 & 0.020 & 0.49 \\[-5pt]
03:32:02.123 & $-$27:43:27.65 & 2.7260 & 24.28 & $-$19.66 & \phantom{$-$}0.84 & 0 & Calzetti (2000) & 2.5 & 6.5 & 9.98 & \phantom{$-$}3.63 & 0.020 & 4.98 \\[-5pt]
03:32:03.036 & $-$27:44:50.08 & 2.5730 & 24.71 & $-$19.01 & \phantom{$-$}0.69 & 0 & Calzetti (2000) & 1.4 & 9.4 & 10.8 & \phantom{$-$}1.52 & 0.020 & 6.39 \\[-5pt]
03:32:03.244 & $-$27:45:18.85 & 3.2171 & 25.14 & $-$21.12 & \phantom{$-$}\nodata & 59 & \nodata & 0.0 & 5.7 & 8.06 & \phantom{$-$}2.65 & 0.020 & 0.66 \\[-5pt]
03:32:03.348 & $-$27:45:24.56 & 2.3170 & 24.74 & $-$20.48 & $-$0.09 & 0 & \nodata & 0.0 & 7.7 & 8.93 & $-$1.01 & 0.004 & 4.97 \\[-5pt]
03:32:03.489 & $-$27:43:40.52 & 3.1320 & 25.01 & $-$20.86 & \phantom{$-$}0.30 & 0 & \nodata & 0.0 & 9.3 & 9.65 & \phantom{$-$}0.79 & 0.004 & 3.62 \\[-5pt]
03:32:04.938 & $-$27:44:31.74 & 3.4739 & 24.45 & $-$21.93 & \phantom{$-$}\nodata & 62 & \nodata & 0.0 & 6.0 & 8.39 & \phantom{$-$}2.67 & 0.020 & 1.59 \\[-5pt]
03:32:05.022 & $-$27:46:12.67 & 3.9170 & 25.22 & $-$21.13 & \phantom{$-$}0.62 & 0 & \nodata & 0.0 & 8.3 & 8.88 & \phantom{$-$}0.98 & 0.004 & 6.5 \\[-5pt]
03:32:07.998 & $-$27:46:57.28 & 2.6123 & 25.25 & $-$19.73 & \phantom{$-$}0.23 & 27 & Calzetti (2000) & 0.7 & 8.4 & 9.13 & \phantom{$-$}1.15 & 0.020 & 0.53 \\[-5pt]
03:32:08.266 & $-$27:41:53.56 & 2.4700 & 25.93 & $-$19.48 & $-$0.08 & 30 & Calzetti (2000) & 0.1 & 9.0 & 9.2 & $-$0.09 & 0.020 & 0.51 \\[-5pt]
03:32:08.455 & $-$27:44:38.85 & 4.1486 & 26.02 & $-$20.49 & \phantom{$-$}0.55 & 0 & \nodata & 0.0 & 7.6 & 8.14 & \phantom{$-$}0.89 & 0.004 & 11.1 \\[-5pt]
03:32:09.447 & $-$27:48:06.80 & 2.8280 & 20.93 & $-$24.50 & \phantom{$-$}0.17 & 23 & Calzetti (2000) & 1.0 & 6.5 & 10.4 & \phantom{$-$}4.02 & 0.020 & 1.92 \\[-5pt]
03:32:09.797 & $-$27:43:08.64 & 2.3021 & 24.09 & $-$20.34 & \phantom{$-$}0.12 & 0 & MW & 0.6 & 7.6 & 9.68 & $-$1.09 & 0.004 & 8.01 \\[-5pt]
03:32:10.189 & $-$27:44:16.30 & 2.3036 & 23.94 & $-$20.55 & \phantom{$-$}0.09 & 0 & Calzetti (2000) & 0.6 & 7.7 & 9.49 & \phantom{$-$}0.25 & 0.004 & 0.32 \\[-5pt]
03:32:11.595 & $-$27:48:50.14 & 2.6061 & 24.84 & $-$20.72 & \phantom{$-$}0.06 & 0 & SMC & 0.2 & 7.0 & 8.35 & \phantom{$-$}1.23 & 0.004 & 2.57 \\[-5pt]
03:32:12.394 & $-$27:48:16.55 & 2.8121 & 24.74 & $-$20.90 & \phantom{$-$}0.05 & 0 & \nodata & 0.0 & 8.8 & 9.29 & \phantom{$-$}0.89 & 0.020 & 1.74 \\[-5pt]
03:32:12.982 & $-$27:48:41.03 & 4.2830 & 25.52 & $-$20.83 & \phantom{$-$}0.63 & 0 & \nodata & 0.0 & 8.2 & 8.7 & \phantom{$-$}0.87 & 0.004 & 0.98 \\[-5pt]
03:32:13.462 & $-$27:47:10.66 & 2.6167 & 24.08 & $-$21.07 & \phantom{$-$}0.14 & 0 & MW & 0.3 & 9.0 & 10 & \phantom{$-$}1.44 & 0.050 & 0.46 \\[-5pt]
03:32:14.788 & $-$27:50:46.49 & 3.2078 & 24.88 & $-$21.12 & \phantom{$-$}0.12 & 0 & \nodata & 0.0 & 8.8 & 9.3 & \phantom{$-$}0.91 & 0.004 & 3.16 \\[-5pt]
03:32:14.992 & $-$27:44:07.60 & 2.6730 & 24.46 & $-$19.65 & \phantom{$-$}0.20 & 0 & Calzetti (2000) & 1.3 & 7.7 & 10 & $-$2.07 & 0.004 & 3.28 \\[-5pt]
03:32:15.653 & $-$27:44:34.44 & 2.6160 & 25.26 & $-$20.05 & \phantom{$-$}0.08 & 0 & Kriek \& Conroy (2013) & 0.5 & 7.5 & 8.59 & \phantom{$-$}1.28 & 0.008 & 0.31 \\[-5pt]
03:32:15.785 & $-$27:41:45.53 & 3.6433 & 25.81 & $-$20.34 & \phantom{$-$}0.40 & 0 & \nodata & 0.0 & 9.0 & 9.19 & \phantom{$-$}0.59 & 0.004 & 1.1 \\[-5pt]
03:32:17.216 & $-$27:47:54.37 & 3.6520 & 25.50 & $-$20.80 & \phantom{$-$}0.29 & 0 & \nodata & 0.0 & 8.0 & 8.53 & \phantom{$-$}0.90 & 0.004 & 5.34 \\[-5pt]
03:32:17.414 & $-$27:44:39.99 & 2.6503 & 25.54 & $-$20.54 & $-$0.15 & 49 & \nodata & 0.0 & 6.3 & 7.81 & \phantom{$-$}1.76 & 0.020 & 3.56 \\[-5pt]
03:32:17.945 & $-$27:49:30.11 & 3.1673 & 25.04 & $-$20.61 & \phantom{$-$}0.18 & 0 & Calzetti (2000) & 0.3 & 8.7 & 9.3 & \phantom{$-$}1.03 & 0.004 & 1.83 \\[-5pt]
03:32:18.289 & $-$27:51:30.35 & 3.6571 & 26.83 & $-$19.29 & \phantom{$-$}0.19 & 0 & \nodata & 0.0 & 8.3 & 8.14 & \phantom{$-$}0.25 & 0.004 & 5.41 \\[-5pt]
03:32:18.726 & $-$27:43:51.67 & 2.4160 & 23.77 & $-$20.36 & \phantom{$-$}0.27 & 0 & SMC & 0.2 & 8.1 & 9.72 & \phantom{$-$}0.63 & 0.050 & 1.05 \\[-5pt]
03:32:18.830 & $-$27:51:35.46 & 3.6609 & 25.58 & $-$20.17 & \phantom{$-$}\nodata & 37 & Calzetti (2000) & 0.7 & 8.4 & 10.3 & $-$8.18 & 0.020 & 0.59 \\[-5pt]
03:32:18.912 & $-$27:41:37.48 & 2.3060 & 25.08 & $-$19.80 & \phantom{$-$}0.00 & 0 & Calzetti (2000) & 0.3 & 7.6 & 8.89 & $-$1.88 & 0.004 & 5.79 \\[-5pt]
03:32:19.028 & $-$27:41:28.92 & 2.4352 & 24.97 & $-$19.10 & \phantom{$-$}0.27 & 0 & Kriek \& Conroy (2013) & 0.6 & 8.1 & 9.45 & \phantom{$-$}0.73 & 0.050 & 0.06 \\[-5pt]
03:32:19.606 & $-$27:48:40.01 & 3.7080 & 25.91 & $-$20.28 & \phantom{$-$}0.35 & 0 & \nodata & 0.0 & 8.5 & 8.69 & \phantom{$-$}0.61 & 0.004 & 1.29 \\[-5pt]
03:32:20.166 & $-$27:49:24.34 & 3.6580 & 26.11 & $-$20.02 & \phantom{$-$}0.20 & 0 & \nodata & 0.0 & 9.0 & 9.01 & \phantom{$-$}0.47 & 0.004 & 2.53 \\[-5pt]
03:32:20.202 & $-$27:50:17.30 & 2.8108 & 25.06 & $-$20.24 & \phantom{$-$}0.18 & 0 & SMC & 0.1 & 8.8 & 9.22 & \phantom{$-$}0.84 & 0.050 & 1.46 \\[-5pt]
03:32:20.876 & $-$27:49:16.08 & 2.4845 & 25.28 & $-$20.39 & $-$0.03 & 0 & Calzetti (2000) & 0.1 & 7.0 & 7.96 & \phantom{$-$}0.84 & 0.020 & 2.65 \\[-5pt]
03:32:21.578 & $-$27:52:21.34 & 3.2051 & 25.91 & $-$20.00 & \phantom{$-$}0.07 & 0 & \nodata & 0.0 & 9.2 & 9.3 & \phantom{$-$}0.44 & 0.004 & 2.18 \\[-5pt]
03:32:22.594 & $-$27:51:17.99 & 3.6600 & 26.94 & $-$19.16 & \phantom{$-$}0.51 & 0 & \nodata & 0.0 & 9.2 & 8.85 & \phantom{$-$}0.11 & 0.004 & 5.96 \\[-5pt]
03:32:22.656 & $-$27:51:22.22 & 3.7060 & 27.08 & $-$19.08 & \phantom{$-$}0.58 & 0 & \nodata & 0.0 & 9.2 & 8.82 & \phantom{$-$}0.08 & 0.004 & 17.2 \\[-5pt]
03:32:22.892 & $-$27:41:15.09 & 3.3565 & 25.56 & $-$20.31 & \phantom{$-$}0.22 & 0 & \nodata & 0.0 & 9.0 & 9.4 & \phantom{$-$}0.54 & 0.004 & 5.35 \\[-5pt]
03:32:23.240 & $-$27:51:57.86 & 3.4700 & 25.69 & $-$19.92 & \phantom{$-$}0.61 & 0 & Calzetti (2000) & 0.5 & 9.1 & 10.1 & \phantom{$-$}0.87 & 0.004 & 0.74 \\[-5pt]
03:32:23.335 & $-$27:51:56.85 & 3.4700 & 23.99 & $-$21.87 & \phantom{$-$}0.60 & 0 & \nodata & 0.0 & 8.5 & 10.1 & \phantom{$-$}0.97 & 0.008 & 0.52 \\[-5pt]
03:32:24.196 & $-$27:42:57.55 & 2.2980 & 23.60 & $-$19.91 & \phantom{$-$}0.12 & 10 & Calzetti (2000) & 0.7 & 8.8 & 10.2 & \phantom{$-$}1.09 & 0.020 & 0.61 \\[-5pt]
03:32:24.348 & $-$27:49:57.66 & 2.6885 & 25.01 & $-$20.08 & \phantom{$-$}0.16 & 0 & Calzetti (2000) & 0.5 & 8.8 & 9.49 & \phantom{$-$}1.11 & 0.050 & 0.92 \\[-5pt]
03:32:24.668 & $-$27:45:33.72 & 2.3115 & 25.03 & $-$20.62 & $-$0.07 & 0 & \nodata & 0.0 & 8.1 & 8.65 & \phantom{$-$}0.75 & 0.004 & 20.1 \\[-5pt]
03:32:26.477 & $-$27:50:44.14 & 3.0604 & 25.07 & $-$20.37 & \phantom{$-$}0.40 & 0 & \nodata & 0.0 & 8.1 & 9.55 & $-$1.63 & 0.020 & 0.83 \\[-5pt]
03:32:27.120 & $-$27:44:43.98 & 2.3230 & 24.78 & $-$19.38 & \phantom{$-$}0.08 & 0 & Calzetti (2000) & 0.8 & 7.7 & 9.35 & $-$1.51 & 0.004 & 0.7 \\[-5pt]
03:32:27.282 & $-$27:48:45.82 & 2.6310 & 25.07 & $-$20.32 & \phantom{$-$}0.05 & 0 & \nodata & 0.0 & 9.0 & 9.24 & \phantom{$-$}0.70 & 0.050 & 2.46 \\[-5pt]
03:32:28.279 & $-$27:44:03.50 & 3.2560 & 24.26 & $-$20.45 & \phantom{$-$}\nodata & 31 & Calzetti (2000) & 3.0 & 9.0 & 11.8 & \phantom{$-$}2.84 & 0.020 & 1.23 \\[-5pt]
03:32:28.426 & $-$27:48:19.13 & 2.6270 & 24.68 & $-$20.80 & \phantom{$-$}0.06 & 0 & LMC & 0.5 & 7.0 & 8.75 & \phantom{$-$}1.48 & 0.004 & 1.02 \\[-5pt]
03:32:28.946 & $-$27:44:11.71 & 3.3680 & 25.11 & $-$21.04 & \phantom{$-$}0.25 & 0 & \nodata & 0.0 & 8.7 & 9.19 & \phantom{$-$}0.89 & 0.004 & 4.43 \\[-5pt]
03:32:29.139 & $-$27:48:52.63 & 3.5970 & 25.04 & $-$21.00 & \phantom{$-$}0.40 & 0 & \nodata & 0.0 & 9.2 & 9.59 & \phantom{$-$}0.85 & 0.004 & 1.42 \\[-5pt]
03:32:29.187 & $-$27:40:22.52 & 3.3083 & 25.58 & $-$20.50 & \phantom{$-$}0.17 & 0 & \nodata & 0.0 & 8.8 & 9.07 & \phantom{$-$}0.66 & 0.004 & 1.2 \\[-5pt]
03:32:29.932 & $-$27:49:28.26 & 3.2293 & 24.88 & $-$20.92 & \phantom{$-$}0.30 & 0 & \nodata & 0.0 & 9.1 & 9.85 & \phantom{$-$}0.80 & 0.008 & 0.12 \\[-5pt]
03:32:31.612 & $-$27:40:08.46 & 3.0762 & 24.30 & $-$21.36 & \phantom{$-$}0.15 & 0 & \nodata & 0.0 & 8.6 & 9.57 & \phantom{$-$}1.08 & 0.050 & 1.47 \\[-5pt]
03:32:31.863 & $-$27:49:50.13 & 2.5781 & 24.63 & $-$20.26 & \phantom{$-$}0.20 & 0 & Calzetti (2000) & 1.5 & 6.7 & 9.3 & \phantom{$-$}2.81 & 0.004 & 1.4 \\[-5pt]
03:32:32.294 & $-$27:41:26.36 & 2.3200 & 23.73 & $-$21.33 & \phantom{$-$}0.05 & 0 & Calzetti (2000) & 0.6 & 7.1 & 8.96 & \phantom{$-$}1.41 & 0.008 & 8.39 \\[-5pt]
03:32:32.357 & $-$27:42:48.09 & 2.4300 & 24.53 & $-$19.48 & \phantom{$-$}0.15 & 0 & Calzetti (2000) & 2.1 & 6.6 & 9.64 & \phantom{$-$}3.34 & 0.004 & 2.16 \\[-5pt]
03:32:32.486 & $-$27:48:52.08 & 2.8179 & 24.88 & $-$20.45 & \phantom{$-$}0.14 & 0 & SMC & 0.1 & 9.3 & 9.87 & \phantom{$-$}0.92 & 0.050 & 1.06 \\[-5pt]
03:32:33.302 & $-$27:42:01.69 & 2.4500 & 24.39 & $-$19.67 & \phantom{$-$}0.33 & 57 & Calzetti (2000) & 1.5 & 6.4 & 8.98 & \phantom{$-$}2.87 & 0.020 & 0.32 \\[-5pt]
03:32:33.325 & $-$27:50:07.33 & 3.7910 & 25.34 & $-$20.93 & \phantom{$-$}0.32 & 0 & Calzetti (2000) & 0.1 & 7.0 & 8.18 & \phantom{$-$}1.06 & 0.004 & 4.18 \\[-5pt]
03:32:33.778 & $-$27:48:14.35 & 2.6182 & 25.00 & $-$20.55 & $-$0.05 & 0 & LMC & 0.3 & 6.9 & 8.28 & \phantom{$-$}1.25 & 0.020 & 3.78 \\[-5pt]
03:32:34.230 & $-$27:49:35.54 & 3.5705 & 25.85 & $-$20.43 & \phantom{$-$}0.11 & 0 & \nodata & 0.0 & 7.0 & 7.9 & \phantom{$-$}0.63 & 0.004 & 9.1 \\[-5pt]
03:32:34.348 & $-$27:48:55.79 & 4.1420 & 25.33 & $-$20.66 & \phantom{$-$}1.06 & 0 & LMC & 0.2 & 8.8 & 9.93 & \phantom{$-$}1.01 & 0.020 & 0.45 \\[-5pt]
03:32:34.677 & $-$27:43:41.06 & 2.2760 & 25.25 & $-$19.42 & \phantom{$-$}0.10 & 0 & LMC & 0.4 & 7.6 & 8.99 & $-$1.78 & 0.004 & 6 \\[-5pt]
03:32:34.903 & $-$27:41:52.83 & 2.3129 & 25.26 & $-$20.25 & $-$0.06 & 0 & Calzetti (2000) & 0.2 & 7.1 & 8.12 & \phantom{$-$}0.57 & 0.008 & 17.9 \\[-5pt]
03:32:34.949 & $-$27:41:06.90 & 2.7708 & 25.16 & $-$20.41 & \phantom{$-$}0.09 & 0 & \nodata & 0.0 & 8.8 & 9.12 & \phantom{$-$}0.73 & 0.050 & 1.27 \\[-5pt]
03:32:35.051 & $-$27:48:23.19 & 2.7970 & 24.82 & $-$19.73 & \phantom{$-$}0.31 & 0 & MW & 0.9 & 7.9 & 10.3 & $-$3.35 & 0.008 & 1.77 \\\hline\pagebreak
\caption{Continued}\\
\toprule \\ [-19pt]
\colhead{RA (J2000)} & \colhead{Dec (J2000)}  & \colhead{$z_{\mathrm{spec}}$} & \colhead{$m_{J,{\rm AB}}$} & \colhead{$M_{\!1500\mathrm{\mathring{A}}}^{\mathrm{AB}}$} & \colhead{($V\!-\!I$)} & \colhead{$\mathrm{f_{AGN}}$} & \colhead{Dust\ Law} & \colhead{$\mathrm{A_V}$} & \colhead{Age} & \colhead{log($\mathrm{M_{\star}/M_{\odot}}$)} & \colhead{$\mathrm{SFR}$} & $Z$ & $\chi^2$ \\ [-5pt]
\colhead{[hh:mm:ss]}  & \colhead{[dd:mm:ss]} & \colhead{\null} & 
 \colhead{[mag]} & \colhead{[mag]} & \colhead{[mag]} & \colhead{[\%]} & \colhead{\null} & \colhead{[mag]} & \colhead{[log(yr)]} & \colhead{\null} & \colhead{[log($\mathrm{M_{\odot}/yr}$)]} & \colhead{\null} & \colhead{\null} \\ [2pt]
\midrule  \\ [-19pt]
03:32:35.717 & $-$27:49:16.01 & 2.5760 & 25.89 & $-$18.75 & \phantom{$-$}0.19 & 16 & Calzetti (2000) & 0.3 & 9.1 & 9.9 & \phantom{$-$}0.25 & 0.020 & 0.51 \\[-5pt]
03:32:35.814 & $-$27:44:54.92 & 2.5658 & 24.76 & $-$20.69 & $-$0.04 & 0 & Kriek \& Conroy (2013) & 0.6 & 6.8 & 8.71 & \phantom{$-$}2.18 & 0.004 & 1.53 \\[-5pt]
03:32:35.956 & $-$27:41:49.97 & 3.6180 & 24.72 & $-$21.50 & \phantom{$-$}0.30 & 0 & \nodata & 0.0 & 8.6 & 9.26 & \phantom{$-$}1.09 & 0.004 & 0.42 \\[-5pt]
03:32:36.310 & $-$27:44:34.48 & 3.4173 & 24.52 & $-$21.18 & \phantom{$-$}0.43 & 0 & \nodata & 0.0 & 8.3 & 9.89 & \phantom{$-$}0.10 & 0.008 & 0.1 \\[-5pt]
03:32:36.824 & $-$27:45:58.01 & 3.7970 & 25.21 & $-$20.87 & \phantom{$-$}0.53 & 0 & \nodata & 0.0 & 8.5 & 9.36 & \phantom{$-$}0.72 & 0.008 & 0.35 \\[-5pt]
03:32:37.783 & $-$27:42:32.46 & 2.9752 & 25.17 & $-$20.28 & \phantom{$-$}0.23 & 0 & SMC & 0.1 & 8.7 & 9.25 & \phantom{$-$}0.85 & 0.050 & 1.34 \\[-5pt]
03:32:38.139 & $-$27:44:33.63 & 2.3277 & 26.29 & $-$18.08 & \phantom{$-$}0.12 & 0 & MW & 0.5 & 8.1 & 9.22 & $-$0.37 & 0.020 & 1.71 \\[-5pt]
03:32:38.869 & $-$27:43:21.57 & 2.4490 & 24.24 & $-$20.65 & \phantom{$-$}0.08 & 0 & Calzetti (2000) & 0.3 & 7.6 & 9.23 & \phantom{$-$}0.08 & 0.020 & 0.1 \\[-5pt]
03:32:39.535 & $-$27:48:28.45 & 2.8218 & 25.05 & $-$20.22 & \phantom{$-$}0.18 & 0 & SMC & 0.1 & 8.6 & 9.38 & \phantom{$-$}0.80 & 0.050 & 0.43 \\[-5pt]
03:32:39.739 & $-$27:47:17.41 & 2.8414 & 25.03 & $-$20.43 & \phantom{$-$}0.12 & 0 & \nodata & 0.0 & 8.8 & 9.48 & \phantom{$-$}0.71 & 0.050 & 0.72 \\[-5pt]
03:32:39.897 & $-$27:49:11.34 & 3.4192 & 27.04 & $-$18.99 & \phantom{$-$}0.31 & 0 & \nodata & 0.0 & 9.2 & 9.28 & \phantom{$-$}0.00 & 0.004 & 12.4 \\[-5pt]
03:32:40.319 & $-$27:42:40.86 & 2.5845 & 24.71 & $-$20.55 & \phantom{$-$}0.04 & 0 & \nodata & 0.0 & 7.4 & 8.76 & $-$0.20 & 0.050 & 1.28 \\[-5pt]
03:32:40.384 & $-$27:44:30.98 & 4.1200 & 26.00 & $-$20.46 & \phantom{$-$}0.44 & 0 & Calzetti (2000) & 0.3 & 6.2 & 8.23 & \phantom{$-$}2.32 & 0.004 & 2.46 \\[-5pt]
03:32:40.699 & $-$27:49:36.69 & 4.0279 & 27.42 & $-$18.81 & \phantom{$-$}0.47 & 0 & \nodata & 0.0 & 8.8 & 8.35 & \phantom{$-$}0.00 & 0.004 & 21 \\[-5pt]
03:32:40.945 & $-$27:49:25.58 & 2.6918 & 25.81 & $-$19.38 & \phantom{$-$}0.06 & 0 & \nodata & 0.0 & 8.0 & 8.86 & $-$0.14 & 0.050 & 3.04 \\[-5pt]
03:32:41.866 & $-$27:43:59.87 & 2.5760 & 25.76 & $-$19.62 & \phantom{$-$}0.23 & 1 & Calzetti (2000) & 0.5 & 7.2 & 8.36 & \phantom{$-$}0.45 & 0.020 & 7.26 \\[-5pt]
03:32:42.788 & $-$27:48:56.91 & 2.6933 & 25.97 & $-$19.01 & \phantom{$-$}0.32 & 0 & Calzetti (2000) & 0.7 & 8.7 & 9.2 & \phantom{$-$}0.89 & 0.050 & 1.57 \\[-5pt]
03:32:42.836 & $-$27:47:02.51 & 3.1930 & 25.04 & $-$20.21 & \phantom{$-$}\nodata & 97 & \nodata & 0.0 & 9.3 & 9.41 & $<\!-$4 & 0.020 & 9.06 \\[-5pt]
03:32:43.314 & $-$27:49:47.06 & 3.3329 & 25.07 & $-$20.88 & \phantom{$-$}0.19 & 0 & \nodata & 0.0 & 9.2 & 9.54 & \phantom{$-$}0.80 & 0.004 & 0.86 \\[-5pt]
03:32:43.387 & $-$27:47:10.52 & 2.6915 & 24.99 & $-$20.65 & $-$0.05 & 0 & \nodata & 0.0 & 7.6 & 8.59 & \phantom{$-$}0.59 & 0.020 & 1.47 \\[-5pt]
03:32:43.556 & $-$27:49:54.45 & 2.6500 & 24.78 & $-$20.60 & \phantom{$-$}0.08 & 0 & Kriek \& Conroy (2013) & 0.1 & 8.6 & 9.11 & \phantom{$-$}0.95 & 0.050 & 0.61 \\[-5pt]
03:32:43.633 & $-$27:43:47.71 & 2.3168 & 24.49 & $-$20.12 & \phantom{$-$}0.16 & 0 & Calzetti (2000) & 0.8 & 7.7 & 9.6 & $-$0.34 & 0.004 & 1.29 \\[-5pt]
03:32:43.686 & $-$27:46:46.34 & 2.4680 & 24.64 & $-$20.12 & \phantom{$-$}0.14 & 0 & MW & 1.4 & 6.8 & 9.77 & \phantom{$-$}3.18 & 0.004 & 1.95 \\[-5pt]
03:32:43.872 & $-$27:42:18.97 & 2.3682 & 25.07 & $-$20.48 & $-$0.09 & 0 & \nodata & 0.0 & 8.0 & 8.65 & \phantom{$-$}0.61 & 0.004 & 2.95 \\[-5pt]
03:32:43.881 & $-$27:44:05.67 & 2.4284 & 24.41 & $-$20.41 & \phantom{$-$}0.32 & 0 & MW & 0.4 & 8.0 & 9.95 & $<\!-$4 & 0.004 & 4.26 \\[-5pt]
03:32:44.598 & $-$27:48:35.87 & 2.5860 & 25.01 & $-$19.97 & \phantom{$-$}0.38 & 0 & SMC & 0.2 & 9.1 & 9.64 & \phantom{$-$}0.90 & 0.050 & 3.3 \\[-5pt]
03:32:44.902 & $-$27:48:45.55 & 2.8038 & 25.47 & $-$19.41 & \phantom{$-$}0.28 & 0 & MW & 0.7 & 8.1 & 9.67 & \phantom{$-$}1.22 & 0.050 & 1.3 \\[-5pt]
03:32:45.147 & $-$27:50:28.06 & 3.3924 & 25.71 & $-$19.79 & \phantom{$-$}0.35 & 0 & SMC & 0.4 & 6.5 & 8.07 & \phantom{$-$}1.84 & 0.050 & 2.1 \\[-5pt]
03:32:45.180 & $-$27:49:45.67 & 2.5712 & 26.51 & $-$18.39 & \phantom{$-$}0.22 & 0 & Calzetti (2000) & 1.3 & 7.2 & 8.43 & \phantom{$-$}1.51 & 0.004 & 1.9 \\[-5pt]
03:32:45.391 & $-$27:50:10.59 & 3.3188 & 26.76 & $-$19.02 & $-$0.04 & 0 & Kriek \& Conroy (2013) & 0.1 & 8.6 & 8.45 & \phantom{$-$}0.21 & 0.004 & 5.29 \\[-5pt]
03:32:46.245 & $-$27:48:46.97 & 4.0200 & 25.25 & $-$20.98 & \phantom{$-$}0.74 & 0 & \nodata & 0.0 & 9.1 & 9.57 & \phantom{$-$}0.83 & 0.004 & 1.42 \\[-5pt]
03:32:46.331 & $-$27:50:52.97 & 2.8984 & 24.76 & $-$20.55 & \phantom{$-$}0.36 & 0 & SMC & 0.4 & 7.0 & 8.62 & \phantom{$-$}1.60 & 0.004 & 4.46 \\[-5pt]
03:32:46.937 & $-$27:50:04.43 & 2.9994 & 25.38 & $-$20.38 & \phantom{$-$}0.16 & 0 & \nodata & 0.0 & 8.8 & 9.04 & \phantom{$-$}0.68 & 0.020 & 3.1 \\[-5pt]
03:32:48.625 & $-$27:48:56.23 & 3.7090 & 26.15 & $-$20.12 & \phantom{$-$}0.40 & 0 & \nodata & 0.0 & 8.7 & 8.79 & \phantom{$-$}0.53 & 0.004 & 5.23 \\[-5pt]
03:32:48.785 & $-$27:49:17.97 & 2.6105 & 25.48 & $-$20.02 & \phantom{$-$}0.07 & 0 & LMC & 0.6 & 6.9 & 8.52 & \phantom{$-$}1.91 & 0.004 & 2.63 \\[-5pt]
03:32:52.905 & $-$27:48:31.50 & 3.6496 & 25.95 & $-$20.27 & \phantom{$-$}0.16 & 0 & \nodata & 0.0 & 8.2 & 8.46 & \phantom{$-$}0.65 & 0.004 & 7.94 \\[-5pt]
12:36:22.940 & \phantom{$-$}62:15:26.67 & 2.5920 & 21.25 & $-$24.69 & $-$0.15 & 56 & Calzetti (2000) & 0.1 & 6.8 & 9.52 & \phantom{$-$}2.69 & 0.020 & 7.44 \\[-5pt]
12:36:25.570 & \phantom{$-$}62:13:50.26 & 2.9320 & 24.96 & $-$20.40 & \phantom{$-$}0.24 & 0 & SMC & 0.1 & 8.5 & 9.36 & \phantom{$-$}0.86 & 0.050 & 2.08 \\[-5pt]
12:36:30.558 & \phantom{$-$}62:16:26.34 & 2.4845 & 24.78 & $-$20.69 & \phantom{$-$}0.01 & 0 & Calzetti (2000) & 0.3 & 6.8 & 8.11 & \phantom{$-$}1.28 & 0.050 & 0.19 \\[-5pt]
12:36:41.724 & \phantom{$-$}62:12:38.87 & 2.5890 & 25.56 & $-$19.93 & $-$0.02 & 0 & \nodata & 0.0 & 7.8 & 8.23 & \phantom{$-$}0.56 & 0.050 & 2.72 \\[-5pt]
12:36:59.373 & \phantom{$-$}62:09:31.37 & 2.9900 & 25.46 & $-$20.47 & \phantom{$-$}0.12 & 0 & Kriek \& Conroy (2013) & 0.3 & 7.0 & 8.3 & \phantom{$-$}1.03 & 0.004 & 3.72 \\[-5pt]
12:37:11.334 & \phantom{$-$}62:10:44.44 & 2.5965 & 25.47 & $-$20.34 & \phantom{$-$}0.04 & 0 & SMC & 0.1 & 6.4 & 7.77 & \phantom{$-$}1.55 & 0.050 & 16.8 \\[-5pt]
12:37:16.126 & \phantom{$-$}62:15:26.43 & 2.9560 & 25.73 & $-$19.45 & \phantom{$-$}0.33 & 0 & \nodata & 0.0 & 8.4 & 9.4 & \phantom{$-$}0.04 & 0.050 & 6.88 \\[-5pt]
12:37:28.113 & \phantom{$-$}62:14:40.03 & 2.5480 & 23.65 & $-$21.32 & \phantom{$-$}0.09 & 0 & MW & 0.2 & 8.9 & 10.1 & \phantom{$-$}1.37 & 0.050 & 0.09 \\[-5pt]
12:37:41.868 & \phantom{$-$}62:13:34.10 & 2.8635 & \nodata & $-$18.81 & \phantom{$-$}0.21 & 0 & \nodata & 0.0 & 8.2 & 8.88 & $-$0.30 & 0.050 & 0.66 \\[-3pt]
\bottomrule
\multicolumn{14}{l}{\textbf{Table columns:} The column header $\mathrm{f_{AGN}}$ indicates the percentage of light in the SED from the AGN at 5000$\mathrm{\mathring{A}}$ and log($\mathrm{M_{\star}/M_{\odot}}$) indicates the}\\
\multicolumn{14}{l}{stellar mass of the galaxy that produces it's best-fitting BC03 SED.}
\vspace*{3pt}
\end{longtable}

\end{document}